\documentstyle[12pt,epsfig,uthesis]{report}
\def\prl{{\em Phys. Rev. Lett. }}
\def\prc{{\em Phys. Rev. {\bf C} }}
\def\prd{{\em Phys. Rev. {\bf D} }}

\def\nima{{\em Nucl. Instr. and Meth. Phys. {\bf A} }}
\def\npa{{\em Nucl. Phys. {\bf A}}}
\def\npb{{\em Nucl. Phys. {\bf B}}}
\def\epjc{{\em Eur. Phys. J. {\bf C}}}
\def\plb{{\em Phys. Lett. {\bf B}}}
\def\mpla{{\em Mod. Phys. Lett. {\bf A}}}
\def\pr{{\em Phys. Rep.}}
\def\zpc{{\em Z. Phys. {\bf C}}}
\def\zpa{{\em Z. Phys. {\bf A}}}
\def\jpg{{\em J. Phys. G: Nucl. Part. Phys. }}
\def\cpc{{\em Comput. Phys. Commun.}}
\def\rpp{{\em Rep. Prog. Phys.}}
\def\ijmpa{{\em Int. J. Mod. Phys.{\bf A}}}
\input{psfig.sty}
\catcode`\@=11
\@addtoreset{equation}{chapter}
\@addtoreset{table}{chapter}

\renewcommand\thetable{\thechapter.\@arabic\c@table}
\@addtoreset{figure}{chapter}

\begin{document}
%\baselineskip 12pt
% version 24.8.96
\title{\bf 
TRANSVERSE ENERGY MEASUREMENT AND FLUCTUATION STUDIES \\ 
IN ULTRA-RELATIVISTIC HEAVY ION COLLISIONS}

\author{\sl \bf  RAGHUNATH SAHOO}
\dept{\bf UTKAL UNIVERSITY}
\copyrightfalse
\figurespagefalse 
\tablespagefalse
\beforepreface
\prefacesection{\bf CERTIFICATE}
This is to certify that the thesis entitled {\bf `` Transverse energy measurement and fluctuation studies in ultra-relativistic heavy ion collisions''}
which is being submitted by {\bf Mr. Raghunath Sahoo} in partial
fulfillment of the degree of 
{\bf Doctor of Philosophy in Science (Physics) of Utkal University,
Bhubaneswar} is a record of his own research work carried out by
him. He has carried out his investigations for the last five and half
years on the subject matter of the thesis under my guidance at
{\bf Institute of Physics, Bhubaneswar}. The matter embodied in the
thesis has not been submitted for the award of any other degree
by him or by anybody else.

\vspace{1.5in}
\begin{center}
${\overline {Prof.\  D.P. \; Mahapatra}}$\\
%Professor\\
Institute of Physics\\
Bhubaneswar 751005
\end{center}

\prefacesection{\bf DECLARATION}
I, Raghunath Sahoo hereby declare that the thesis entitled {\bf `` Transverse energy measurement and fluctuation studies in ultra-relativistic heavy ion collisions''}
which is being submitted by me in partial fulfillment of the degree of 
{\bf Doctor of Philosophy in Science (Physics) of Utkal University,
Bhubaneswar} is a record of my own research work carried out by
me. I have carried out my investigations for the last five and half
years on the subject matter of the thesis under the guidance of 
Prof. D. P. Mahapatra, at {\bf Institute of Physics, Bhubaneswar}. 
The matter embodied in the thesis has not been submitted to any other University/Institute
 for the award of any other degree by me.

\vspace{1.5in}
\begin{center}
${\overline {Raghunath \; Sahoo}}$\\
Institute of Physics\\
Bhubaneswar 751005
\end{center}

%%\prefacesection{}

%\begin{center}
%${}$ \\
%\vspace{3.5cm}
%{\Large \it Dedicated \\
%\vspace{0.5cm}
%To \\
%\vspace{0.5cm}
%My Parents }
%\end{center}
%%\tableofcontents

\prefacesection{Acknowledgments}

I take this opportunity to thank all of my friends, well-wishers and
family members for their constant support, advice and encouragements. 
This thesis would not have been possible without the support of these
people. 

First, I would like to express my deep sense of gratitude to 
Prof. D.P.~Mahapatra, my thesis supervisor at Institute of Physics, 
for his invaluable guidance, encouragement and his constant support 
throughout my research period. He has been with me like a family member.

I also place on record my sincere thanks to Dr. Y.P.~Viyogi, the PMD
project leader and Director of Institute of Physics, for his support 
and useful advices throughout my research period. I have learnt a lot
from him, specially his patience and his continuous effort till one gets
results. It is a pleasure and joyous experience in working with him 
during the ALICE PMD test beam-time in CERN and PMD installation period 
at BNL, USA.

It is my pleasure to thank my collaborators: Dr. Subhasis Chattopadhyaya, 
VECC, Kolkata for giving me his valuable time and useful suggestions during
the analysis work. He was very instrumental in pushing this analysis for
a success. I am greatly indebted to him for his free sharing of insights
and constructive criticism during my research. I have learnt a lot on the 
technical details of the Barrel Electromagnetic Calorimeter from 
Prof. A.A.P. Suaide and Dr. (Mrs) Marcia Maria de Moura of the university 
of Sao Paulo, Bazil. They have been always a source of inspiration and help 
for me, during the time of data analysis and whenever needed.

I would like to express my heartiest thanks to Prof. B.K. Nandi, IITB;  
who has been always with me like my elder brother with his all time support 
and encouragements. He has stood with me in all my bad phases during my
research period, giving me courage to stand on my own. I have learnt a
lot from him, starting from the basics of ROOT and modular programming
to any dipper physics issues. I also express my sincere thanks to 
Dr. T.K.~Nayak for his support and encouragement he has extended for me. 
I thank Dr. Bedangadas Mohanty for his timely helps through out my Ph.D. 
period. His sincerity in work and pursuing any analysis in hand without 
giving it up, is of worth learning. He has been a source of inspiration for me.

I am indebted to Prof. S.N.~Behera and Dr. R.K.~Choudhury, past directors 
of IOP for their help and supports in all respects. I would like to express 
my sincere thanks to Prof. Bikash Sinha, Director VECC and SINP for allowing 
me to work at VECC and use the facilities there, whenever needed. 
His hospitality and encouragements are always to remember. My special thanks 
goes to Dr. D.K.~ Srivastava, Head Physics Group, VECC, Kolkata for his 
advice, support, constant encouragements, useful discussions on my analysis 
results and good wishes for me. I thank Prof. J. Cleymans, University of 
Cape Town, for useful discussions. I also thank Prof. S.C.~Phatak  and 
Prof. A.M. Srivastava for many interesting discussions I had with them. 
I will be incomplete without thanking Prof. R. Varma, IITB. He has 
extended his all time 
encouragements and suggestions during my interaction with him. He has 
been more like a friend-cum-teacher for me. I am also very thankful to 
Dr. Jan-e Alam, VECC, Kolkata for giving useful suggestions and helping 
me in better understanding of QGP and related physics. It was pleasant 
and joyful in interacting with him both from physics and personal fronts. 

I would like to thank the whole STAR collaboration and the PMD group in 
particular for providing a place for me to work and learn. I have learnt many
things from all of our PMD collaborators. It is a great pleasure in 
interacting in the Spectra Physics Working Group while presenting the 
analysis results. Working in such a big collaboration has made my 
capabilities many-folds in terms of presenting myself to bigger and diverging
communities. It was a pleasant experience to work with Dr. Anand Kumar Dubey 
and Dr. Dipak Kumar Mishra, my seniors and collaborators at IOP. I remember 
the moments I have spent with Dipak bhai discussing various problems 
while working in the collaboration. I would also like to thank Mr. Ajay Kumar 
Dash and Mr. Chitrasen Jena for their company while working at IoP.

It was equally interesting to work closely with other members of the
PMD group. It includes Dr. M.M. Aggarwal, Dr. A.K. Bhati and their 
group in Panjab University, Chandigarh; Dr. Sudhir Raniwala and
Dr. Rashmi Raniwala from University of Rajasthan, Jaipur;
Dr. S.K. Badyal, Dr. L.K. Mangotra, Dr. Saroja K. Nayak, Dr. P.V.K.S. Baba 
and their group at Jammu University, Jammu; Dr. Zubayer Ahmed,
Mr. G.S.N. Murthy, Mr. R.N.~Singaraju, Dr. P. Ghosh, Mr. S. Pal and Mr. M.R. 
Dutta Majumdar of VECC Kolkata. I feel I was very fortunate to work with 
a number of research scholar friends in the collaboration from the above 
institutions. I thank them for their help, co-operation and good wishes.

I take this opportunity to thank my seniors, scholar friends and others at IoP.
They have made my stay in the beautiful campus of IOP a memorable and 
pleasing one. It will be impossible for me to forget the picnics, pujas, 
cricket matches, tennis matches and dinner, lunch, breakfast and tea-time 
interactions with them. I thank them for all their direct and indirect help 
throughout my research period at IOP. Special thanks to my friend-cum-elder
brother Hara bhai for his company and friendship.

I must mention that, one of the most interesting, knowledgeable and
difficult course I have attend is the Pre-Doctoral course at IOP.
I thank all my teachers who taught me during the pre-doctoral course
and other faculty members of IOP.

I am also thankful to all the staff of the beautiful IOP library, the 
computer center and all the administrative and non-academic staff of 
IOP, for their co-operation and help at every stage.  

My sincere regards and due respect to my teachers, Prof. N. Barik, 
Prof. L. P Singh, Prof. S. Mohanty, Prof. L. Maharana,  Prof. D.K. Basa,
Prof. N.C. Mishra, Prof. (Mrs) P. Khare, Dr. K. Maharana, and 
Dr. (Mrs) S. Mahapatra at Utkal University, Dr. A.R. Panda and 
Dr. Sk.Samsur, Dr. J.P. Roy, Purusottama Das who taught me at Kendrapara 
college. It is because of their teachings, constant encouragement and 
their blessings, that I have come this far in my research career. 
I am very pleased in getting a family friend like Mr. Baya Prasad Sahoo, 
who has been with me in all situations.

My humble regards, respect and thanks goes to my grandfather, 
late grandmother, parents and in-laws for their patience and  
encouragement during my research period. It is because of their love, 
affection and blessings that I could complete this thesis. 
I also feel delighted to note the love and affections I have got from my
brothers and sisters. The person who has been with me at all times 
giving me her love, affection, encouragements and co-operations is no 
one except my wife, Gayatri. She has been a driving force in me.

\vspace{1.1in}
{\bf Date:}
\hspace{2.8in}
{\bf (Raghunath Sahoo)}

%\prefacesection{List of Publications}
%\begin{enumerate}

%\item[\large{*}1.]
%\normalsize {\it {\bf Event-by-Event Fluctuations in Particle Multiplicities and 
 %        Transverse energy Produced in 158 A GeV Pb + Pb collisions.}}\\
 %       M. M. Aggarwal et. al., (WA98 Collaboration),\\ 
  %      Submitted to Physical Review, e-print : nucl-ex/0108029.

%\end{itemize}
%\end{enumerate}
%\end{itemize}

% {\it * Results from these publications are included in this thesis.} \\

%\noindent \Large {\bf Publications in Conferences, Symposia and Internal notes}\\
%\normalsize

\afterpreface
\listoffigures
\listoftables
{\bf}\chapter{Introduction to Physics of Quark Gluon Plasma}
\markboth{nothing}{\it Introduction to Physics of QGP}

\section{Introduction}

The quest to understand the origin of mass, the fundamental particles and
their interactions have been a driving force in particle physics research.
The nature is governed by four fundamental interactions -  the strong, the 
weak, the electromagnetic and the gravitational interactions. Each of these 
fundamental interactions are mediated by an exchange particle and the 
relative strength of the interaction is determined by a characteristic 
coupling. The most interesting interaction is the strong force which is 
mediated by color charged gluons unlike the electromagnetic force which is 
mediated by charge-less photons. Thus the theory of strong interaction, the 
Quantum Chromodynamics (QCD) is more challenging and exciting than the theory 
of Quantum Electrodynamics (QED). The structure of nucleons or more generally 
the structure of hadrons is understood in terms of elementary particles called
``quarks'' that interact so strongly that they could only be observed in 
``color-neutral'' groups of two (the mesons), three (the baryons) and perhaps 
five of them. The complicacy of the theory comes through the self interaction 
of gluons which is also strong in nature.

In 1975, Collins and Perry \cite{collins} predicted that at high density 
$\sim$ few times normal nuclear matter density, the properties of nuclear 
matter are not governed by the hadronic degrees of freedom but by quark 
degrees of freedom. Furthermore, lattice QCD calculations suggest that at 
high temperature ($\sim 150-200$) MeV, there will be a phase transition from 
the confined state of hadrons to the deconfined state of quarks, anti-quarks 
and gluons, called the Quark Gluon Plasma (QGP). It is expected that 
nucleus-nucleus collisions at ultra-relativistic energies will be able to 
create such a deconfined state of matter.

QGP is a thermalized state of matter with overall color neutrality and it's 
properties are governed by quarks, anti-quarks and gluons which are normally 
confined within hadrons. In QGP, the inter-particle interaction is much less 
than the average kinetic energy of the particles. This is one of the properties
 of a weakly interacting plasma.

\section{QCD, Deconfinement and QGP}

QCD, the non-abelian gauge theory of colored quarks and gluons is the theory of 
strong interaction. The color charge is associated with the non-abelian gauge 
group $SU(3)_c$ with the quarks carrying three color charges. The gluons come 
in eight colors and the gluon fields exhibit self-interaction. QCD exhibits two 
remarkable features. At low energies and large length scales, QCD is a 
non-perturbative field theory. The effective coupling constant $\alpha_s(Q^2)$ is 
large and the quarks and gluons are permanently confined inside hadrons. This 
feature is usually explained by the postulate of ``color confinement'', which 
implies that all observable states are color singlets or colorless objects. Its 
behavior asymptotically approaches that of a non-interacting free-field theory 
in the high-energy (large momenta, $Q$) or short-distance scale limit where, 
$\alpha_s(Q^2)$ decreases logarithmically. This is commonly called 
``asymptotic freedom'' and it imply that quarks and gluons are weakly interacting 
at high energies and hence perturbative QCD can be a useful tool at this domain. 
\begin{figure}
\begin{center}
\epsfig{figure=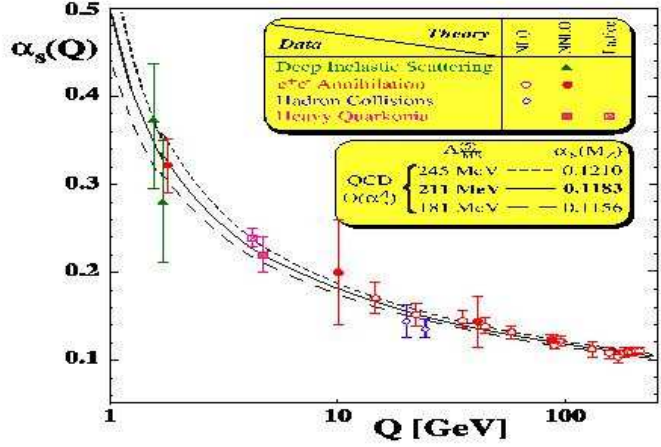,width=10cm}
\end{center}
\caption{\label{qcd}(a)The QCD running coupling constant $\alpha_s(Q^2)$ vs $Q$.}
\end{figure}
The QCD Lagrangian is given by
\begin{equation}
{{\mathcal L}_{QCD} = i\bar\psi\gamma^\mu(\partial_\mu - igA_\mu (x))\psi - 
m\bar\psi\psi-\frac{1}{4}F_{\mu\nu}^aF_a^{\mu\nu}}
\label{qcdL}
\end{equation}
where, the color potential $A_\mu(x)$ is a $3 \times 3$ matrix and can be 
represented by a linear combination of the 8 Gell-Mann matrices 
(the ``generators of $SU(3)$ group''):
\begin{equation}
{A_\mu (x) = \frac{1}{2}\sum_{a=1}^8 A_\mu^a(x)\lambda_a.}
\label{gelmann}
\end{equation}
The gauge field $A_{\mu}(x)$ at the space-time point $x$ can be considered as 
having eight degrees of freedom in color space with eight components: $A_{\mu}^a(x)$,
a=1,2, .....,8. This color potential is introduced to make the Lagrangian invariant 
under rotations of the color co-ordinate frame (three dimensional) at the same 
space-time point $x$ (called local gauge invariance; local in space-time). The 
eight component field strength tensor is given by
\begin{equation}
{F_{\mu\nu}^a = \partial_{\mu}A_{\nu}^a -\partial_{\nu}A_{\mu}^a + 
gf_{abc}A_{\mu}^bA_{\nu}^c,}
\label{field}
\end{equation}
where $f_{abc}$ are the antisymmetric structure constants for the Lie group SU(3). The
product $F_{\mu\nu}^aF_a^{\mu\nu}$ is also invariant under a local color gauge
transformation.
\begin{figure}
\begin{center}
\epsfig{figure=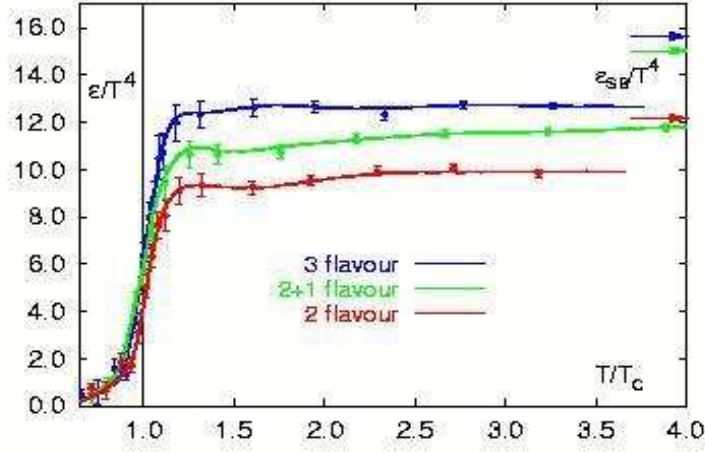,width=10cm}
\end{center}
\caption{\label{qcd1}Lattice calculations of the dependence of the density versus 
temperature.}
\end{figure}
The running coupling constant for strong interaction behaves like
\begin{equation}
{\alpha_s(Q^2) = \frac{1}{\beta_0 ln(Q^2/\Lambda^2)}}
\label{coupling}
\end{equation}
where $\Lambda$ is a dimensional parameter introduced in the renormalization
process and the energy scale where $\alpha_s(Q^2)$ diverges to infinity. $\beta_0$ 
is a constant that depends on the number of active quark flavors. As shown in 
Fig.~\ref{qcd}, the above formula explains the ``asymptotic freedom'' 
and ``quark confinement'' nature of QCD.
\begin{figure}
\begin{center}
\includegraphics[width=5.5in]{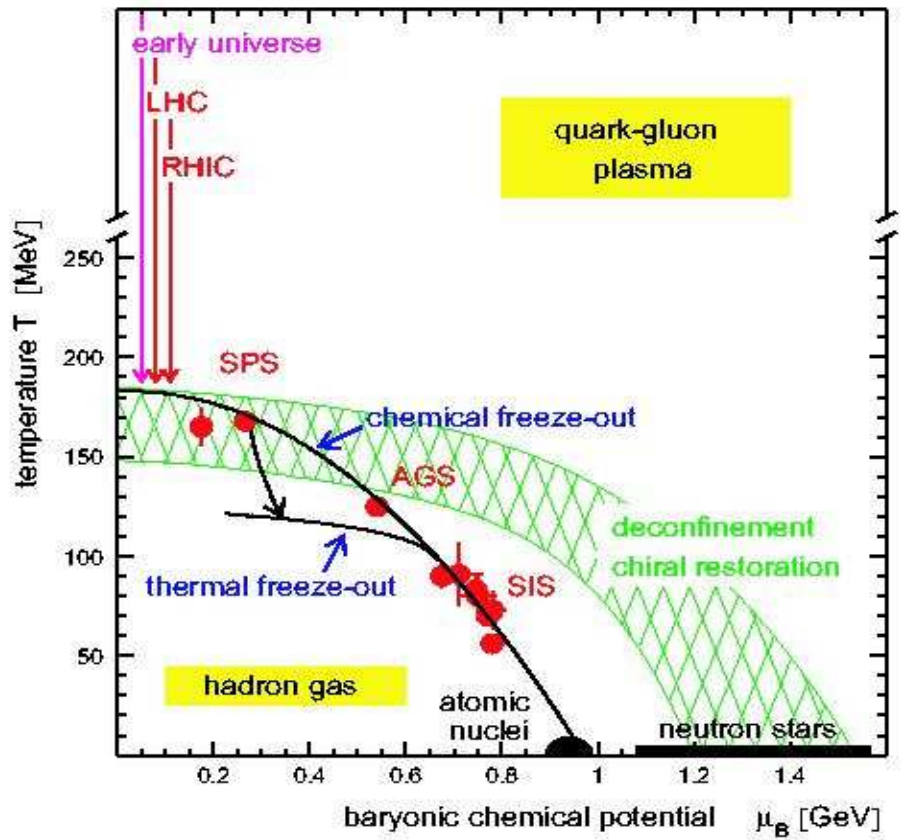}
\end{center}
\caption{\label{phaseDiagram}The phase diagram of hadronic matter, showing hadron gas
and QGP regions. The temperature T and the baryochemical potential $\mu_B$ data are
derived from particle yield ratios. The solid curve through the data points represents
the chemical freeze-out of hadronic-matter. Figure reference \cite{qcdPhase}.}
\end{figure}

What is the energy density and temperature required for the formation of QGP?

The energy density in a nucleon \cite{halzen} is
\begin{equation}
{\rho_E^N = \frac{m_N}{\frac{4}{3}\pi R^3} \simeq 0.5~GeV/fm^3},
\label{density}
\end{equation}
with $R\simeq 0.8 fm.$. In nucleus-nucleus collisions at relativistic
energies, large density results from heating through individual NN
interactions as well as the compression of the nuclei when traversing 
one another. In the collision of two nuclei at this energy, a quark 
will not be associated with its parent nucleon if the energy per quark
in the struck part (plasma) exceeds the energy per quark inside a free
nucleon. Therefore, quarks and gluons can't be identified anymore
with individual nucleons and we will have a transition to partonic
matter called QGP. According to this estimate, the energy density for 
QGP formation should be more than $0.5 GeV/fm^3$ ($\sim~ 3-4$ times the 
energy density corresponding to normal nuclear matter).

Lattice QCD calculations, considering two light quark flavors (Fig.\ref{qcd1}), 
predict a phase transition from a confined state of hadronic matter to a 
deconfined state of quarks and gluons (QGP), at a temperature around 
$\sim 150-200$ MeV \cite{karsch}. As shown in the phase diagram 
(Fig.\ref{phaseDiagram}), the phase transition is expected to take place at 
either high temperature or at large baryon chemical potential. The figure 
shows regions probed by different beam energies as have been obtained from 
various heavy ion accelerators. The non-zero baryon chemical potential and 
the high temperature region is the one probed by the highest energy collisions
at the Relativistic Heavy Ion Collider (RHIC) located at the Brookhaven
National Laboratory (BNL), USA. The region where $\mu_B = 0$ is believed to 
be one in which the early universe existed, and is also accessible to
numerical simulations of QCD on a lattice.

The deconfinement in QGP is also explained by the Debye screening. The mesons 
and the hadrons could be defined as the bound states of $q$'s and $\bar{q}$'s. 
In the presence of a large number of $q$'s and $\bar{q}$'s, however the mutual 
$q-\bar{q}$ or $q-q$ interactions are screened and weakened. This is analogous 
to the case corresponding to the screening of Coulomb interaction in the 
presence of large number of electrons. When the temperature is increased, the 
number of $q$'s and $\bar{q}$'s increases and, consequently, the screening 
radius decreases. At sufficiently high temperature, if the screening radius 
becomes shorter than the hadron radius itself, then the $q-\bar{q}$ system can 
no longer be bound and thus, is deconfined. This is also the reason, why the 
system is called a plasma.

QCD predicts the existence of a Quark-Gluon-Plasma \cite{shuryak, gavai} at 
extreme nuclear densities or at extremely large temperature. In the new phase, 
hadrons dissolve, strong interactions become very weak and an ideal 
color-conducting plasma of quarks and gluons is formed. In QGP, the long 
range color force is Debye screened due to collective effects in the same way 
as it happen in the case of electromagnetic plasma. Thus the quarks in QGP 
can only interact via a short-range effective potential and they become almost
free and deconfined. Although the quarks can still be sensitive to the 
perturbative part of the interaction, non-perturbative effects almost 
disappear. One can think heuristically of weak- and strong-coupling domains 
of QCD as two phases: the high energy phase of color-conducting QGP and the 
low energy phase of color insulators i.e. hadrons \cite{cpSingh}. 

Ultra relativistic nuclear collisions have given us a method to search for QGP
formation in the laboratory \cite{kajantieM}. A large number of particles produced 
in a finite volume of the collision suggests the existence of a large energy 
density. 

\section{Possible Signatures of QGP}

The central problem connected with ultra relativistic heavy ion collisions is 
to search for the evidence for the QGP formation and thus to deduce the energies, 
entropies and temperatures of the system formed in central collisions. Although 
the existence of a hot and dense phase of QCD is of no doubt, its observability 
in nuclear collisions is a matter of intense debate. As the hot and dense matter 
is initially formed for a brief time interval, it subsequently expands and cools 
substantially beyond the confining point $T_c$ before freeze-out, when the 
particles leave the fireball and reach the detectors. The problem becomes more 
intriguing because hot, excited hadronic matter essentially consists of a 
quark-gluon system (QGS). After a phase transition, it turns to QGP involving
collective behavior. So the observation of QGP essentially means making a 
distinction between signals arising from QGP and hot QGS respectively.
\begin{figure}
\begin{center}
\epsfig{figure=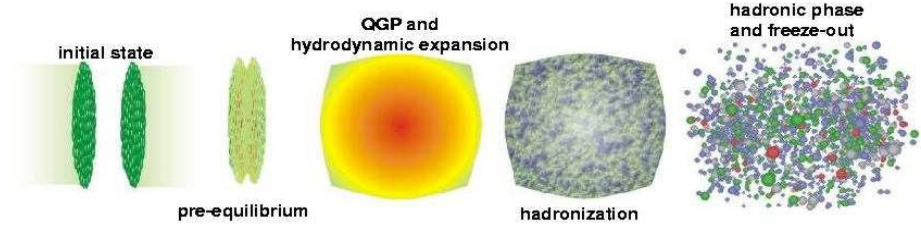,width=14cm}
\end{center}
\caption{\label{collision}Cartoon showing the initial state of ultra-relativistic
nuclear collision and time evolution of the fireball.}
\end{figure}
There are many proposed possible experimental signatures of QGP. These range
from plotting of $<p_T>$ as a function of particle multiplicity to more current
searches for modification of particle properties (enhancement, suppression, 
medium-induced modification of mass or width) and statistical studies of 
fluctuations in observables such as multiplicity, charge, transverse energy, 
E$_T$ etc. on an event-by-event basis. There are probes which give information 
about ``surface effects'' like $p_T$ distributions of hadrons, strangeness 
production and particle interferometry which reveal final state information. 
Some other probes deal with deeper ``volume effects'' which are sensitive to 
early times after the collision. These include hard processes (large momentum 
transfer) like heavy flavor production ($J/\psi, D-mesons, \Upsilon$), jets 
and high $p_T$ particle production. However, the emphasis has been mainly on 
the volume probes which provide direct information from the hot and dense 
phase of the reaction and are not influenced much by the number of hadrons 
produced in the collision. But it is essential to study the modification of the 
proposed signals by non-QGP, nuclear effects. It's is also worth noting that 
all the different results must be investigated as a function of the associated 
particle multiplicity or equivalent probe giving information on global 
characteristics.

One would also try to understand the onset of all the signals in terms of 
different handles available to us. Experimentally we can vary the collision 
centrality (by selecting multiplicity or transverse energy), collision species 
(a controlled variation of system size) and the center of mass energy. The 
experimental data at the end helps to constrain the theoretical model parameters 
and inputs which include a) equation of state, b) expansion dynamics
and collective flow, c) size and life time of the system for hydrodynamical 
models d) initial parton densities, e) parton mean-free path and cross-section, 
f) nuclear shadowing of initial parton distributions and g) the amount of 
parton energy loss in the plasma. We will discuss briefly some of the 
experimental signals which have been proposed to probe the matter created in 
heavy ion collisions \cite{cpSingh,bass,muller}.

\subsection{Strangeness Enhancement}

Strangeness enhancement is one of the important probes of QGP formation 
\cite{rafel,pkoch}. There is no strangeness in the initial state of the 
collision- the strangeness (which is created during the collisions) thus 
provides information on the reaction dynamics. Moreover, this creates a 
link between the partonic and hadronic phases. In hadronic reactions, the 
production of particles containing strange quarks is normally suppressed 
due to the high mass of strange quark ($m_s \simeq 60-170$ MeV) compared 
to the same for $u$ and $d$ quarks. In the presence of QGP, the temperature 
is of the order of $s$-quark mass and the rapid filling of the phase 
space available for $u$ and $d$ quarks should favor the production of 
$s\bar{s}$ pairs in interactions of two gluons ($gg \rightarrow s\bar{s}$) 
\cite{rafel,pkoch}. This should be reflected in an enhanced production 
of multi-strange baryons and strange anti-baryons in the QGP phase compared 
to a purely hadronic scenario, at the same temperature. The important 
observables in this respect are the yields of strange as well as multi-strange
hadrons and the ratios of the number of strange hadrons to non-strange
hadrons produced in the collision process. To account for incomplete chemical 
equilibration, a strangeness fugacity $\gamma_s$ is introduced in a 
thermo-chemical approach. The particle ratios can be calculated assuming 
either a hadron gas or a QGP scenario and a comparison can be made of the 
values thus extracted in conjunction with other model parameters such as $T$, 
$\mu_B$ and entropy. As the strange hadrons interact strongly, their final 
state interactions must be modeled in details before predictions and 
comparisons of strange particle yields can be made \cite{manuel}.
The enhancement in the $K/\pi$, $\phi/\omega$, $\bar{\Lambda}/\Lambda$ and
$\bar{\Xi}/\Xi$ ratios are the experimental observables in this sector \cite{cpSingh}.

Point to note here is kaons account for about $70\%$ of overall strangeness 
production. The enhancement in $K/\pi$ ratio from pp, pA to AA is very distinct 
(factor $\sim 2$). The fact that strangeness/entropy ratio stays the same for all 
three systems (S+S, S+Ag and Pb+Pb), suggests that the system has reached some 
kind of saturation in the $s$ and $\bar{s}$ yields in S+S reaction. This seems 
to rule out the possible interpretation of strangeness enhancement as a 
consequence of hadronic re-interactions \cite{odyniec}.

\subsection{$J/\psi$ Suppression and Open charm Enhancement}

$J/\psi$ suppression was suggested by Matsui and Satz \cite{matsui} as a 
clean signal of QGP. The color charge of a quark in QGP, is screened due to 
the presence of quarks, anti-quarks and gluons in the plasma. If we place a 
$J/\psi$ particle, which is a bound state of charm quark $c$ and a charm 
anti-quark $\bar{c}$, the Debye screening will weaken the interaction
of $c$ and $\bar{c}$. Furthermore, in QGP the quarks and gluons are deconfined and
the string tension between $c$ and $\bar{c}$ vanishes. In QGP, the screening radius
$r_D$ becomes less than the binding radius $r_H$. Because of these two effects
a $J/\psi$ particle placed in QGP at high temperature will dissociate leading 
to suppression of its production in high energy nucleus-nucleus collisions.

To understand the effect of QGP on a  $J/\psi$ particle, let's consider $J/\psi$ 
as a two body system of a charm quark interacting with a charm anti-quark in the 
absence of QGP. We place the $c$ quark with a color charge $q > 0$ at the origin 
and the $\bar{c}$ quark with a color charge $-q$ at {\bf r}. The color potential 
between $c$ and $\bar{c}$ is given by the Coulomb potential
\begin{equation}
{V_0({\bf r}) = \frac{q}{4\pi r}}
\label{coulomb}
\end{equation}
There is also a confining linear potential between $c$ and $\bar{c}$ which 
increases with their separation,
\begin{equation}
{V_{linear}({\bf r}) = kr},
\label{linear}
\end{equation}
where $k$ is the string tension coefficient. The potential energy for the 
$c\bar{c}$ system is 
\begin{equation}
{V({\bf r}) = (-q)\frac{q}{4\pi r} + kr},
\label{potential}
\end{equation}
The Hamiltonian for the $c\bar{c}$ system is given by
\begin{equation}
{H = \frac{{\bf p}^2}{2\mu} -\frac{\alpha_{eff}}{r} + kr},
\label{hamiltonian}
\end{equation}
where $\mu = m_c/2$, is the reduced mass of the $c\bar{c}$ system and 
$\alpha_{eff} = q^2/4\pi$.

Now, if we put the $c\bar{c}$ system in QGP, the system will be affected in the 
following way. The string tension depends on the temperature. The finite 
temperature of the quark matter therefore alters the string tension coefficient 
$k$ between $c$ and $\bar{c}$. Secondly, the presence of quark matter also leads 
to rearrangement of the densities of quarks, anti-quarks and gluons around $c$ 
and $\bar{c}$. This rearrangement leads to the screening of $c$ from $\bar{c}$ 
and vice versa. As a consequence, the interaction between $c$ and $\bar{c}$ is 
modified from a long range Coulomb interaction Eq~(\ref{coulomb}) into a 
Yukawa-type short range (given by Debye screening length $\Lambda_D$) interaction.

Quark confinement occurs when the string tension doesn't vanish. However, the absence
of the string tension doesn't automatically mean that $c$ and $\bar{c}$ can't form a 
bound state. They remain interacting with each other with the Coulomb interaction
$-\alpha_{eff}/r$ which in tern is modified by the Debye screening. However, at very 
high temperature, the screening is so high that $c$ and $\bar{c}$ hardly form any bound
state and the $c\bar{c}$ systems dissociate ($\Lambda_D \propto \frac{1}{T}$). The $c$
quark and $\bar{c}$ anti-quark subsequently hadronize by combining with light quarks
and light anti-quarks to emerge as ``open charm'' mesons such as $D(c\bar{u}$, and 
$c\bar{d}$), $\bar{D}(\bar{c}u$ and $\bar{c}d$), $D_s(c\bar{s})$, and 
$\bar{D_s}(\bar{c}s)$. Hence in nucleus-nucleus collisions, in case of the formation 
of a QGP phase, $J/\psi$ production will be suppressed which will result in the 
enhanced production of open charm states.

There have been several experimental investigations to study $J/\psi$ production 
in high energy heavy-ion collisions both at CERN (by NA38 and NA50 collaborations) 
\cite{na38,na50,na50_1,na50_2} and RHIC \cite{rhic_jpsi,phenixJ}. There are also 
nuclear effects, such as the break up of $J/\psi$ by hadronic co-movers, which 
can also suppress the measured $J/\psi$ cross-section in nucleus-nucleus collisions 
\cite{capella}.

\subsection{Jet Quenching}

Jets produced by high-energy quarks and gluons in ultra-relativistic heavy ion 
collisions can also provide a potential probe for the formation of a QGP. High 
energy partons coming from the initial hard collisions lose energy by traversing 
through the dense matter. This energy loss is greater in AA collisions, as 
compared to pp and pA collisions \cite{eLoss}. This phenomenon is called jet 
quenching. Jet quenching results from the energy loss $(-dE/dx)$ of a high-$p_T$ 
parton as it propagates in the dense matter. In case of hadronic matter the 
partons are decelerated due to the string tension ($(-dE/dx)_{had} \simeq 1$)
GeV/fm, whereas in a QGP, they lose energy by collisions with the thermal quarks 
and gluons (elastic scattering) and by bremsstrahlung radiation. It appears that 
when both the contributions are added, the stopping power of QGP is comparable 
to that of hadronic matter. However the radiative energy loss is proportional 
to the square of Debye momentum ($\mu_D \approx \lambda_D^{-1}$) which, according 
to lattice calculations, decreases rapidly close to the phase transition. Thus 
in the vicinity of the deconfinement transition, there might be a region where the 
stopping power of strongly interacting matter decreases with growing energy 
density. Therefore in the mixed phase, in which the system is expected to spend 
most of its time, variations of jet quenching may provide a signature for the
phase transition \cite{cpSingh}. 
\begin{figure}
\begin{center}
\epsfig{figure=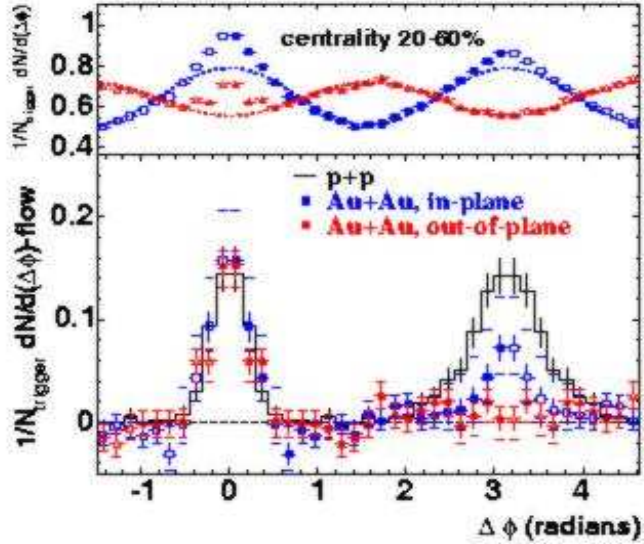,width=9cm}
\end{center}
\caption{\label{jetAsym}Upper panel: Azimuthal distributions of associated 
particles, for in-plane trigger particles (squares) and out-of-plane trigger 
particles (triangles) for Au+Au collisions at centrality $20\%-60\%$. Open 
symbols are reflections of solid symbols around $\Delta\phi =0 $ and 
$\Delta\phi = \pi$. Elliptic flow contribution is shown by dashed lines. Lower 
panel: Distributions after subtracting elliptic flow, and the corresponding 
measurement in $p+p$ collisions (histogram) Fig. Ref. \cite{starJet}.}
\end{figure}
In addition a quark or gluon jet traversing through a dense medium will also be 
deflected due to the scattering from the constituents of the QGP and also the 
hadronic matter. This angular deflection will introduce jet acoplanarity or an 
azimuthal asymmetry. One can then perform the angular correlations among 
high-$p_T$ particles to study the energy loss effects of the partons in the 
medium. Fig.~\ref{jetAsym} shows the STAR collaboration measurement of back-to-back 
(around $\Delta\phi = \pi$) correlations in Au+Au collisions. In-plane trigger 
particles (high-$p_T$) are suppressed compared to $p+p$ and even more suppressed 
for out-of-plane trigger particles. This shows jet quenching in $Au+Au$ collisions 
compared to $p+p$ collisions.

The  nuclear effects that influence strongly the absolute yield of the 
moderate-$p_T$ and high-$p_T$ hadrons are gluon shadowing and jet quenching. 
Both the effects are of fundamental interest as they pertain to nuclear structure 
at the partonic scale and the energy-loss mechanism in dense matter. To measure 
these nuclear effects in relativistic heavy ion collisions, a comparison of 
hadron $p_T$ spectrum (from nucleus-nucleus collisions) with that of $pp$ 
collisions at the same energy is needed. A properly defined ratio of the two 
is called ``nuclear modification factor'' and is defined as,
\begin{equation}
{R_{AB}(p_T) = \frac{d^2N/dp_Td\eta}{T_{AB}d^2\sigma^{pp}/dp_Td\eta},}
\label{rab}
\end{equation}
where, $d^2N/dp_Td\eta$ is the differential yield per event in the nuclear 
collision A+B, $T_{AB} = <N_{bin}>/\sigma^{pp}_{inel}$ describes the nuclear 
geometry, and $d^2\sigma^{pp}/dp_Td\eta$ for $p+p$ inelastic collisions is 
determined from the measured $p+p$ differential cross-section. In the absence 
of nuclear effects such as shadowing, the Cronin effect, or gluon saturation, 
hard processes are expected to scale with the number of binary collisions and 
$R_{AB}(p_T) = 1$.

The high-$p_T$ hadron suppression in central Au+Au collisions also could be studied by
comparing the hadron spectra in central and peripheral Au+Au collisions. It's measured
by the observable, $R_{CP}$, defined by,
\begin{equation}
{R_{CP}(p_T) = \frac{<N_{bin}^{peripheral}>d^2N^{central}/dp_Td\eta}
{<N_{bin}^{central}>d^2N^{peripheral}/dp_Td\eta},}
\label{rcp}
\end{equation}

\subsection{Photons and Dileptons}

Photons and dileptons are regarded as the ``penetrating'' probes of a hot and dense
matter such as QGP because they are produced in the early thermal stage of the process 
and they are not affected by the subsequent hadronization of the system. They interact
only electromagnetically and their mean free paths are much larger than the transverse
size of the collision volume created in nuclear collisions. As a result, high-energy
photons and dileptons, produced in the interior of the plasma, usually pass through the
surrounding matter without interacting, carrying information directly from wherever 
they were formed to the detector. Contrary to this, the situation for hadronic particles 
are completely different, as their abundances and momentum distributions are changed by
re-scattering in the expansion phase. Dileptons and photons are thus called 
``thermometers'', because they can reveal information about the temperature of the
primordial matter. 

The main difficulty in reading out the history of the final-state evolution from the
measured photon and dilepton spectra is the existence of several sources which produce
photons and dileptons at all stages of the fireball creation and evolution. In the
mass spectrum, single leptons are not considered because weak decays of hadrons 
(especially strange and charm hadrons) produce a strong background. If one considers
mass spectrum of opposite sign dileptons, several vector mesons are seen to appear
in the spectrum because most of them decay into lepton pairs. They essentially give
information about expansion dynamics as well as the effects of the dense medium on the
hadrons. Hence in order to study dilepton signals, one should (1) have a good 
understanding of the thermal production rate of dileptons, and (2) compare these 
rates with other sources. The most significant channel for the production of dileptons 
in the QGP sector is the annihilation of quark-antiquark pairs. In the hadronic sector, 
it is, $\pi^+\pi^-$ annihilation. Other sources of dilepton production are, 
$\pi N \rightarrow {\it l \bar{l}}+ \chi$ and $N\bar{N} \rightarrow {\it l \bar{l}}+ \chi$. 
The final signal is a mixture of thermal emission of lepton pairs from the plasma and 
the later hadron gas, with a background from Drell-Yan processes in which a quark and 
anti-quark pair from the colliding systems annihilates. Charm production also provides 
a potential background because charm decays possess a large semileptonic branching 
ratio \cite{cpSingh}. Hence, the dilepton signal in the invariant mass spectrum is 
a convolution of several such complicated backgrounds on top of it. A study of the 
$p_T$ dependence of various mass windows might perhaps help to disentangle the 
different contributions to the spectrum.

It is shown in Ref.\cite{sinha}, that the ratio of dilepton production rate from the
hadronic sector to that of from the QGP, becomes independent of the chemical potential
and temperature if the plasma temperature goes beyond $T_c > 200$ MeV. It is also 
proposed \cite{kapu} that the ratio of the dilepton rapidity density to the square 
of the charged pion rapidity density may provide a measurement of phase transition 
temperature, provided we get some abrupt change in the slope of the ratio.

Hard and direct photons from the QGP can arise mainly from Compton scattering
($qg \rightarrow q\gamma, ~\bar{q}g \rightarrow \bar{q}\gamma$) and annihilation
($q\bar{q} \rightarrow g\gamma$). Unfortunately, a thermal hadron gas with the
Compton scattering reaction $\pi\rho \rightarrow \gamma\rho$ and pion annihilation
$\pi\pi \rightarrow \gamma\rho$ have been shown to shine as bright as a QGP 
\cite{jkapu}. There are also other channels of photon production like, $\pi^0 \rightarrow
\gamma\gamma, ~\omega \rightarrow \pi^0\gamma, ~\rho^0 \rightarrow \pi^+\pi^-\gamma$. 
However, clear signal of photons from a very hot QGP could be visible at $p_T$ in 
the range 2-5 GeV/c \cite{dks, strick}. However, the flow effects can prevent a 
direct identification of the temperature and the slope of $p_T$ distribution. 
WA98 has observed a direct photon signal in Pb+Pb collisions at SPS \cite{wa98}. 
Compared to the results to $pA$ data, there seems to be an enhancement for central
collisions, suggesting a modification of the photon production mechanism. 
There are also measurements on direct photon at RHIC \cite{phenixDPp, phenixDPA}.

There are suggestions \cite{jkapu} to measure $T_c$ by the photons in the $p_T$ range
from 1 to 3 GeV. The technique used is simply to fit the $p_T$ distribution with a 
thermal distribution- the temperature from the fit is approximately equal to the transition
temperature.

\subsection{Medium Effects on Hadron Properties}

The widths and masses of the $\rho, \omega$ and $\phi$ in the dilepton pair invariant
mass spectrum are sensitive to medium-induced changes, especially to possible drop of
vector meson masses preceding the chiral symmetry restoration transition. The CERES
data from S+Au and Pb+Au collisions at SPS showed an excess of dileptons in the low-mass
region $0.2 < M < 1.5 ~GeV/c^2$, relative to $pp$ and $pA$ collisions \cite{ceres1,ceres2}.
Although the CERES data can be explained by a hydrodynamic approach assuming the creation
of a QGP \cite{hydroQGP}, alternative scenarios have also provided explanations. These have
included for instance, microscopic hadronic transport models incorporating mass shifts
of vector mesons and calculations involving in-medium spectral functions (coupling the
$\rho$ with nucleon resonances) without requiring a shift in the $\rho$ mass \cite{rapp}.
When the life time of a resonance particle is comparable with the evolution time scale 
of the phase transition in nucleus-nucleus collisions, the measured properties (mass, 
width, branching ratio, yield and $p_T$-spectra) associated with the resonances will 
depend on the collision dynamics and chiral properties of the medium at high energy density
and temperature \cite{zxu}.

\subsection{Equation of state and Flow}

The system formed in heavy ion collision, although small in size, is sufficiently
large for statistical physics to be applicable. The thermodynamical properties of the
system in statistical equilibrium are described by an equation of state (EoS). One of
the important goals of heavy-ion program is to look for the phase transition between
hadronic matter and the QGP. The order of the phase transition is not clear till date.
Lattice gauge theory calculations using three or more flavors show the phase transition
to be first order, while calculations with two light quarks lead to a second order phase
transition. For first order phase transition, the pressure remains constant in 
the region of phase co-existence. This results in vanishing velocity of sound 
$c_s = \sqrt{\partial p/\partial\epsilon}$. The expansion of the system or collective
flow is driven by the pressure gradient, therefore expansion depends crucially on $c_s^2$.
Matter in the mixed phase expands less rapidly than a hadron gas or a QGP at the same
energy density and entropy. In case of rapid changes in the EoS without phase transition,
the pressure gradients are finite, but still smaller than for an ideal gas EoS, and 
therefore the system expands more slowly \cite{kapusta, gersdorff}. This reduction of
$c_s^2$ in the transition region is commonly referred as ``softening'' of EoS. The
respective region of energy densities is called a soft region. Here the flow
will temporarily slow down (or possibly even stall). Consequently a time delay is
expected in the expansion of the system. This prevents the deflection of spectator
matter (the bounce-off) and, therefore, causes a reduction of the directed transverse
flow in semi-peripheral collisions. The softening of EoS should be observable in the 
excitation function of the transverse directed flow of baryons. 

Due to it's direct dependence on the EoS, $P(\rho,T)$, flow excitation functions can
provide unique information about phase transition: the formation of abnormal nuclear
matter, e.g., yields a reduction of the collective flow \cite{hoffman}. A directed flow
excitation function as a signature of phase transition into the QGP has been proposed
by several authors \cite{stocker,amelin}. 
\begin{figure}
\begin{center}
\epsfig{figure=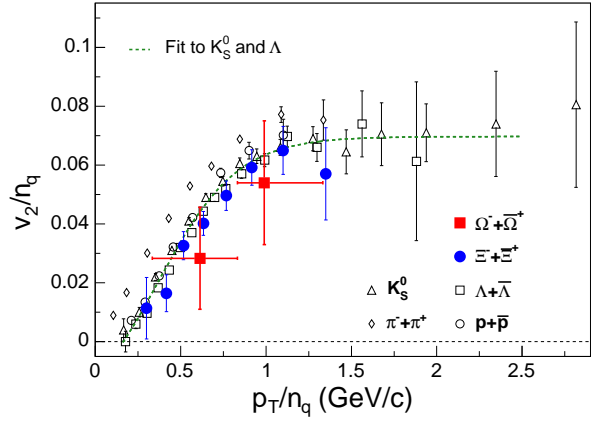,width=8cm}
\end{center}
\caption{\label{flowMSB}(color online). Number of quarks ($n_q$) scaled $v_2$ as a
function of scaled $p_T$ for $\Xi^- + \bar{\Xi^+}$ (solid circles) and $\Omega^- +
\bar{\Omega^+}$ (solid squares). Same distributions also shown for $\pi^+ + \pi^-$ 
(open diamonds), $p+\bar{p}$ (open triangles), $K_s^0$ (open circles) and $\Lambda 
+ \bar{\Lambda}$ (open squares). All data are from 200 GeV Au+Au minimum-bias
collisions. The dashed line is the scaled result of the fit to $K_s^0$ and $\Lambda$.
Fig. Ref. \cite{flowMS200}.}
\end{figure}
In experiments, for peripheral nucleus-nucleus collisions, an event in the plane 
perpendicular to the beam axis exhibits an azimuthal anisotropy in the particle
distributions. this happens because the pressure gradient drives the emission of
particles in the peripheral collisions. The initial state spatial anisotropy is
converted to the final state momentum anisotropy. The azimuthal distribution
of particles in momentum space can be expanded in terms of a Fourier series
\begin{equation}
{E\frac{d^3N}{dp^3} = \frac{1}{2\pi}\frac{d^2N}{p_Tdp_Tdy}
(1+\sum_{n=1}^{\infty}2v_ncos[n(\phi-\psi_r)])}
\label{flow}
\end{equation}
where, $\psi_r$ denotes the reaction plane angle. The Fourier coefficients $v_n$ stands 
for the $n$th harmonic of the event azimuthal anisotropy. The first harmonic coefficient
$v_1$ is called $directed ~flow$ and the 2nd harmonic coefficient $v_2$ is called the
$elliptic~ flow$.

There are several flow measurements at RHIC, taking identified charge particles and 
multi-strange baryons etc. \cite{flow130, flow130I, flowMS}. Multistrange baryons, due
to their large mass and small cross sections should be less sensitive to hadronic 
re-scattering in the later stages of the collision and therefore a good probe of the
early stage of the collision. Their transverse flow would then primarily reflect the
partonic flow. Fig.~\ref{flowMSB} shows the STAR measurement of $n_q$-scaled (number
of constituent quarks) $v_2$ versus the $n_q$-scaled $p_T$ distribution. 
The constituent quark scaling reflects the partonic collectivity at RHIC. 
Furthermore, this supports the idea that the partonic flow of $s$ quark is similar 
to that of $u,d$ quarks.

\subsection{Global Variables}

Global observables like transverse energy $E_T$, particle multiplicities ($N_{\gamma}, 
N_{ch}$ etc.), $p_T$-spectra of the produced particles and their pseudo-rapidity
distributions ($dE_T/d\eta, dN/d\eta$), with mass number and beam energy provide
insight about the dynamics of the system and regarding the formation of QGP 
\cite{bjorken, kataja}. It is also proposed that the correlation of
transverse momentum $p_T$ and the multiplicity of the produced particles may serve as
a probe for the EoS of hot hadronic matter \cite{vanHove}. According to Landau's
hydrodynamic model \cite{landau}, the rapidity density ($dN/dy$), reflects the entropy
and the mean transverse momentum ($<p_T>$) the temperature of the system. Except at the
phase transition points, the rapidity density linearly scales with $<p_T>$. If the
phase transition is of first order, then the temperature remains constant at the
coexistence of the hadron gas and the QGP phase, thereby increasing the entropy density.
So $<p_T>$ will show a plateau with increase of entropy. Hence the global observables
like $dN/dy$ and $<p_T>$ will give indication of QGP phase and the order of phase 
transition. $dE_T/d\eta$ gives the maximum energy density produced in the collision
process which is necessary to understand the reaction dynamics. The formation of QGP
may also change the shape of the pseudo-rapidity distribution \cite{sarkar, dumitru}.

The event multiplicity distribution gives information of the centrality and energy density
of the collision. The scaling of multiplicity with number of participant nucleons 
($N_{part}$) reflects the particle production due to soft processes (low $p_T$). Whereas,
at high energy when hard processes (high-$p_T$) dominate, it's expected that the 
multiplicity will scale with number of elementary nucleon-nucleon collision ($N_{coll}$).
There are models \cite{kharzeev} to explain the particle production taking a linear 
combination of $N_{part}$  and $N_{coll}$ (called a two-component model).

\subsection{Observable Fluctuations}

Lattice QCD calculations find a phase transition in strongly interacting matter which is
accompanied by a strong increase of the number of effective degrees of freedom 
\cite{bernard, boydG}. Although, the nature and order of the phase transition is not
known very well, lattice calculations suggest that QCD has a weak first-order transition
provided that the strange quark is sufficiently light \cite{bernard, boydG}, that is for 
three or more massless flavors. However, when the quark flavors become massive, the QCD 
transition changes to a smooth cross over.

Phase transition being a critical phenomenon, is associated with divergence of 
susceptibilities and hence fluctuations in corresponding observables. Hence, observable 
fluctuations could be used as probes of deconfinement phase transitions.

Fluctuations are very sensitive to the nature of the phase transition. First-order phase 
transition is expected to lead to large fluctuations due to droplet formation 
or more generally density or temperature fluctuations. In case of a second-order phase
transition the specific heat diverges, and this has been argued to reduce the fluctuations
drastically if the matter freezes out at the critical temperature. Even if the transition 
is not of first order, fluctuations may still occur in the matter that undergoes a 
transition. The fluctuations may be in density, chiral symmetry, strangeness or other 
quantities and show up in particle multiplicities. The ``anomalous'' fluctuations depend 
not only on the type and order of the phase transition, but also on the speed by which 
the collision zone goes through the transition, the degree of equilibration, the 
subsequent hadronization process, the amount of re-scattering between hadronization 
and freeze-out \cite{heiselberg}.

There have been efforts to use observable fluctuations like, ratios of charged particles
\cite{jeon, jeon2}, baryon number multiplicity \cite{gavin}, net charge \cite{netCh}, 
mean $p_T$ \cite{meanPt}, transverse energy \cite{etFluct}, strangeness \cite{sFluct},
isospin \cite{isospin} etc., to probe the deconfinement phase transition.

It is necessary to understand the role of statistical fluctuations, in order to extract new
physics associated with fluctuations. The sources of these fluctuations include impact
parameter fluctuations, fluctuations in the number of primary collisions and in the results
of such collisions, fluctuations in the relative orientation during the collision of
deformed nuclei, effects of re-scattering of secondaries and QCD color fluctuations.

\section{Estimation of Initial Energy Density in Heavy-Ion Collisions}
\begin{figure}
\begin{center}
\includegraphics[width=6.0in]{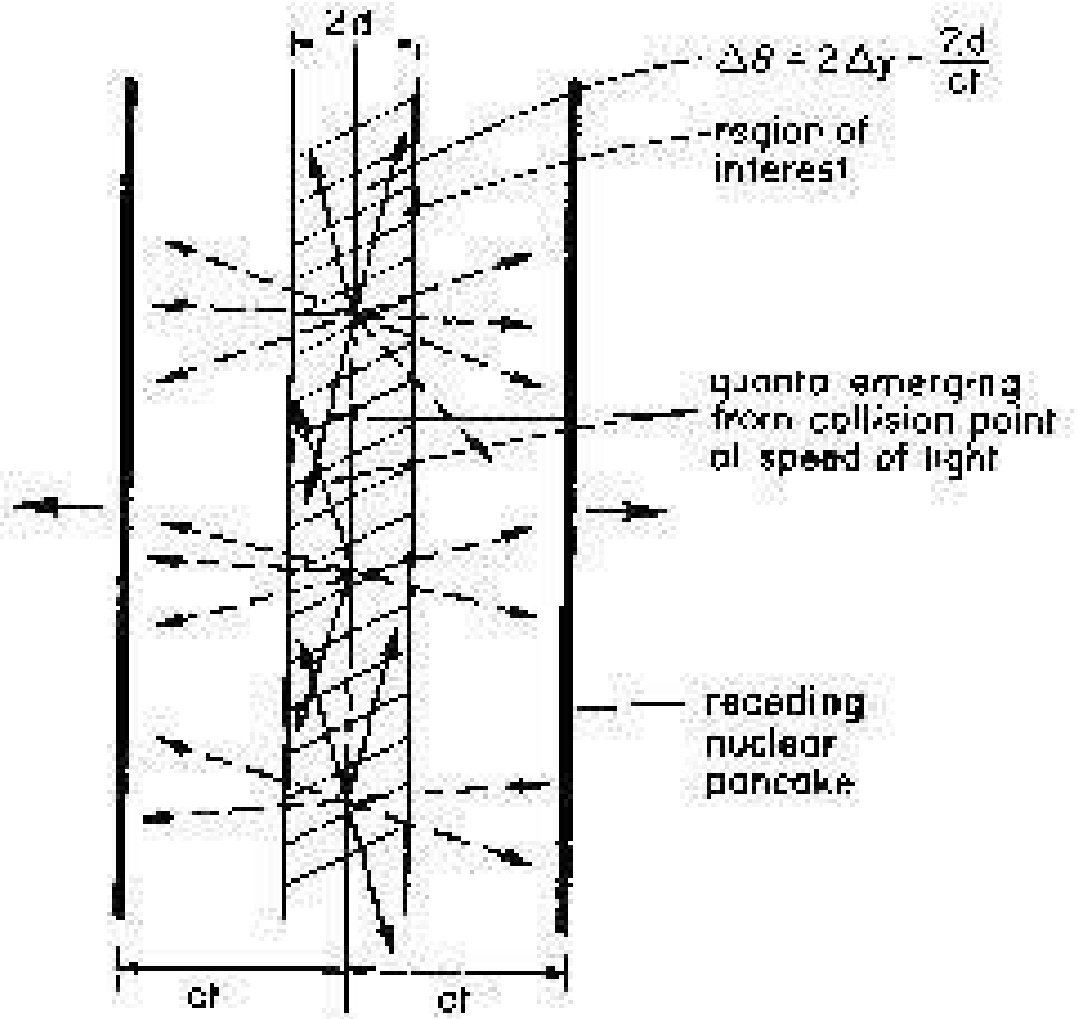}
\caption{\label{bjorken}Geometry for the initial state of centrally produced
plasma in nucleus-nucleus collisions. This picture is valid in any frame in
which the incoming nuclei have very high energies and so are Lorentz contracted.}
\end{center}
\end{figure}
In any frame where the two incoming nuclei have very high energies, the region
when/where the nuclei overlap will be very thin in the longitudinal direction
and very short in duration. In this scenario, it's fair to describe all 
secondary produced particles as having been radiated out from a very thin
``disk'' and that they are all created at essentially at same time. This is 
the Bjorken hydrodynamic picture \cite{bjorken} of nucleus-nucleus collision.

Once the beam ``pancakes'' recede after their initial overlap, the region
between them is occupied by secondaries at intermediate rapidities. We can
calculate the local energy densities of these created particles, if we 
assume that the secondaries can be considered to be formed at some proper
time $\tau_{Form}$, after they are radiated out from the thin source disk.

Our region of interest, in any frame will be a slab perpendicular to the beam
direction, with longitudinal thickness $dz$, with one face of the ``source''
plane in this frame, and the transverse overlap area $A$. The region described 
here corresponds to half the shaded region shown in Fig.~\ref{bjorken}. 
Since $\beta_{\parallel}\simeq 0$ for particles near the source location, 
this is an appropriate region over which we can calculate a meaningful energy '
density. At time $t = \tau_{Form}$, this volume will contain all the (now-formed) 
particles with longitudinal velocities $0 \leq \beta_{\parallel} \leq dz/\tau_{Form}$ 
(since we assume particles can't scatter before they are formed!). Then we can 
write this number of particles as $dN = (dz/\tau_{Form})\frac{dN}{d\beta_{\parallel}}$, 
or equivalently $dN = (dz/\tau_{Form})\frac{dN}{dy}$, where $y$ is longitudinal
rapidity, since $dy = d\beta_{\parallel}$ at $y = \beta_{\parallel} = 0$.
If these particles have an average total energy $<m_T>$ in this frame 
($E = m_T$ for particles with no longitudinal velocity), then the total energy
divided by the total volume of the slab at $t = \tau_{Form}$ is 
\begin{equation}
{\langle \epsilon(\tau_{Form})\rangle = \frac{dN<m_T>}{dz A} = \frac{dN(\tau_{Form})}
{dy} \frac{<m_T>}{\tau_{Form} A} = \frac{1}{\tau_{Form} A} \frac{dE_T(\tau_{Form})}{dy}},
\label{bj}
\end{equation}
where, we have equated $\frac{dE_T}{dy} = <m_T>\frac{dN}{dy} \approx 
<m_T>\frac{3}{2}\frac{dN_{ch}}{dy}$ and emphasized that Eq.~(\ref{bj}) is true for 
the transverse energy density present at time $t = \tau_{Form}$. The factor 3/2 
compensates for the neutral particles.

Eq.~(\ref{bj}) is referred as $Bjorken ~energy ~density, ~\epsilon_{B_j}$. It is a valid
measure of peak energy density in created particles, on very general grounds and
in all frames, as long as two conditions are satisfied: (1) A finite formation time
$\tau_{Form}$ can meaningfully be defined for the created secondaries; and (2) The
thickness/``crossing time'' of the source disk is small compared to $\tau_{Form}$,
that is, $\tau_{Form} >> 2R/\gamma$. Here $R$ is the rest-frame radius of the nucleus 
and $\gamma$ is the Lorentz factor. In particular, the validity of Eq.~(\ref{bj})
is completely independent of the shape of the $dE_T(\tau_{Form})/dy$ distribution
to the extent that $\beta_{\parallel}$ is infinitesimally small in a co-moving frame;
a plateau in $dE_T/dy$ is not required. For practical purposes at RHIC, we will 
consider condition (2) above to be satisfied as long as $\tau_{Form} > 2R/\gamma$
is true.

Historically, $\epsilon_{B_j}$ has been calculated using the final state $dE_T/dy$
and simply inserting a nominal value of 1 fm/c for $\tau_{Form}$. In addition, fixed 
target experiments have been using $dE_T/d\eta$ as an estimate for $dE_T/dy$, which
is a good approximation for these experiments. For collider experiments,
a correction is made for the Jacobian $dy/d\eta$: ($\sqrt{1-m^2/<m_T>^2}\frac{dN}{dy}
= J\frac{dN}{dy}  = \frac{dN}{d\eta}$). However, we can't take 
$\epsilon_{B_j}$ as an exact estimate of energy density without some justification
for the value of 1 fm/c taken for $\tau_{Form}$. Hence, we term it as 
$\epsilon_{B_j}^{Nominal}$.  An indication of potential problems
with this choice arises immediately when considering AGS Au+Au and SPS Pb+Pb collisions,
where the center of mass ``crossing times'' $2R/\gamma$ are 5.3 fm/c and 1.6 fm/c
respectively, which implies that this choice for $\tau_{Form} = 1$ fm/c actually
violates the validity condition $\tau_{Form} > 2R/\gamma$ we set for the use of 
Eq.(\ref{bj}). So we will deprecate the use of $\epsilon_{B_j}^{Nominal}$ as 
a quantitative estimate of actual produced energy density and instead 
treat it only as a compact way of comparing $dE_T/d\eta$ measurements across 
different systems, centralities and beam energies.

\subsection{Realistic $\tau_{Form}$ and $\epsilon_{B_j}$ estimates}

Is it possible to justify a better estimate for $\tau_{Form}$? From general quantum
mechanical grounds, in a frame where it's motion is entirely transverse, a particle
of energy $m_T$ can be considered to have ``formed'' after a time $t = \hbar/m_T$
since it's creation in that frame. To estimate the average transverse mass, we can
use the final-state $dE_T/d\eta$ to estimate $dE_T(\tau_{Form})/dy$ and, 
correspondingly, use the final-state $dN/d\eta$ as an estimate for $dN(\tau_{Form})/dy$
to obtain
\begin{equation}
{\langle m_T \rangle = \frac{dE_T(\tau_{Form})/dy}{dN(\tau_{Form})/dy} \simeq
 \frac{dE_T/d\eta}{dN/d\eta}~~(Final ~state).} 
\label{mt}
\end{equation}
It has been observed experimentally that the ratio of final-state transverse energy 
density to charge particle density, each per unit pseudo-rapidity is constant at
about 0.85 GeV for full energy central Au+Au collisions. This value is constant 
for a wide range of centrality and shows a very little change with beam energy, 
decreasing to 0.7 GeV, when $\sqrt{s_{NN}}$ is decreased by a order of magnitude
down to 19.6 GeV. If we approximate $dN_{ch}/d\eta = (2/3)dN/d\eta$ in the final 
state, then Eq.(\ref{mt}) would imply $<m_T> \simeq 0.57$ GeV and corresponding 
$\tau_{Form} \simeq 0.35$ fm/c, a value shorter than the ``nominal'' 1 fm/c but 
still long enough to satisfy our validity condition $\tau_{Form} > 2R/\gamma$ at RHIC.

It's worth noting that the value of energy density obtained by Eq.~(\ref{bj}) 
represents a conservative lower limit on the actual $<\epsilon(\tau_{Form})>$
achieved at RHIC. This follows from two observations: (1) The final-state measured
$dE_T/d\eta$ is a solid lower limit on the $dE_T(\tau_{Form})/dy$ present at formation
time; and (2) The final-state ratio $(dE_T/d\eta)/(dN/d\eta)$ is a good lower limit
on $<m_T>$ at formation time, and so yields a good upper limit on $\tau_{Form}$. The
justification of these statements could be realized as follows.

There are several known mechanisms that will decrease $dE_T/dy$ as the collision
system evolves after the initial particle formation, while no mechanism is known
that can cause it to increase (for $y = 0$, at least). Therefore, it's final-state
value should be a solid lower limit on its value at any earlier time. A list of
mechanisms through which $dE_T/dy$ will decrease after $t = \tau_{Form}$ includes:
(i) The initially formed secondaries in any local transverse ``slab'' will, in a 
co-moving frame, have all their energy in transverse motion and none in longitudinal
motion; if they start to collide and thermalize, at least some of their $E_T$ will
be converted to longitudinal modes in the local frame; (ii) Should rough local
thermal equilibrium be obtained while the system's expansion will still primarily
longitudinal, then each local fluid element will lose internal energy through $pdV$
work and so its $E_T$ will decrease; (iii) If there are pressure gradients during
a longitudinal hydrodynamic expansion then some fluid elements may be accelerated
to higher or lower rapidities; these effects are complicated to predict, but we can 
state generally that they will always tend to $decrease~ dE_T/dy$ where it has its
maximum, namely at $y = 0$. Given that we have strong evidence that thermalization
and hydrodynamic evolution do occur in RHIC collisions, it's likely that all these
effects are present to some degree, and so we should suspect that final-state
$dE_T/d\eta$ is substantially lower than $dE_T(\tau_{Form})/dy$ at mid-rapidity.

Coming to the estimate of $\tau_{Form}$, the assumption that $\tau_{Form} = \hbar/<m_T>$
can't be taken as exact, even if the produced particles' $m_T$'s are all identical,
since ``formed'' is not an exact concept. However, if we accept the basic validity
of this uncertainty principle argument, then we can see that the approximation in
Eq.~(\ref{mt}) provides a lower limit on $<m_T>$. First, the numerator $dE_T/d\eta$
is a lower limit on $dE_T(\tau_{Form})/dy$, as above. Second, the argument is often
made on grounds of entropy conservation that the local number density of particles
can never decrease~\cite{entropy}, which would make the final-state denominator in 
Eq.~(\ref{mt}) an upper limit on its early-time value.

\section{Event Simulation: HIJING 1.38}

To understand the data in heavy-ion collision experiment, it is essential
to compare the results with some model predictions. Taking all known
physics processes starting from the particle production, their interactions
till they are detected, the real experiment is simulated with the help of
theoretical models. There are various models (called event generators)
with their own physics goals and physics interactions, to study the particle
production and the final state properties. The silent features of the model
HIJING (Heavy Ion Jet INteraction Generator)~\cite{hijing}, we have used 
for the present study are described below.

The HIJING is a Monte-Carlo event generator based on QCD-inspired models 
for multiple jet production to study the jets and associated particle production 
in high energy $pp, pA$ and $AA$ collisions. This model includes mechanisms such
as multiple mini-jet production, soft excitation, nuclear shadowing of
parton distribution functions and jet interaction in dense matter.

In relativistic-heavy ion collisions, mini-jets are expected to dominate
transverse energy production in the central rapidity region. Particle production
and correlation due to mini-jets must be investigated in order to recognize new
physics of QGP formation. Due to the complication of soft interactions, mini-jet
production can only be incorporated in a pQCD (perturbative Quantum Chromodynamics) 
inspired model. In HIJING, multiple mini-jet production is combined together 
with Lund-type model \cite{lund} for soft interactions, based on a pQCD-inspired model. 
Within this model, triggering on large-$p_T$ jet production automatically biases 
toward enhanced mini-jet production. Binary approximation and Glauber geometry 
for multiple interaction are used to simulate
$pA$ and $AA$ collisions. A parametrized parton distribution function inside a 
nucleus is used to take into account parton shadowing. Jet quenching is modeled
by an assumed energy loss $dE/dz$ of partons traversing the produced dense
matter. A simplest color configuration is assumed for the multiple jet system and
Lund jet fragmentation model used for the hadronization.

\section{Thesis Organization}

The work presented in this thesis is mostly related to the transverse energy
measurement and fluctuation studies at $\sqrt{s_{NN}} =$ 62.4 GeV  Au+Au collisions. 
The data sample corresponds to that taken by the STAR detector in Run-IV. Details 
of the procedure of the estimation of transverse energy has been discussed 
along with the results. The thesis is organized as follows.

In Chapter 1, an overview of physics of quark gluon plasma and heavy ion
collisions has already been given along with the results obtained so far. The
signatures of QGP has been discussed as an introduction to the subject.
In Chapter 2, the STAR detector system in RHIC experiment has been presented.
Chapter 3 deals with the estimation procedure of transverse energy and
a discussion on the results and observations. In Chapter 4, transverse
energy fluctuation studies are presented. The summary and conclusion of the work is
presented in Chapter 5. The method used for the estimation of systematic uncertainties
is given in a separate chapter as an appendix to the thesis.

\vfill
\eject

\chapter{The STAR Detector}
\markboth{nothing}{\it STAR Detector}

\section{Introduction to RHIC}

 The RHIC (Relativistic Heavy Ion Collider)~\cite{rhic} at Brookhaven 
National Laboratory (BNL), USA, is a 2.4 miles accelerator ring with 
multipurpose colliding beam facility which provides beams of both 
relativistic heavy ions and polarized protons. RHIC was designed with 
the following unique goals.

\newcounter{pubcs}
\begin{list}{\arabic{pubcs}. }{\usecounter {pubcs}}
  
\item  To simultaneously accelerate different species in each beam.
  
\item  To access a wide range of collision energies from a minimum of 
  $\sqrt{s_{NN}}$ = 20 GeV, for Au+Au collisions, to a maximum of 
  $\sqrt{s_{NN}}$ = 500 GeV for p+p collisions.
  
\item To provide a high luminosity ({\it L}) beam, making the 
measurement of rare processes (small cross-sections) feasible. 
  
\end{list} 

The particle accelerator that satisfies the above first two conditions 
is a synchrotron with two independent beam-pipes. However the last 
of the criteria puts a strict requirement on the luminosity of the 
collider. For a process with a cross section $\sigma_i$, the event 
rate $R_i$ is given by $R_i = \sigma_i . {\it L}$. The luminosity 
{\it L} is given by 
\begin{equation}
{\it L} = fn \frac{N_1 N_2} {A}
\end{equation}
where ${\it N_1}$ and ${\it N_2}$ are the number of particles contained 
in the colliding bunches, {\it A} is the cross-sectional area of the 
overlap between the two colliding beams of particles, {\it f} is the 
revolution frequency, {\it n} being the number of bunches per beam. 
Therefore, high luminosity could be achieved by maximizing {\it f}, {\it n} 
\& {\it N} and minimizing the beam profile {\it A}.  

The collider consists of two quasi-circular concentric accelerator/storage 
rings on a common horizontal plane. The Blue Ring is for clock-wise and 
the Yellow Ring is for counter-clock-wise beams. For each ring there is 
an independent set of bending and focusing magnets as well as radio 
frequency acceleration cavities. The two rings are oriented to intersect 
one another at six locations along their 3.8 km circumference. Each ring 
consist of six arc sections (each $\sim$ 356 m long) six insertion sections 
(each $\sim$ 277 m long) with a collision point at their center. The rings 
are focused for collision at the interaction regions using a common set 
of dipole magnets, the DX and D0 located at 10 m and 23 m respectively. 
\begin{figure}
\begin{center}
\epsfig{figure=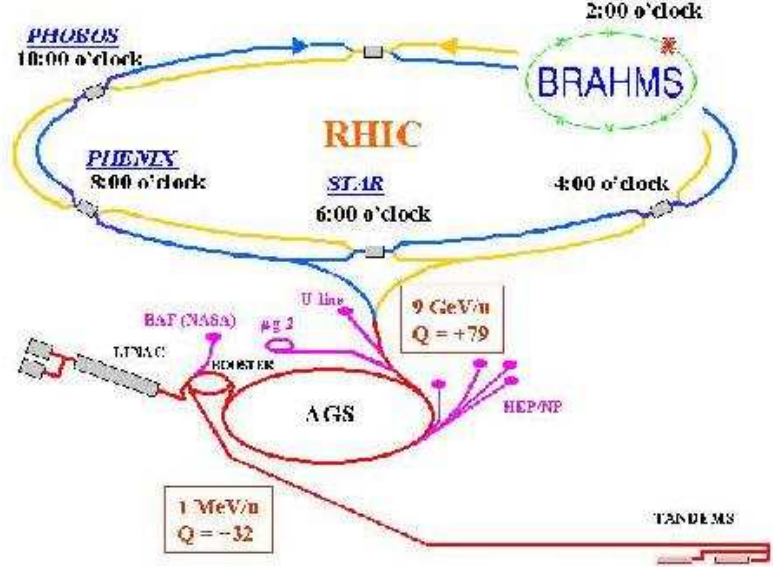,width=12cm}
\caption{\label{RHICring}The Relativistic Heavy Ion Collider (RHIC) Accelerator 
Complex.}
\end{center}
\end{figure}
The basic design parameters of the collider is given in Table ~~\ref{majorparam}.
\begin{table}[htb]
\begin{center}
\begin{tabular}{|l|l|l|}
\hline
 Performance Specifications   &  For Au-Au  &  For p-p \\
\hline
Beam energy  &    100$\rightarrow$ 30 GeV/{\it u}   & 250$\rightarrow$ 30 GeV \\
Luminosity   &    $2\times 10^{26} cm^{-2}s^{-1}$ & $1.4\times 10^{31} cm^{-2}s^{-1}$\\
Number of bunches/ring & $60(\rightarrow 120)$  & $60(\rightarrow 120)$ \\
Number of Particles/Bunch & $1\times10^9$ & $1\times10^{11}$\\
Luminosity life time & $\sim$ 10h & $>$ 10h\\
Bunch width & 30 ns & \\
Ions per beam & $ 1\times 10^{9}$ & \\

\hline
\end{tabular}
\end{center}
\vspace{0.2cm}
\caption{Performance specifications of RHIC.}
\label{majorparam}
\end{table}
The collision systems used include the heavy-ions Au+Au, Cu+Cu, d+Au 
and p+p. For Au+Au systems of center of mass energy 200 GeV, Au ions 
with charge Q = -1 are created using a pulsed sputter ion source. 
They are then accelerated through the Tandem Van de Graaff facility 
and a series of stripping foils ultimately yielding Au ions of kinetic 
energy 1 MeV/nucleon and a net charge of Q = +32. There are two 
Tandem Van de Graaff accelerators that can run exclusively (using 
one as a back up) or in parallel (to accelerate two different species 
simultaneously). The ions are then directed to the booster synchrotron 
through a 550 meter transfer line. The booster accelerates the Au ions 
to an energy of 95 MeV/nucleon. The Au ions leaving the booster are 
further stripped to Q = +77 and are transferred into the AGS, where 
they are accelerated to 8.86 GeV/nucleon and sorted into four final 
bunches. Finally, the ions are transferred from AGS to RHIC and 
stripped to the bare charge state of Q = +79 during the transfer. For 
p+p beam, protons are injected from the 200 GeV Linac directly into
the booster synchrotron followed by acceleration in the AGS and 
injection into RHIC.

\section{The RHIC Detector Systems}

RHIC consists of four experiments namely, STAR ~\cite{star}, PHENIX~\cite{phenix}, 
PHOBOS~\cite{phobos} and BRAHMS~\cite{brahms} at four different interaction points 
to study matter at high temperature and energy density for the possible formation 
of the deconfined state of quarks and gluons, the Quark-Gluon-Plasma (QGP). 

The STAR (Solenoidal Tracker At RHIC) utilizes a solenoidal geometry with a large
Time-Projection Chamber (TPC) installed inside a solenoidal magnet, providing
a close to $4\pi$ solid angle tracking capability for charge particles from the
collisions. With projections on the end sectors giving the {\it x-y} co-ordinates 
and drift time of ionization electrons giving the z-coordinates of track segments, 
the TPC has three dimensional tracking capability. The {\it dE/dx} measurement of 
track segments allow an identification of particles over a significant momentum 
range of interest. In addition, with the barrel and endcap calorimeters, STAR has 
the capability of detecting photons, electrons and measuring their energy. It has
a Silicon Vertex Tracker (SVT) which surrounds the beam pipe resulting in an
improvement in the momentum resolution of the system. It also facilitates the 
detection of decay vertices of short-lived particles.
\begin{figure}
\begin{center}
\includegraphics[width=2.8in]{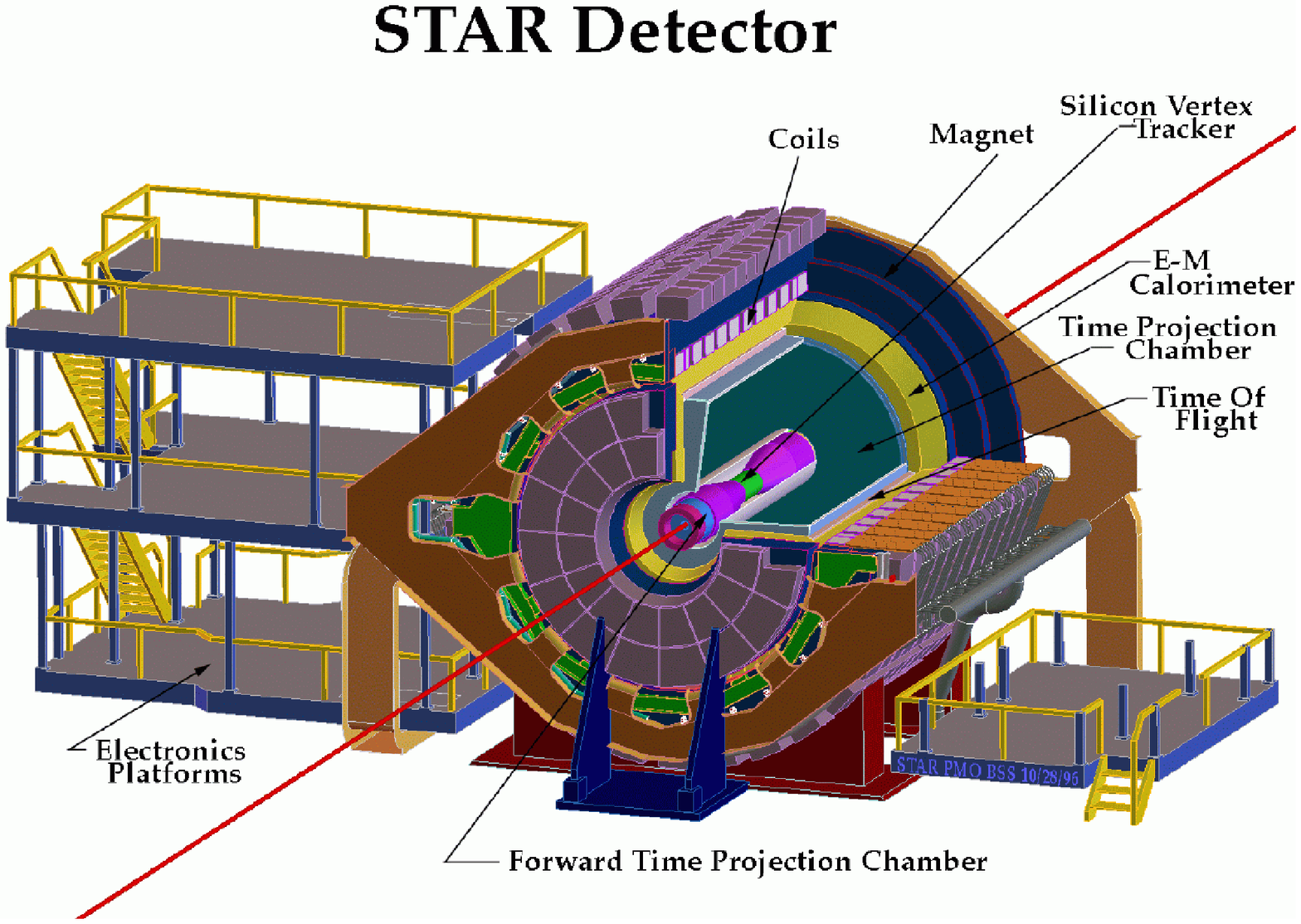}
\includegraphics[width=2.8in]{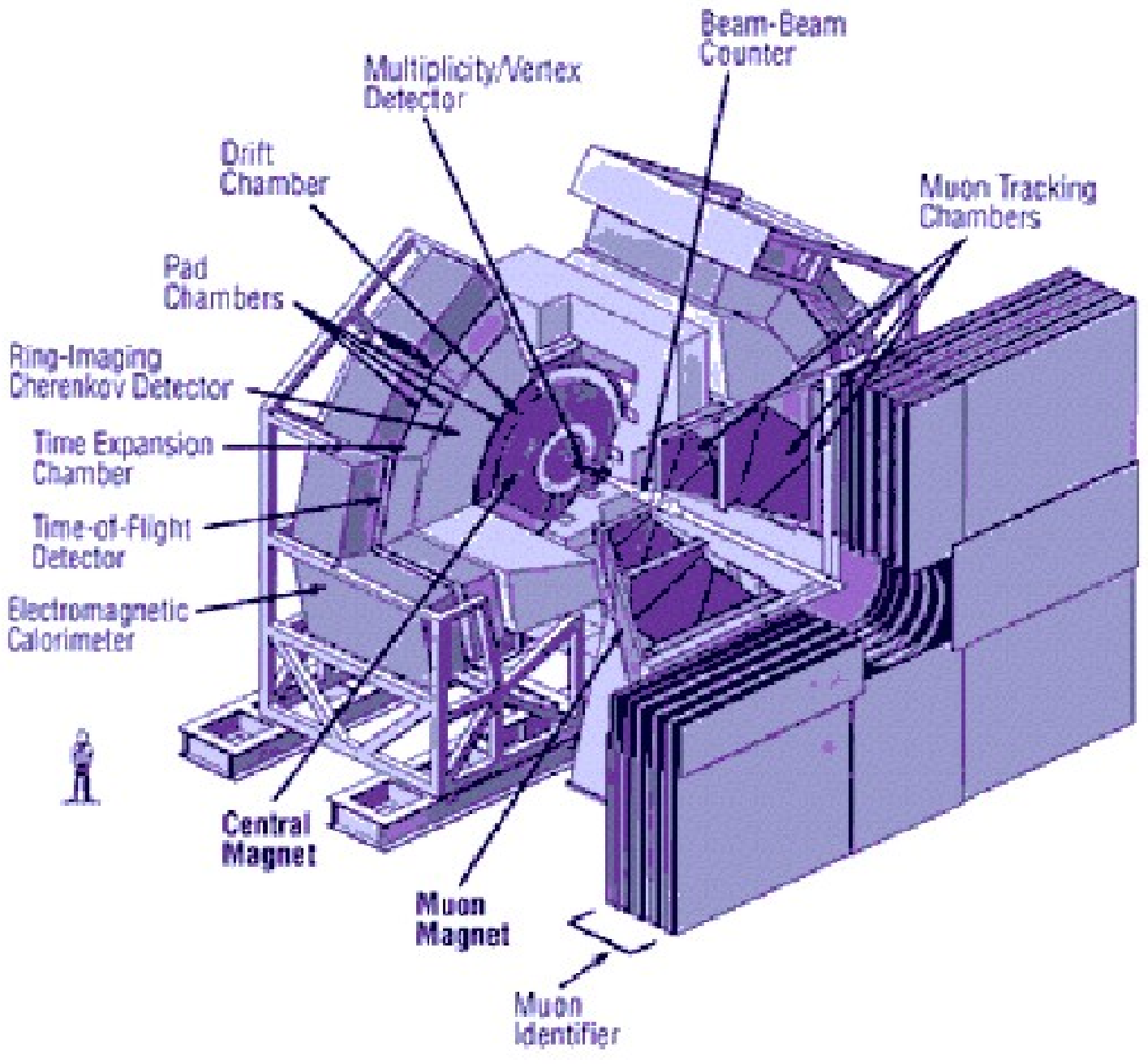}
STAR~~~~~~~~~~~~~~~~~~~~~~~~~~~~~~~~~~~~~~~~PHENIX
\caption{\label{rhicDet1} Schematic pictures of STAR and PHENIX experiments 
 at RHIC, showing different sub-detectors.}
\end{center}
\end{figure}
 The PHENIX (Pioneering High Energy Nuclear Interaction eXperiment) 
(Fig.~\ref{rhicDet1}) consists of three magnetic spectrometers. It has a 
central spectrometer consisting of 
an axial field magnet and two detector arms, one on west and another on east 
side. It has also two muon arms, one on the north and the other on the south 
side of the Central Spectrometer along the direction of beams. The basic 
concept of the Central Spectrometer is to cover selected solid angles with 
quasi-concentric layers of high-speed detectors of various types. 
Detector subsystems in the Central Spectrometer arm include Drift Chamber, 
Pad Chamber, Time Expansion Chamber for tracking and Ring Imaging 
Cherenkov Detector, Time-of-Flight Detector and Electromagnetic 
Calorimeter for particle identification. The east and west central arms are 
centered at zero rapidity and instrumented to detect 
electrons, photons, and charged hadrons. The north and south forward arms have 
full azimuthal coverage and are instrumented to detect muons. The global 
detectors (Zero Degree Calorimeter (ZDC), Beam-Beam Counter (BBC) and the 
Multiplicity Vertex Detector (MVD)) measure the start time, vertex and 
multiplicity of particles produced in the interactions. A pair of ZDCs detect 
neutrons from grazing collisions providing a trigger for the most peripheral 
collisions. A pair of BBCs provide a measure of the ToF of forward particles 
to determine the time of a collision. They also provide a trigger for the more 
central collisions in addition to providing a measure of the collision position 
along the beam axis. The MVD helps in a more precise determination of event 
position, multiplicity and measures fluctuations  of the charge particle 
distributions. It is composed of concentric barrels of silicon-strip detectors 
and end-caps made up of silicon pads. The PHENIX experiment has a high rate 
capability and fine granularity along with excellent energy, momentum and mass 
resolution. The major goal of this experiment has been to measure the spin 
structure of nucleons and to probe each phase of QGP evolution through 
the study of rare processes involving photons, electrons and muons as well 
as the predominant hadron production. It has the unique capability of 
measuring direct photons over a wide range of $p_T$.

\begin{figure}
\begin{center}
\includegraphics[width=2.8in]{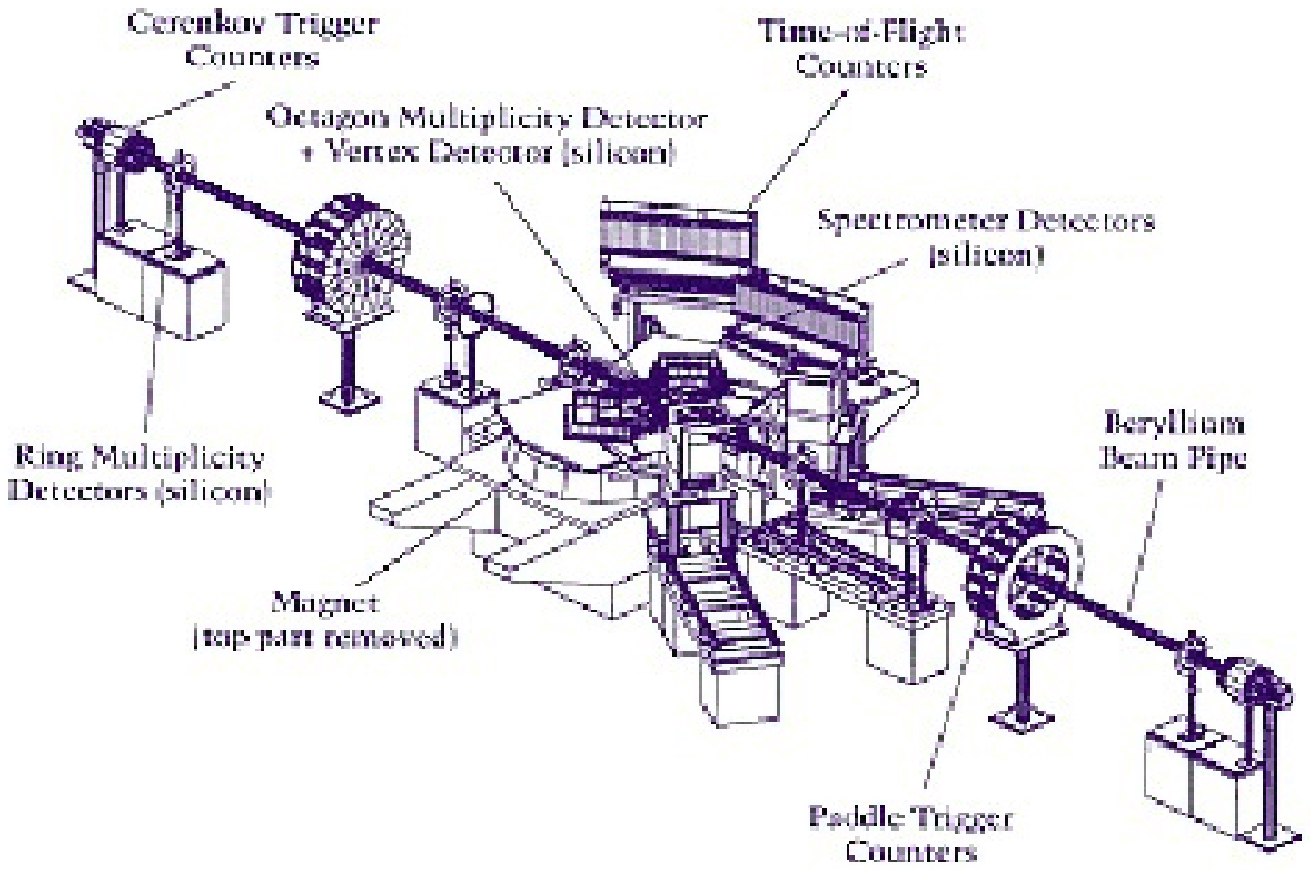}
\includegraphics[width=2.8in]{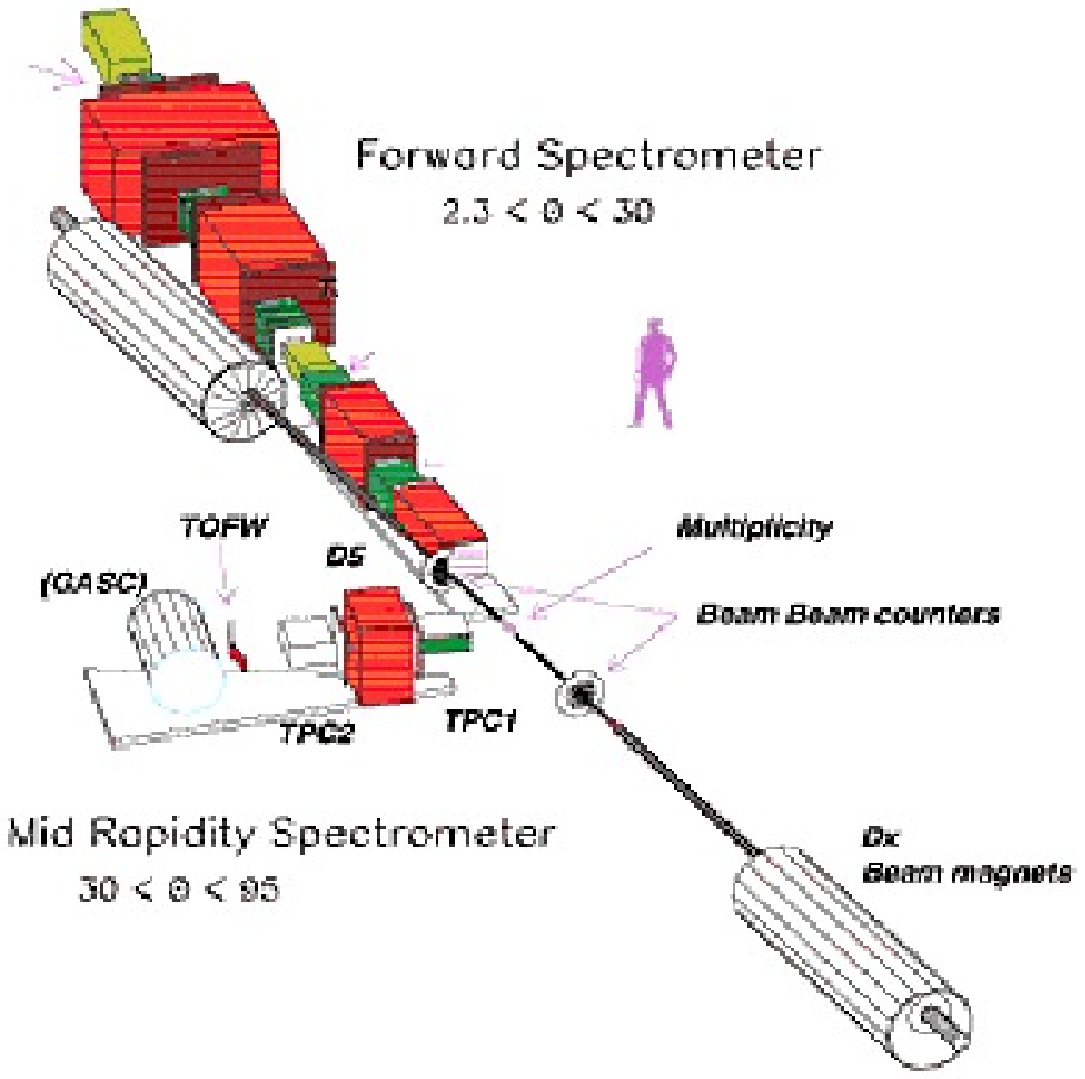}
PHOBOS~~~~~~~~~~~~~~~~~~~~~~~~~~~~~~~~~~~~~~~~BRAHMS
\caption{\label{rhicDet2} Schematic pictures of  PHOBOS and BRAHMS at RHIC, 
showing different sub-detectors.}
\end{center}
\end{figure}

The BRAHMS (Broad RAnge Hadron magnetic Spectrometer) (Fig.~\ref{rhicDet2})
is a two-arm magnetic spectrometer. One of the arms with a small solid 
angle is in the forward direction is for the detection of high momentum 
particles. The other arm is on the side of the collision point for 
measurements in the mid-rapidity region. Both arms are movable to vary the 
settings to cover wide range of kinematical regions. The spectrometer consists of 
room temperature narrow gap dipole magnets, drift chamber planes, other 
tracking devices, Cherenkov counters and the Time-of-Flight (ToF) detectors 
which enables momentum determination and particle identification over a 
wide range of rapidity and transverse momentum. BRAHMS does hadron 
spectroscopy with a wide range of $p_T$ and rapidity to provide essential 
information on reaction dynamics of collision process. In addition, it 
measures the rapidity density distributions for protons and anti-protons 
which is a sensitive measurement for the study of the dynamics of heavy ion 
collisions.
The PHOBOS (named after the famous mission to Mars) (Fig.~\ref{rhicDet2})
experiment consists of a Multiplicity array, a Vertex detector, 
a two-arm magnetic spectrometer including a time-of-flight wall, 
silicon detectors, a set of silicon paddle and cherenkov detector 
arrays for event triggering and centrality selection. 
With its multiplicity array it is capable of detecting charged particles over 
the full solid angle. In addition, it can detect identified charged particles 
near mid-rapidity in its two spectrometer arms with opposite magnetic fields. 
The unique feature of PHOBOS is to measure very low $p_T$ charged particles 
using the silicon detectors.
 
For the universal characterization of heavy ion collisions, all the four 
experiments have a common detector subsystem, namely a pair of ZDCs located 
behind the beam splitting point, out side the dipole magnets. Each of
the ZDCs is a small calorimeter which detects neutron multiplicity  
providing a measure of the collision centrality. The ZDC pair at each 
crossing point is also used as a luminosity monitor. 

\section{The STAR Detector}

STAR~\cite{star} was constructed to investigate the behavior of strongly 
interacting matter at high energy density and to search for signatures of 
quark-gluon plasma (QGP) formation. STAR measures many observables 
simultaneously to study both the soft (non-perturbative) and the hard 
(perturbative) aspects of the possible QGP phase transition and to understand 
the space-time evolution of the collision process in ultra-relativistic 
heavy ion collisions. The primary goal is to obtain a fundamental 
understanding of the microscopic structure of these hadronic interactions 
at high energy density. A large acceptance design was thus chosen to maximize 
the information recorded per collision. Additionally, STAR is instrumented 
with with a high level trigger system that allows real-time selection
of rare processes such as high $p_T$ jet, direct photon and heavy quarkonia 
production. To meet these goals STAR was designed primarily for hadron 
measurements over a large solid angle featuring layered detector subsystems 
for high precision tracking, particle identification, momentum analysis and 
calorimetry about mid-rapidity. STAR's large acceptance makes it unique for 
detection of hadron jets at RHIC.

\begin{figure}
\begin{center}
\epsfig{figure=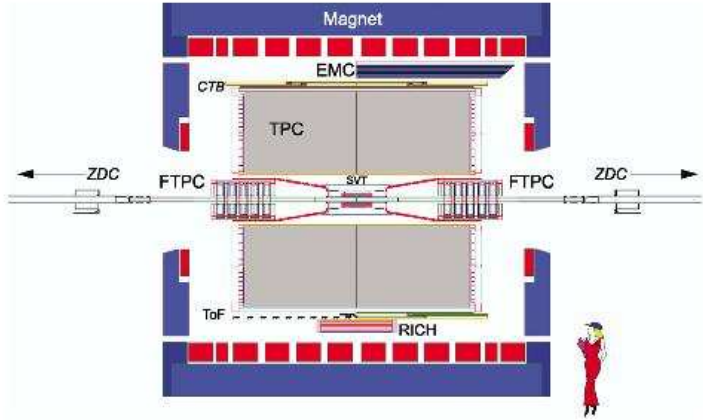,width=10cm}
\caption{\label{starDet}
Cutaway side view of the STAR detector as configured in 2001.}
\end{center}
\end{figure}

The layout of the STAR  experiment is shown in Fig.~\ref{rhicDet1}. 
A cutaway side view of the STAR detector as configured for the RHIC 2001 
run is shown in Fig.~\ref{starDet}. It is a large acceptance cylindrical 
detector with full azimuthal acceptance with a room temperature solenoidal 
magnet~\cite{magnet} with a uniform magnetic field of 0.5 T. This provides 
the charge particle momentum analysis. Time Projection Chamber (TPC)~\cite{tpc} 
provides the main tracking of charged particles in STAR with full azimuthal
acceptance for $|\eta| \leq 1.8$. Close to the beam pipe TPC is augmented 
by a silicon inner tracking system (SVT~\cite{svt} and SSD~\cite{ssd}).
This combined system can yield four radial layers of high precision space 
points, improving the position resolution of the detector and allowing for 
the secondary vertex reconstruction of short lived particles. The silicon 
detectors cover a pseudo-rapidity range $|\eta| \leq 1$ with complete 
azimuthal symmetry. Both SVT and TPC contribute to particle identification 
using ionization energy loss. To extend the tracking to the forward region,
there exists a radial-drift TPC (FTPC)~\cite{ftpc} which covers 
$2.5 < |\eta| < 4$ with complete azimuthal coverage. A Ring Imaging 
Cherenkov detector (RICH)~\cite{rich} ($|\eta| < 0.3$, $\delta\phi = 0.11\pi$) 
and a time-of-flight (TOF)~\cite{tof} patch ($-1 < \eta < 0$, 
$\delta \phi = 0.04 \pi$), present in the STAR detector, can extend 
particle identification to larger momenta, over a limited solid angle at 
mid-rapidity. Outside the TPC is a highly segmented Barrel Electromagnetic 
Calorimeter (BEMC)~\cite{bemc} with full $\phi$-coverage for 
$0 \leq |\eta| \leq 1$. It provides measurement of total energy of 
electromagnetic particles primarily electrons and photons. Along with BEMC, 
the Endcap Electromagnetic Calorimeter (EEMC) ~\cite{eemc} with full 
$\phi$-coverage for $-1 < \eta < 2$. It allows measurement of transverse 
energy, $E_T$, of events in addition to providing a trigger for high 
$p_T$ photons, electrons and electromagnetically decaying hadrons. 
The EMCs (BEMC and EEMC) include a set of Shower Max Detectors (SMD) to 
distinguish high momentum single photons from photon pairs resulting 
from $\pi$ and $\eta$ meson decays. The EMCs also provide prompt charged 
particle signals essential to discriminate against pile up tracks in the 
TPC, arising from other beam crossings, falling within the 40 $\mu$s drift 
time of the TPC, which are anticipated to be prevalent at RHIC pp collision 
luminosities ($\approx 10^{32} cm^{-2} s^{-1}$). In the forward region, 
at a distance of 5.5 m from the interaction point, STAR has a highly 
granular pre-shower Photon Multiplicity Detector (PMD)~\cite{pmd} for 
counting photons and their spatial distributions on an event-by-event 
basis. It covers a pseudo-rapidity region of $-3.8 \leq \eta \leq -2.4$ 
with full azimuthal coverage. Along with FTPC, PMD can address different 
physics issues related to charge-to-neutral fluctuations (DCC-like 
signatures), azimuthal anisotropy (Flow), multiplicity fluctuations
(in photons) etc. A Forward Pion Detector (FPD)~\cite{fpd} sits in the 
forward region at about 7.5 meter from the interaction point.
Additionally, there are several detector systems which are used for event 
selection purposes. These include, Central Trigger Barrel (CTB)~\cite{ctb} 
($|\eta| < 1$, $\delta \phi = 2\pi$) two Zero-Degree Calorimeters (ZDCs)~
\cite{zdc} located at $\theta < 2$ mrad to the beam axis, at 18 meters 
on both sides of the interaction point, a Beam Beam Counter 
(BBC) at high $\eta$ region and a Multi Wire Proportional Counter 
(MWPC). The CTB measures charge particle multiplicity for trigger 
purposes while the BBC is used for normalizing event rates in the pp program. 
The ZDCs measure the neutron multiplicity in a small solid angle near zero 
degree for generating an interaction signal for RHIC operation as well as
providing a hadronic minimum bias trigger.

\subsection{Time Projection Chamber (TPC)}

TPC is the main tracking device in STAR. The TPC records the tracks of
particles, measures their momenta and identifies the particles by measuring 
their ionization energy loss ($dE/dx$). It haz full azimuthal coverage
with $|\eta|\leq 1.8$. Particles are identified over a momentum range 
from 100 MeV/c to greater than 1 GeV/c and momenta are measured over a 
range of 100 MeV/c to 30 GeV/c.
\begin{figure}
\begin{center}
\epsfig{figure=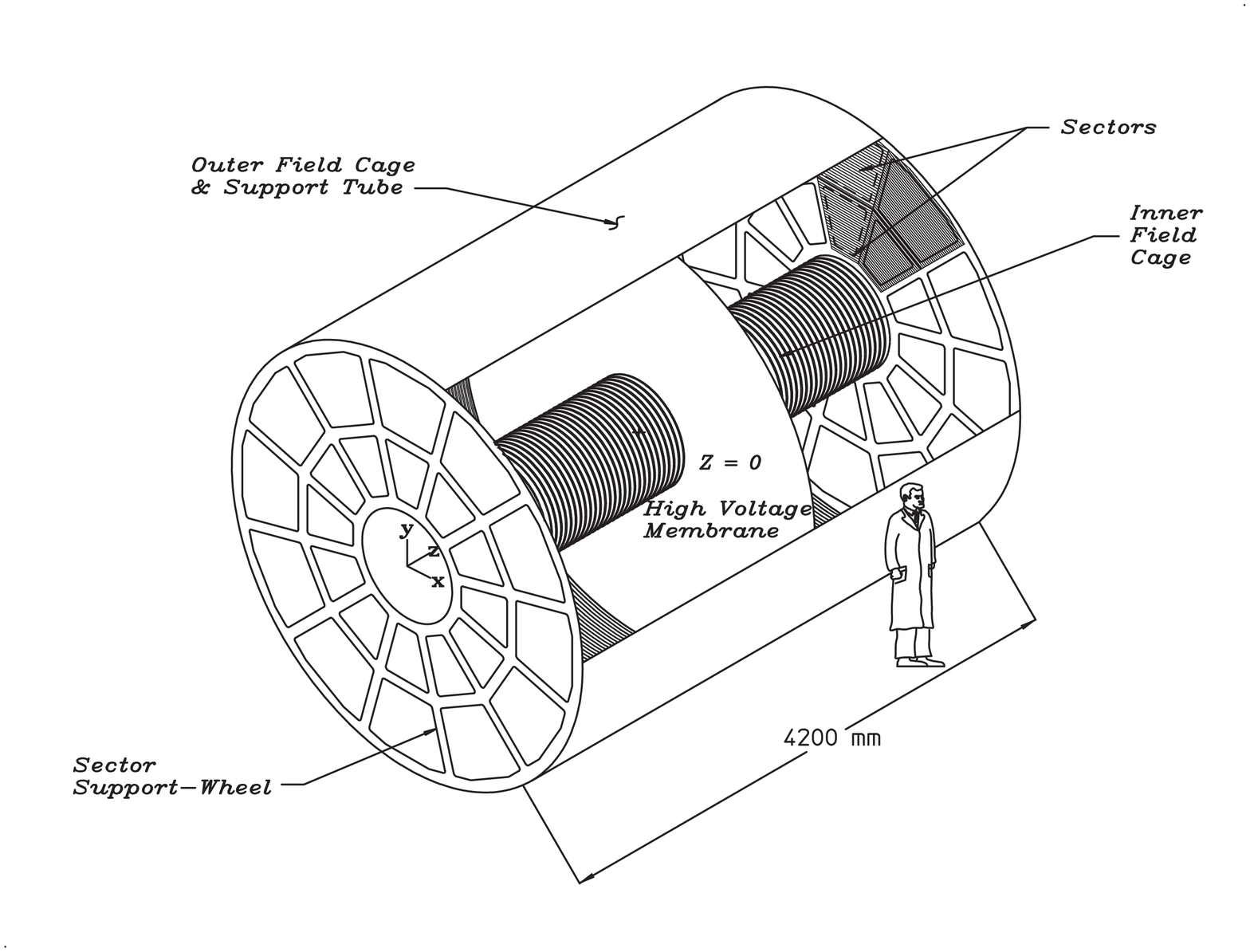,width=10cm}
\caption{\label{tpcDet}
The STAR TPC surrounds a beam-beam interaction region at RHIC.}
\end{center}
\end{figure}
The STAR TPC is shown schematically in Fig.~\ref{tpcDet}. It's surrounded by 
a large solenoidal magnet which operates at 0.5 T. The TPC is 4.2 m in length 
and 4 m in diameter. The cylinder is concentric with the beam axis. The inner 
and outer radii of the active volume are 0.5 m and 2.0 m respectively. TPC 
covers a pseudo-rapidity interval that ranges from $-1.8 < \eta < 1.8$ for the 
inner radius and $-1 < \eta < 1$ for the outer radius (see Fig.~\ref{tpcC}). 
It's an empty volume of gas ($P10: Ar+10\% ~CH_4$) at $\sim 2$ mbar above 
atmospheric pressure, in a well-defined, uniform electric filed of $\approx$ 
135 V/cm. The paths of primary ionizing particles passing through the gas 
volume are reconstructed with high precision from the released secondary 
electrons which drift in the uniform electric field to the readout end caps 
at the ends of the chamber. Two co-ordinates are determined by the location 
where the electron is detected. The third co-ordinate is reconstructed using 
the time taken for the electron to reach the wire chamber (time bin) and the 
electron drift velocity in the gas.
\begin{figure}
\begin{center}
\epsfig{figure=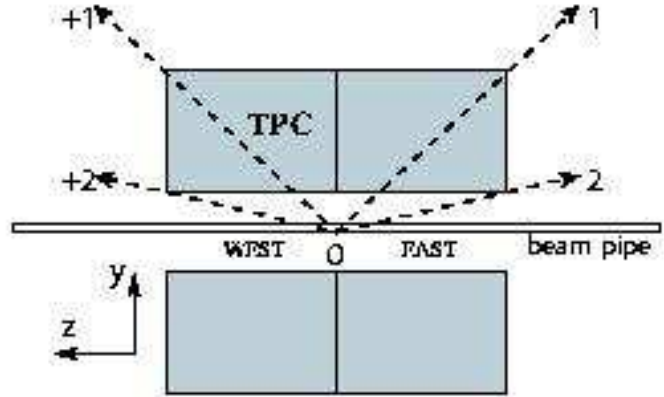,width=10cm}
\caption{\label{tpcC}Schematic view of the TPC illustrating the co-ordinate system and 
pseudo-rapidity coverage. The x-axis points out of the page of the paper. }
\end{center}
\end{figure}
 The secondary electrons drift inside the electric field to two detection planes, 
one on each end of the chamber ($z\simeq \pm210 cm$). A large diaphragm made up 
of carbon coated Kapton (central membrane), having a thickness of 70 $\mu$m, is 
stretched between the inner and outer field cages at the center of the TPC 
($z = 0 cm$). The central membrane is maintained at a high voltage with respect 
to the detection planes. The liberated electrons thus drift away from the central 
membrane to the closest end cap of TPC, where their position in the ($r,\phi$) 
plane is determined as a function of time. The mean drift time constitutes a 
measurement of the electron's ionization point along the z-axis, yielding the 
third dimension.

\begin{figure}
\begin{center}
\epsfig{figure=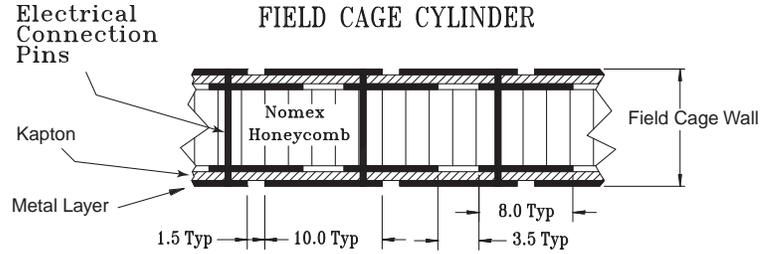,width=10cm}
\caption{\label{tpcF}An IFC showing the construction and composition of the cylinder wall. 
Dimensions are in mm.}
\end{center}
\end{figure}

The requirement of the presence of uniform electric field ($\vec{E}$) inside 
TPC is achieved by a set of field cages. The field cage design consists 
of two concentric cylinders which define the active volume of the TPC. A highly 
uniform electric filed is created along the axis by a series of equipotential 
rings placed on the surfaces of the inner and outer field cages. The field 
magnitude is the greatest at the central membrane having a bias voltage of -31 
kV and decreases in
a steady manner to zero voltage at the ground wires located on either end of TPC.
Irregularities in the spacing of the rings or in the rings themselves will result 
in radial field components and consequently lead to a degradation in the momentum
resolution. The schematic picture of the field cage design is shown in 
Fig.~\ref{tpcF}.

\begin{figure}
\begin{center}
\includegraphics[width=5.5in]{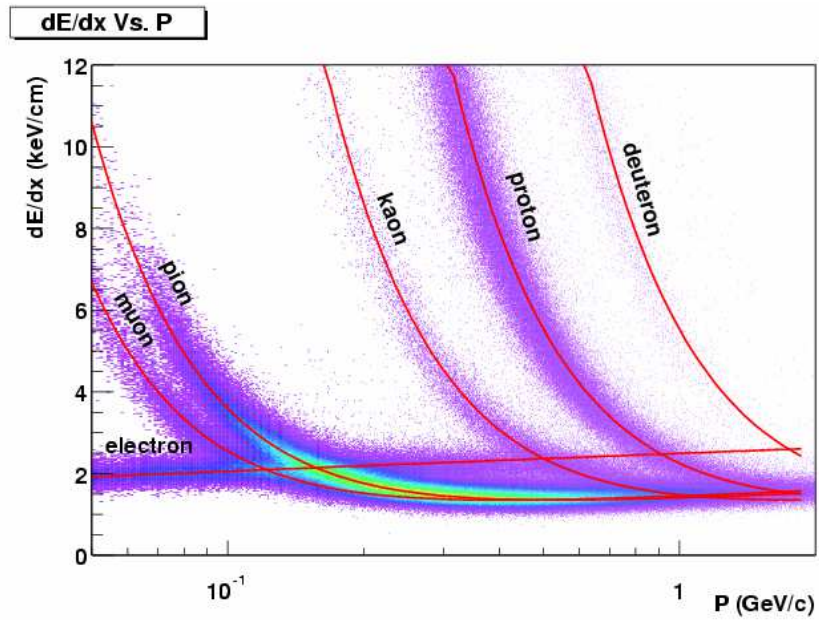}
\caption{\label{dedx}The energy loss distribution of particles in the 
STAR TPC as a function of the momentum of primary particles. The curves
are the Bethe-Bloch function for different particle species. The magnetic 
field was 0.25 T.}
\end{center}
\end{figure}

The electric field and the gas conditions determine the drift velocity ($v_{drift}$)
of the electrons in TPC. $v_{drift}$ is measured in various ways, but a primary
calibration makes use of the mirrored laser system where a single beam is split
into many beams of known locations. The beam ionize the P10 gas and the electrons 
are detected in the MWPC. The known locations of the beams allow for an absolute 
calibration of $v_{drift}$. The drift velocity was calculated to be $5.44 \pm 0.01
cm/\mu s$, with typical time dependent variations of the order of $\sim 6\%$.\\
{\bf Particle Identification Method}\\
Charge particles passing through TPC lose energy through ionization. The total
charge collected from each hit of a track, due to ionization is proportional to
the energy loss of the particle. The mechanism of energy loss is an efficient
tool for identifying particle species in TPC. This method works fine for low
momentum particles but as the particle energy increases, the energy loss becomes
less mass dependent and it's hard to distinguish particles with velocity
$v > 0.7c$. The STAR was designed to separate pions and protons up to 1.2 GeV/c.
For a charge track crossing the entire TPC we obtain 45 $dE/dx$ values from the
energy loss at 45 pad rows. The length over which the energy loss is measured, 
is too short to average out ionization fluctuations. In fact particles loose 
energy while going through the gas in frequent collisions with atoms where a 
few tens of eV are released. Hence it is not possible to measure accurately 
the average $dE/dx$. Indeed, the most probable energy loss is measured. We do
this by removing the largest ionization clusters. The energy loss is distributed
according to the Landau probability distribution. One of the properties of this
distribution is that it's tail dies off very slowly and the dispersion of values
around the mean is very large (theoretically, it's infinite). A typical procedure
to reduce fluctuations from the long Landau tails is to truncate the distribution.
In STAR, we have used $70\%$ truncation, i.e. the highest $30\%$ ionization values
were discarded. Using the remaining values, a truncated mean is computed and this
becomes the basis of any analysis using identified particles in TPC. The measured
truncated mean for primary and secondary charged particles is shown as a function 
of momentum in Fig.~\ref{dedx}.
 The continuous curves are the Bethe-Bloch parameterization used in the analysis 
for different particle hypotheses \cite{bethe}. We see that at the lowest momentum, 
the pions have a greater
ionization energy loss than the electrons which are already in the saturation region
of the curve. The pions cross over the electron band around $ 0.15 GeV/c$ reaching a
minimum at about $0.3 GeV/c$. The pions in their relativistic rise merge with the 
kaons, which are still in the $1/\beta^2$ region, at about $1 GeV/c$.

The Bethe-Bloch formula for energy loss of a particle of charge $z$ (in units of $e$)
and speed $\beta (= v/c)$, passing through a medium of density $\rho$, is given by

\begin{eqnarray}
-\frac{dE}{dx} = 2 \pi N_a {r_e}^2 m_e c^2 \rho \frac{Z}{A} \frac{z^2}{\beta^2}
  \left[ln \left(\frac{2 m_e \gamma^2 v^2 W_{max}}{I^2}\right) - 2 \beta^2 - \delta 
- 2 \frac{C}{Z}\right]
\end{eqnarray}
where\\
$N_a$: Avogadro's number\\
$m_e$: mass of electron\\
$r_e (= e^2/m_e)$: classical electron radius\\
$Z$: atomic number of the absorbing material\\
$A$: atomic weight of the absorbing material\\
$\gamma = 1/\sqrt{1-\beta^2}$\\
$I$: mean excitation potential\\
$W_{max}$: maximum energy transfer in a single collision\\
$\delta$: density correction\\
$C$: shell correction\\

\subsection{Zero Degree Calorimeter (ZDC)}

\begin{figure}
\begin{center}
\epsfig{figure=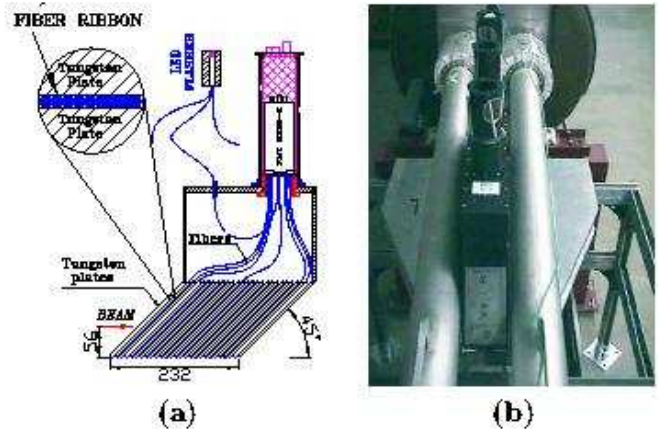,width=10cm}
\caption{\label{zdc}(a) Diagram of ZDC layout. (b) Picture of mounted ZDC 
between the RHIC beam 
lines.}
\end{center}
\end{figure}
The very basic step in collider experiments is to decide when to read out 
the data from the detector. This is called triggering and the scheme may 
be basic or complex depending on the topology of the collisions to be 
selected. At the most basic level, it is required to know if a collision 
has occurred. Such a scheme is called minimum bias trigger. At the most 
complex level one may want to ignore all the events that do not satisfy 
the topology of, e.g. a top quark event. At RHIC a common minimum bias 
trigger scheme was developed for heavy ion running for all the four 
experiments. The trigger is based on two ZDCs, one on each side of the 
interaction point. An inelastic heavy ion collision is accompanied by the 
emission of beam remnant neutrons at high energies and small angles 
($\sim 2$ mrad) with respect to the beam. ZDC is a hadronic calorimeter 
that is designed to detect neutrons. The ZDCs are centered at zero degree 
approximately at $\pm 18$ meters downstream of the interaction point 
and subtend 2.5 mrad. ZDCs are located beyond the DX dipole magnets which 
bend the beams back into their respective orbits. The DX magnets additionally 
act to sweep away charged fragments, so only neutral fragments can reach the 
ZDC. Fig.~\ref{zdc} shows the layout and location of ZDCs in the beam 
line.

The ZDCs employ layers of tungsten absorbers together with Cherenkov fibers for
sampling. The light generated in the fibers is sent to a set of three PMTs (Photo
Multiplier Tube) with the summed analog output of the PMTs used to generate the
ZDC signals. The hadronic Au+Au minimum bias trigger used by STAR requires a 
coincidence between the two ZDCs, with each ZDC signal having a summed analog PMT
output corresponding to $\sim 40 \%$ of a single neutron signal. The readout
electronics used in each of the experiments are identical in design. The signal
from each ZDC is split in two, with one signal being sent to the RHIC for luminosity
monitoring and the other used as input for the experiment trigger.

\subsection{Central Trigger Barrel (CTB)}

An additional design requirement of STAR was the ability to select events in 
real time based on charged particle multiplicity at mid-rapidity. This is 
achieved via a collection of scintillating tiles arranged in a cylindrical 
fashion around the radial exterior of the TPC (see Fig.~\ref{starDet}). This 
collection of tiles is called CTB, the Central Trigger Barrel. Charge 
particles traversing through a CTB tile, generate scintillation light which 
is collected via a PMT. The corresponding output is proportional to the 
number of charged particles that traversed through the slat. CTB is a fast 
detector and along with ZDC, it allows for a powerful charged particle 
multiplicity trigger.
\begin{figure}
\begin{center}
\includegraphics[width=3.5in]{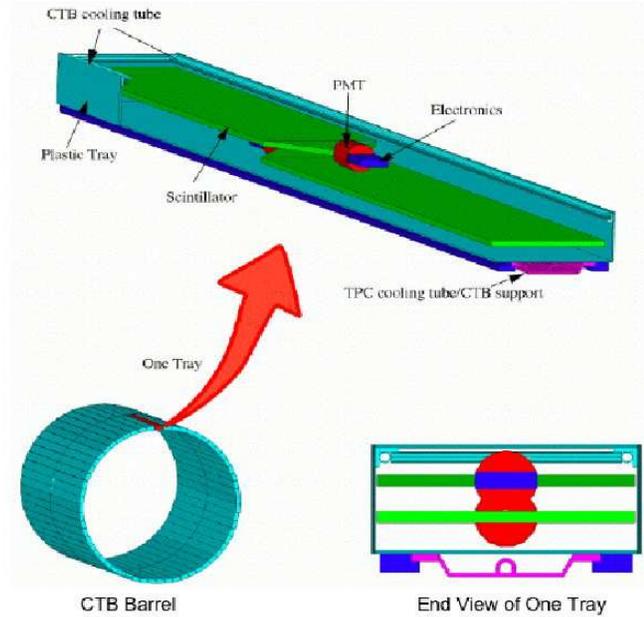}
\caption{\label{ctb}CTB cylinder and the details of tray and slat.}
\end{center}
\end{figure}
The CTB consists of 240 slats of plastic scintillator, 4m in length, with 
$ |\eta| \leq 1$ and full $\phi$-coverage. The slats are housed in aluminum 
trays with two slats per tray. Each slat has one radiator, one light guide 
and one PMT. Fig.~\ref{ctb} shows a segment with two slats. The slat closest 
to the center contain tiles of $112.5 \times 21 \times 1~cm^3 $ and the slat 
away from the center is $130 \times 21 \times 1 ~cm^3$. A single slat 
covers $\Delta \phi = \pi/30$ radians and $\Delta \eta = 0.5$. The PMT 
signals generated by the slat are sent to digitizer boards, each having 
16 inputs. Within each digitizer, the signals are sent to an integrator, 
an 8-bit ADC and then to a discriminator. The output of the discriminator 
can be summed over the barrel and used in the trigger logic. The CTB 
calibration yields an average of 5 ADC counts for a minimum ionizing 
particle.
\begin{figure}
\begin{center}
\includegraphics[width=3.5in]{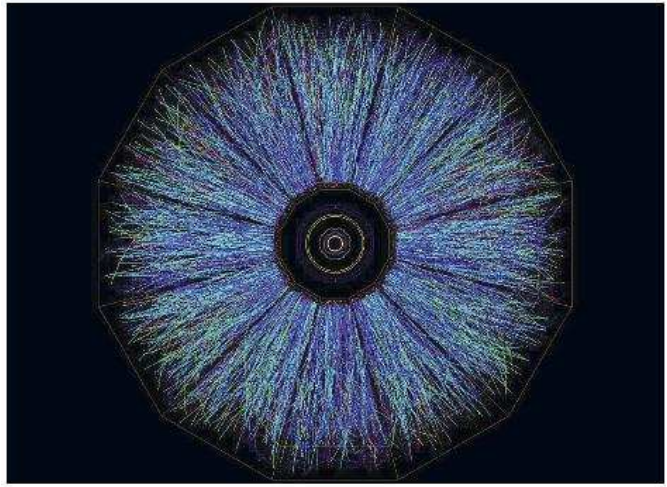}
\caption{\label{event}Beam's eye view of a central event in the STAR TPC. 
This event was drawn by the STAR level-3 on line display.}
\end{center}
\end{figure}

\subsection{Beam Beam Counter (BBC)}

\begin{figure}
\begin{center}
\epsfig{figure=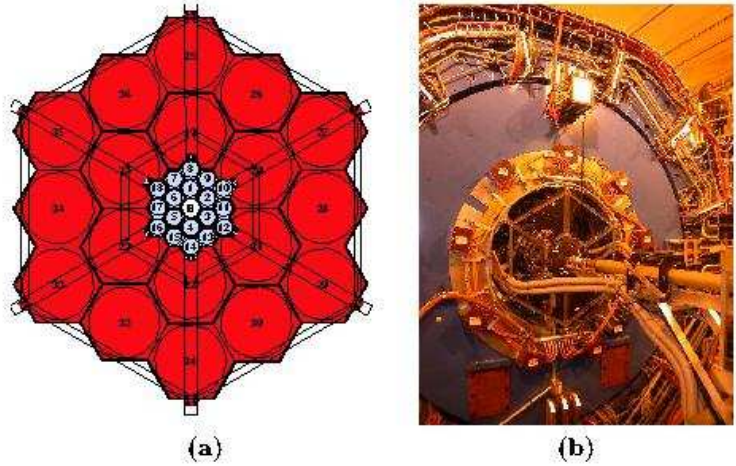,width=10cm}
\caption{\label{bbc}(a) Diagram of BBC layout. (b) Picture of the BBC frame 
mounted on the STAR end cap with the RHIC beam line.}
\end{center}
\end{figure}
The ZDC and CTB are specifically designed for triggering in Au+Au collisions. The 
topology of p+p collisions is vastly different from Au+Au collisions in terms
of multiplicity or the number of tracks per event. In Au+Au collisions, the 
event multiplicity is very high and thus a minimum bias trigger can be implemented
based on many mid-rapidity tracks and spectator neutrons, which p+p collisions lack.
To solve this problem of acceptance gap and for the p+p triggering, Beam Beam Counters
(BBCs) were introduced. Non-Singly Diffractive (NSD) inelastic interactions are 
characterized by the breakup of both the incoming protons. The hard scattered 
partons are realized in the final state as particles near mid-rapidity, while the 
remnant partons produce two ``beam-jets''. The beam-jets are groups of high energy 
hadrons that are focused in the high (near beam) rapidity region. 
A traditional 
trigger for NSD interactions is therefore a set of two scintillating disks that 
are sensitive to the beam jet region. 

The BBCs are annular disk shaped scintillator detectors which are sensitive to charge 
particles. They are situated at $\pm3.5$m from the interaction point as two end caps 
of the TPC. Fig.~\ref{bbc} shows a schematic diagram of BBCs and a picture of BBC 
frame mounted in STAR. Each BBC disk is composed of scintillating tiles arranged in 
a hexagonal closed pack structure. The RHIC beam line passes through the center of 
the BBCs with a 1 cm annular clearance.

Eight PMTs are used for the 18 inner tiles. Each tile has four wavelength shifting
(WLS) optical fibers inserted into circular grooves inscribed within the hexagonal
scintillator to collect scintillation light. The scintillation light is then sent
to the PMTs and digitized via an ADC. The tiles are grouped to allow for radial and
azimuthal segmentation of the readout. The grouping is 1, 2-3, 4, 5-6, 7-9, 10-12,
13-15, 16-18. The fine segmentation of the BBCs was not used in the p+p minimum bias
trigger. Rather the trigger sums, the output of all the tiles on a BBC and a coincidence
of both BBCs firing above noise threshold was required within a time window of 
$\Delta t \equiv |t_{east}^{BBC} - t_{west}^{BBC}| < 17$ ns, which is determined by
the time resolution of the detector.

\subsection{Silicon Vertex Tracker (SVT)}

\begin{figure}
\begin{center}
\epsfig{figure=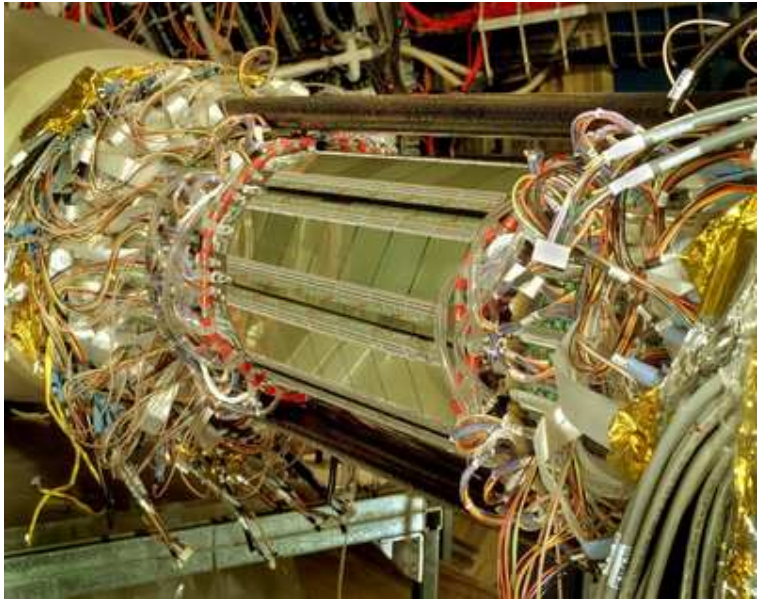,width=10cm}
\caption{\label{svt}Picture of SVT instrumented in RHIC beam line.}
\end{center}
\end{figure}
The Silicon Vertex Tracker (SVT) is a solid state drift detector located just 
outside the beam pipe \cite{svt}. It is designed to improve the primary vertexing, 
track identification to a higher precision and better energy-loss measurement 
for particle identification. It aims to enhance the physics capabilities of 
the TPC. It also helps in the reconstruction of very short-lived particles 
through secondary vertexing close to the interaction zone. In addition, it 
expands the kinematical acceptance for primary particles to very low momentum
by using independent tracking in the SVT alone for charged particles that do 
not reach the active volume of TPC due to the applied magnetic field. The 
total active length of SVT is 44.1 cm. It has full azimuthal coverage with 
$|\eta| < 1$. The detector is made up of 216 independent wafers with a 
dimension of 6.3 cm $\times$ 6.3 cm, containing over 13 million pixels 
multiplexed onto 1300 readout channels. They are arranged in three cylindrical 
layers at distances of about 7, 11 and 15 cm from the beam axis. A ``pixel'' in 
a drift detector is defined by the anode segmentation in one co-ordinate
and the drift velocity divided by the sampling frequency in the drift direction
co-ordinate. The pixel-like readout of the silicon drift detector makes it 
a good choice for the high multiplicity environment in heavy ion reactions 
at RHIC. Since it has three layers, a minimum of three space points are required 
for the determination of the tracking parameters. It has a lower momentum cut 
off  $\sim$ 50 MeV/c, compared to the TPC for which the corresponding value is
$\sim$ 150 MeV/c. The SVT thus enhances the capability of studying low-$p_T$ 
physics.

Position resolution of 20$\mu$m in both co-ordinates as well as energy loss 
(dE/dx) measurements with a resolution of about $7\%$ were achieved with 
STAR-SVT. In high multiplicity environments SVT could be used for improving 
the vertex finding resolution. The overall tracking efficiency for SVT is 
approximately $80\%$. It is useful for the study of heavy-flavor physics as
well.

\subsection{Silicon Strip Detector (SSD)}

\begin{figure}
\begin{center}
\epsfig{figure=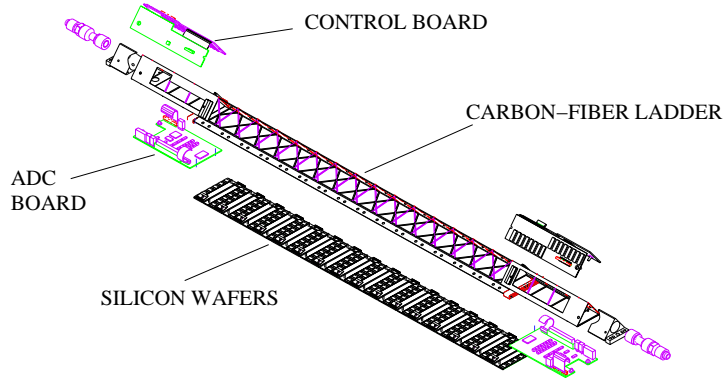,width=10cm}
\caption{\label{ssdL}An SSD ladder showing separately it's components}
\end{center}
\end{figure}
The STAR Silicon Strip Detector (SSD) \cite{ssd} constitutes the fourth layer 
of the inner tracking system. Installed between the SVT and the TPC, the SSD 
enhances the tracking capability of the STAR detector by measuring accurately 
two-dimensional hit position and energy loss for charged particles. It aims 
specifically at improving the extrapolation of TPC tracks through SVT hits and 
increasing the average number of space points measured near the collision, thus 
increasing the detection efficiency of long-lived meta-stable particles. 

The SSD is placed at a distance of 230 mm from the beam axis, covering a 
pseudo-rapidity range of $|\eta| < 1.2$ which leads to a total silicon surface 
close to $1 m^2$. The design of the SSD is based on two clamshells, each containing 
10 carbon-fiber ladders. Each ladder (shown in Fig.~\ref{ssdL}) supports 16 wafers 
using double-sided silicon strip technology (768 strips per slide) and connected to 
the front-end-electronics by means of the Tape Automated Bonded (TAB) technology. 
The ladders are tilted with respect to their long axis, allowing the overlap of 
the detectors in the transverse plane for better hermiticity and alignment
performances. A bus cable transports the analog signals along the ladder to two 
10 bits ADC boards installed at both the ends. 
After digitization, the signals are sent to readout boards
which are linked to the DAQ system through Giga-link optical fibers.

\subsection{Forward Time Projection Chamber (FTPC)}

\begin{figure}
\begin{center}
\includegraphics[width=4.5in]{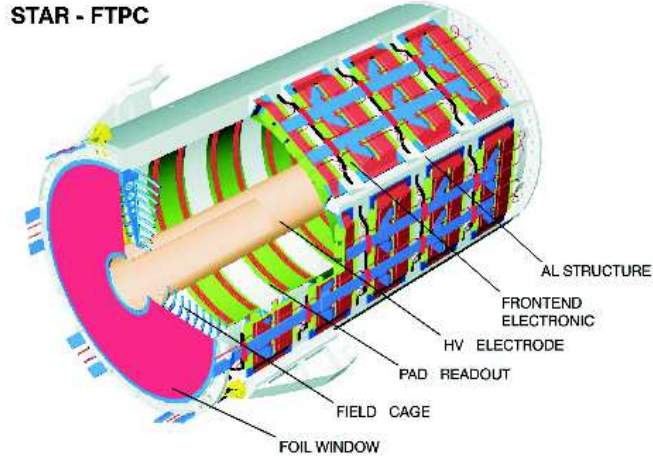}
\caption{\label{ftpc}Layout of Forward Time Projection Chamber}
\end{center}
\end{figure}
The two Forward Time Projection Chambers (FTPCs) are radial drift chambers 
constructed to extend the phase space coverage of the STAR experiment to 
$2.5 < |\eta| < 4.0$. The FTPCs are located along the beam line at 210 cm 
from the center of the TPC. They measure momenta and production rates of 
charged particles as well as neutral strange particles. The FTPC \cite{ftpc}
concept was determined firstly by the high particle density with tracks 
under small angles with respect to the beam direction and secondly by 
the restricted available space inside the TPC. The 
layout of FTPC is shown in Fig.~\ref{ftpc}. It is a cylindrical structure, 
120 cm long and 75 cm in diameter, with a radial drift field and readout 
chambers located in five rings on the outer cylinder surface. Each ring 
has two pad rows and is subdivided azimuthally into six readout chambers. 
The radial drift configuration was chosen to improve the two-track separation 
in the region close to the beam pipe where the particle density is the 
highest. The field cage is formed by the inner HV-electrode, a thin 
metalized plastic tube and the outer cylinder wall at ground potential.
The field region at both ends is closed by a planner structure of 
concentric rings, made up of thin aluminum pipes. The front end electronics 
(FEE), which amplifies, shapes and digitizes the signals, is mounted on the 
back of the readout chambers. Each particle trajectory is sampled up to 10 
times. During the passage of energetic particle, the ionization electrons 
drift to the anode sense wires inducing signals on the adjacent cathode 
surface which are read out by 9600 pads (each $1.6 \times 20 ~mm^2$). The 
above design has some new features.

1. The electrons drift in a radial electric field perpendicular to the 
solenoidal magnetic field.

2. Curved readout chambers are used to keep the radial field as ideal as 
possible.

3. A two-track separation of 1-2 mm is expected, which is an order of 
magnitude better than all previously build TPCs with pad readout.

The FTPCs use a mixture of Ar and $CO_2$ with 50:50 ratio. The FTPC track 
reconstruction is effected by calculating the track points from the charge 
distributions measured by the readout electronics. The grouped track points 
(up to 10 position measurements per track) and the magnetic field map are 
then used to estimate the momentum. The FTPCs helps in complete event 
characterization at forward rapidity.

Each FTPC has a position resolution of 100 $\mu m$ and a two track separation 
of 1 mm. The momentum resolution is between 12 to 15$\%$ with an overall 
reconstruction efficiency lying between 70 to 80$\%$.

\subsection{Time Of Flight (TOF)}

\begin{figure}
\begin{center}
\includegraphics[width=5.0in]{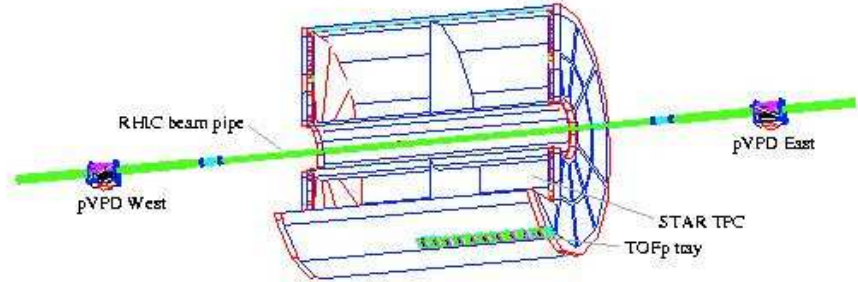}
\caption{\label{tof1}A scale drawing of the locations of pVPD and TOFp 
detectors in relation to the STAR TPC and the RHIC beam pipe. Here TPC 
view is cut away and the STAR magnet as well as other systems are not drawn.}
\end{center}
\end{figure}
Time of Flight (TOF) detector aims for the identification of hadrons 
produced in heavy ion collisions. The STAR TOF consists of two separate 
detectors namely, Pseudo Vertex Position Detector (pVPD) (the 'start' 
detector) and the Time of Flight Patch (TOFp) (the 'stop' detector). The 
TOFp has a phase space coverage of $-1 < \eta < 0$ with $\Delta \phi = 0.11\pi$. 
It extends particle identification up to $p_T \sim 3 GeV/c$ for both p and 
$\overline{p}$. The pVPD consists of two identical detector assemblies
that are positioned very close to the beam pipe, out side the STAR magnet 
on both the sides. The TOFp sits inside the STAR magnet just outside the 
TPC. The location of the collision vertex along the beam pipe is determined
by measuring the arrival time of the forward particle pulses at pVPD and TOFp.
The average of these two arrival times is the event start time. This, with 
the TOFp stop time, provides time interval measurements. The design of the 
pVPD is based on plastic scintillator readout using photomultiplier tubes 
with CAMAC-based digitization. There are three pVPD detector elements on 
each side of STAR at a distance of about 5 m from the interaction region. 
The start resolution attained by the pVPD is around 24 ps, implying a 
pVPD single detector resolution of 58 ps. The total time resolution of the 
system averaged over all detector channels is about 87 ps. This allows a 
$\pi/K/p$ discrimination for momenta up to $\sim 1.8$ GeV/c and direct 
$(\pi + K)/p$ discrimination up to $\sim 3$ GeV/c.
\begin{figure}
\begin{center}
\includegraphics[width=5.0in]{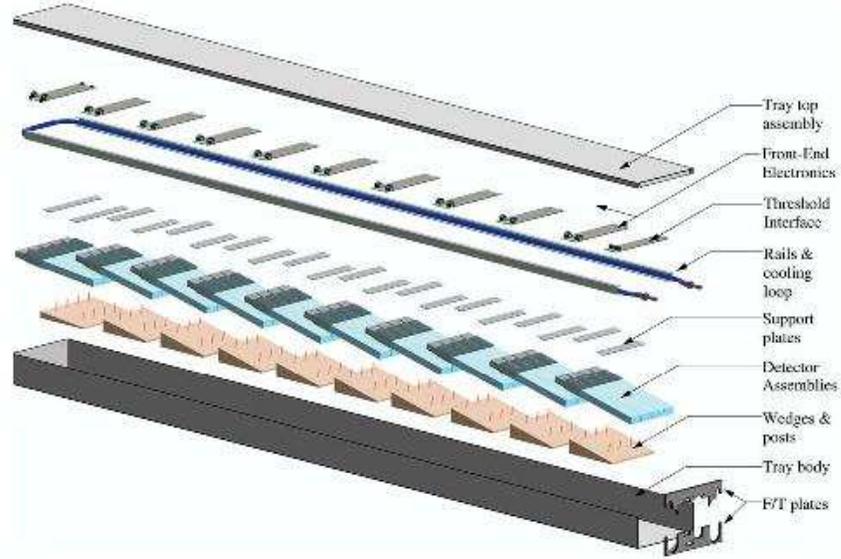}
\caption{\label{tof2}An exploded view of TOFp tray}
\end{center}
\end{figure}
The STAR TOF tray (TOFr) is based on multi-gap resistive plate chamber (MRPC) 
technology. It consists of a highly segmented cylindrical detector immediately
surrounding the TPC. It covers a pseudo-rapidity range of $-1 < \eta < 0$ and
$\Delta \phi = \pi/30$. For the full coverage in STAR, there are 120 trays, 60 
each on east and west sides. Each individual tray is 2.4 m long, 21.3 cm wide
and 8.5 cm deep. Each tray corresponds to 33 MRPCs having 6 readout 
channels~\cite{tof}.

\subsection{Forward Pion Detector (FPD)}

The STAR Forward Pion Detector (FPD) is placed on the east side at about 
7.5 m along z-direction from the interaction region, at a radial distance 
of about 50 cm from the beam line. The FPD consists of a prototype of the 
Endcap Electromagnetic Calorimeter (pEEMC) together with a Pb-glass detector 
array. FPD measures single-spin transverse asymmetry for leading $\pi^0$s 
coming from p+p collisions. It confirms if the colliding beams are polarized 
and can lead to information on the polarization vector at the STAR collision
point.

The pEEMC  part of the FPD is a lead sampling calorimeter comprised of 21 
layers of 5 mm thick Vulcan lead sheets interleaved with 24 layers of 5 mm 
thick Kuraray SCSN-81 plastic scintillator sheets. The total material is 
of approximately 21 radiation length. The layers are machined into 12 
optically isolated tiles in a 3 $\times$ 4 pattern and thus forming 12
towers. The collection and transportation of scintillation light is done 
using 0.83 mm diameter wavelength shifting fibers inserted into 
``sigma grooves'' machined in the scintillator. The other part of the FPD,
called the Shower Maximum Detector (SMD), sits behind the sixth layer of 
pEEMC with about 5 radiation length of pEEMC material in front of it. It is 
comprised of two orthogonal planes of finely segmented scintillator strips.
There are 60 horizontal and 100 vertical strips. Each strip has a 
transverse profile resembling an equilateral triangle with an apex-to-base 
height of 5 mm. Optical isolation was achieved by wrapping individual 
triangular strips with 50 $\mu$m of aluminized mylar. The strips are arranged
parallel to each other such that any two adjacent ones have a cross section 
in the form of a parallelogram~\cite{fpd}.
\begin{figure}
\begin{center}
\includegraphics[width=3.0in]{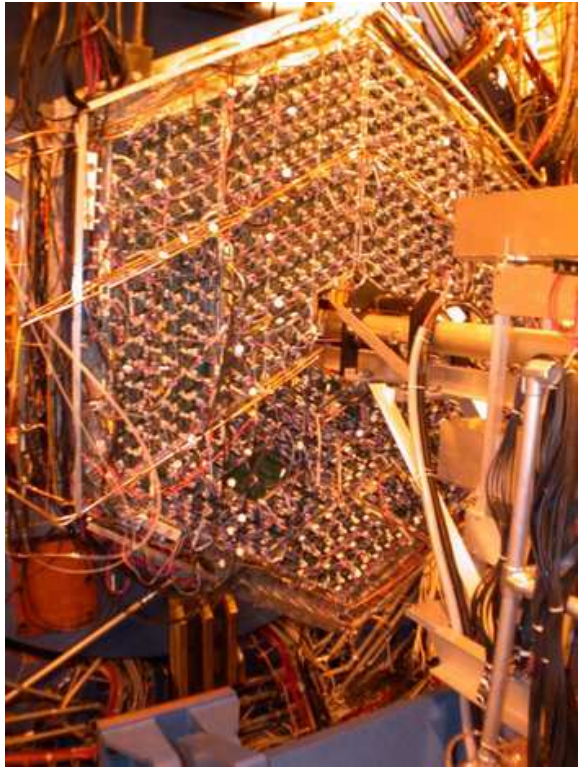}
\caption{\label{pmd}Beam view of STAR PMD installed in RHIC beam line (view from tunnel).}
\end{center}
\end{figure}

\subsection{Photon Multiplicity Detector (PMD)}

The STAR Photon Multiplicity Detector (PMD) is a highly granular pre-shower 
detector aiming at detecting photons in the forward rapidity region, where 
high particle density precludes the use of a calorimeter. It is situated at 
a distance of 5.4 m from the interaction region on the east side of the STAR 
experiment, covering a pseudo-rapidity interval of $-3.7 \leq \eta \leq -2.4$ 
with full azimuthal coverage~\cite{pmd}. PMD measures the multiplicity and 
spatial distribution of photons on an event by event basis. Using similar
information of charged particles from other detectors like FTPC, PMD can 
address various physics issues like: (a) determination of reaction plane and 
the probes of thermalization via studies of azimuthal anisotropy and flow 
~\cite{pmdFlow}, (b) critical phenomena near the phase boundary leading to 
fluctuations in global observables like multiplicity and pseudo-rapidity 
distributions ~\cite{bMohantyTH, pmdPRL} and (c) signals of chiral symmetry 
restoration \cite{bMohantyTH}.

The STAR PMD consists of a pre-shower and a charge particle veto (CPV) plane. 
Each plane consists of a large array of hexagonal cells (41,472 in each plane)
which are tiny gas proportional counters. A mixture of $Ar$ and $CO_2$ in the 
ratio of 70:30 is used as the sensitive medium. The cells are physically 
separated from each other by thin metallic (copper) walls to contain 
$\delta$-electrons. A honeycomb of $24 \times 24$ cells form a unitmodule in 
the form of a rhombus. A set of unitmodules are enclosed in a gas tight 
chamber called a supermodule. The number of unitmodules in a supermodule 
varies from 4 to 9. Each detector plane consists of 12 such supermodules. A 
5 mm thick steel support plate and a 15 mm thick lead plate together form a 
converter of thickness $3X_0$ which is sandwiched between the CPV and the 
pre-shower planes. For a supermodule, the metallic walls of the honeycomb form
a common cathode kept at a large negative potential. The individual anode 
wires (gold coated tungsten) in the cells are kept at ground potential and 
are connected to the readout electronics. GASSIPLEX chips \cite{gassiplex} 
have been used in the front end electronics (FEE) with C-RAMS based readout.
\begin{figure}
\begin{center}
\includegraphics[width=6.0in]{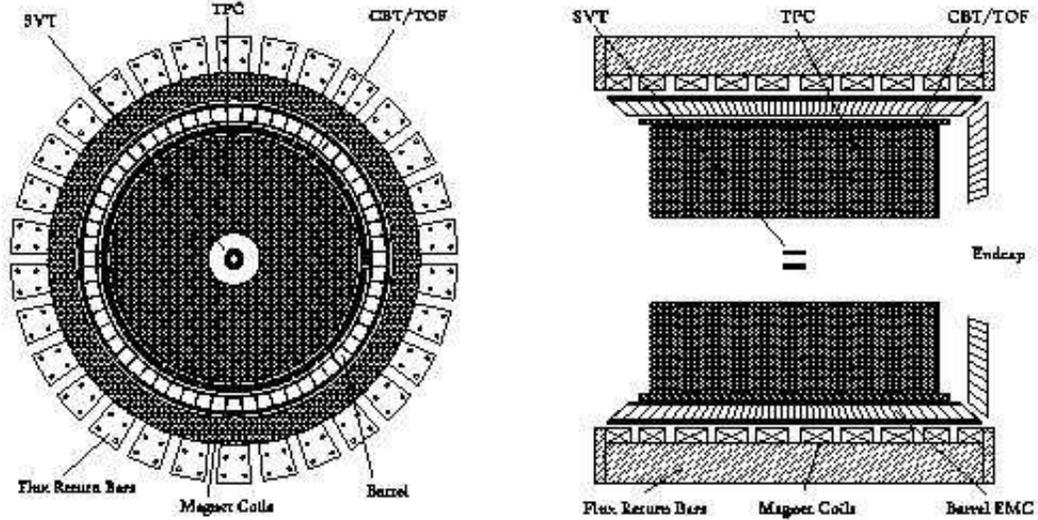}
\caption{\label{bemc1}Cross-sectional views of STAR detector showing BEMC}
\end{center}
\end{figure}

In PMD, a photon passing through the lead converter produces an electromagnetic 
shower. These shower particles produce signals in several cells of the sensitive 
volume of the detector. Charged hadrons usually affect only one cell and 
produce a signal resembling that of Minimum Ionizing Particles (MIPs). The 
thickness of the converter is optimized for high conversion probability of 
photons, limiting the transverse spread of showers to minimize their overlap 
in a high multiplicity environment. In order to have better hadron rejection 
capability, the charge particle veto detector is placed before the lead converter. 

\subsection{Barrel Electromagnetic Calorimeter (BEMC)}

\begin{figure}
\begin{center}
\includegraphics[width=6.8in]{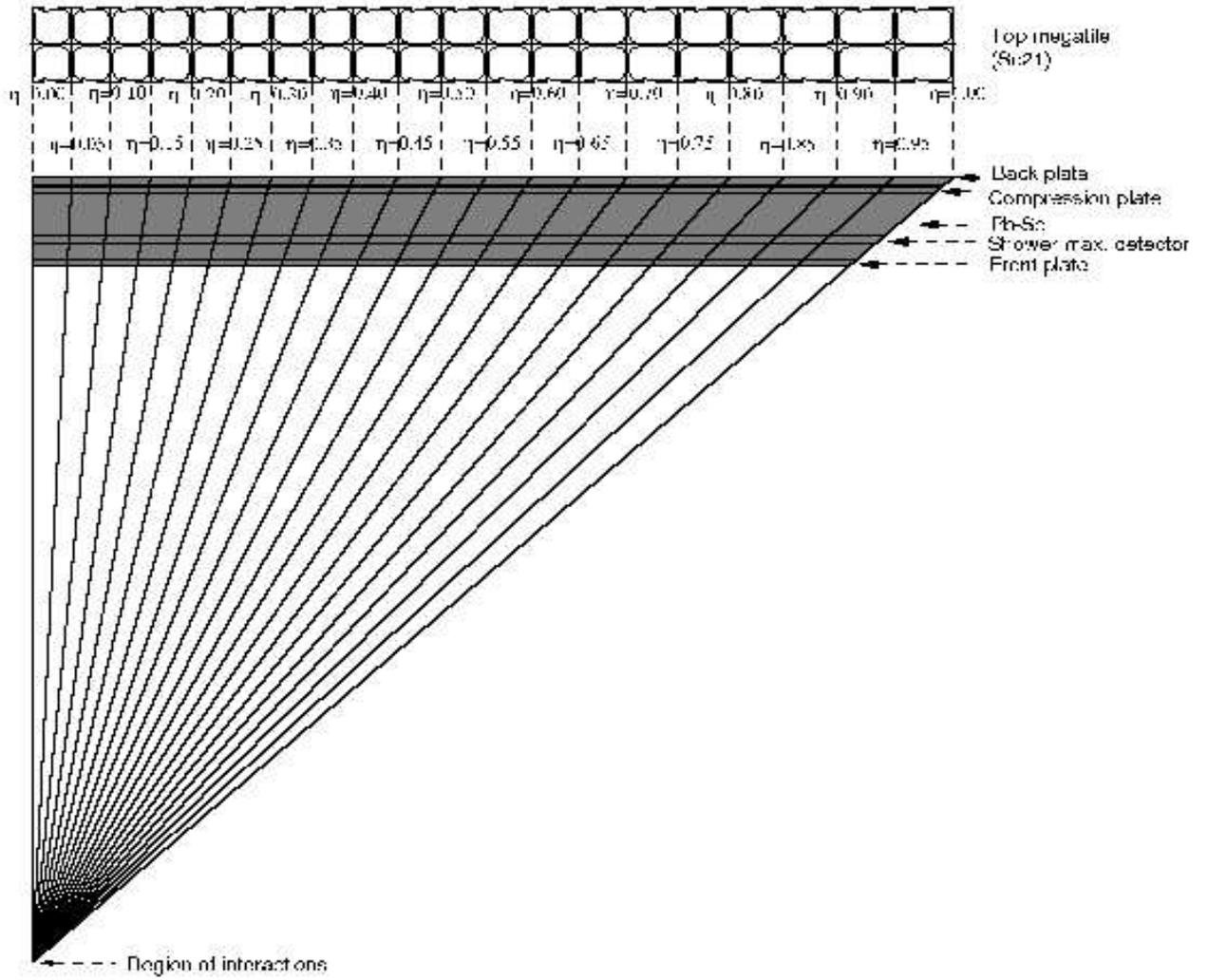}
\caption{\label{bemc2}Side view of a calorimeter module showing the projective 
nature of the towers. The 21st mega-tile is also shown in plan view.}
\end{center}
\end{figure}

\begin{figure}
\begin{center}\includegraphics[width=4.0in]{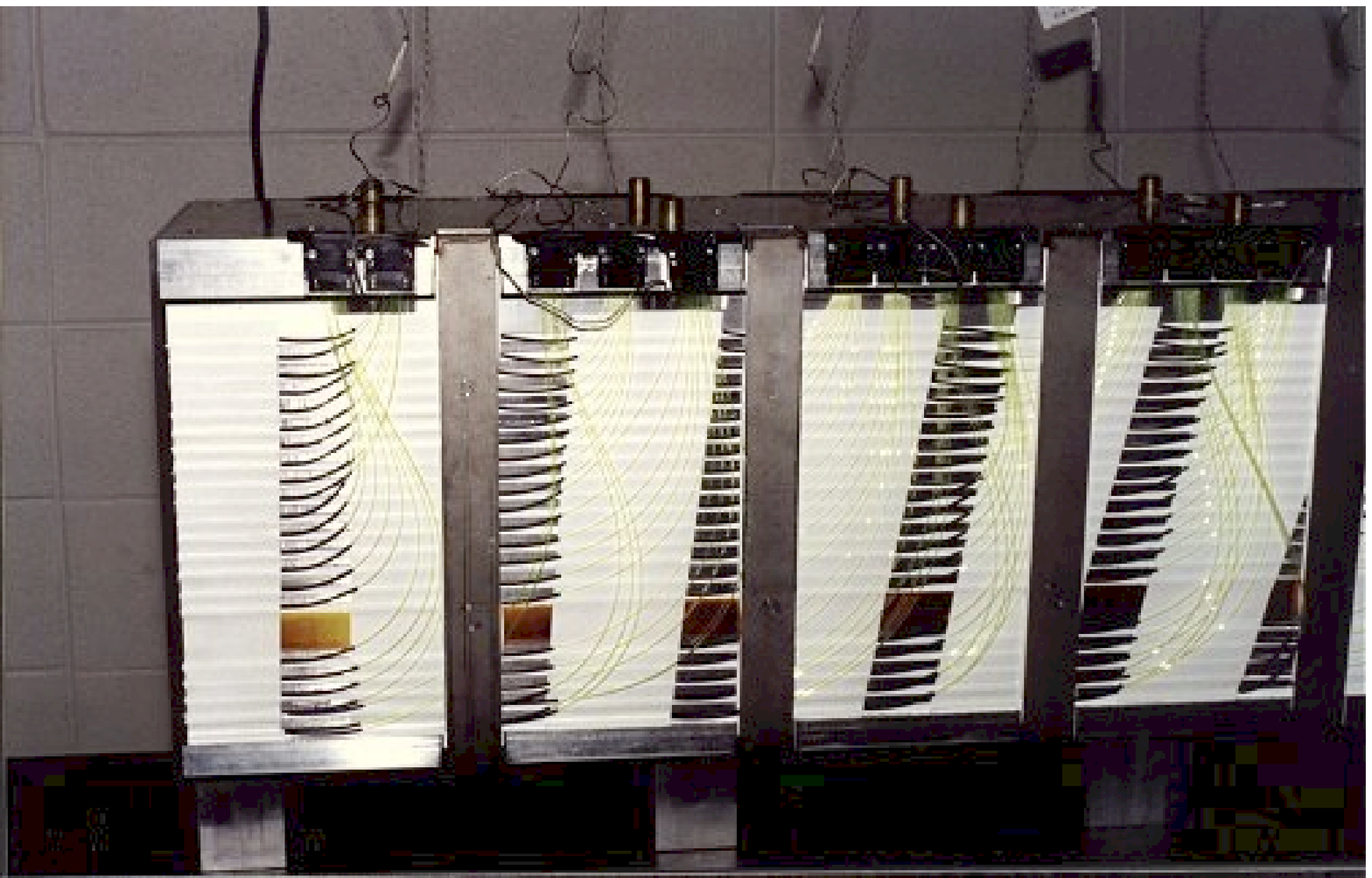}
\caption{\label{bemc3}Photograph of a BEMC module taken near the $\eta = 0$ end 
showing the projective towers and the WLS fiber routing pattern along the side 
of the modules. The WLS fibers terminate in 10 pin optical connectors mounted 
on the back (top in the photo) plate of the module.}
\end{center}
\end{figure}
The STAR Barrel Electromagnetic Calorimeter (BEMC) triggers on and used to
study rare, high $p_T$ processes (jets, leading hadrons, direct photons, heavy 
quarks) with a large acceptance for photons, electrons, $\pi^0$ and $\eta$ mesons 
in a variety scenarios (polarized pp through AuAu collisions). The BEMC permits 
the reconstruction of $\pi^0$'s and isolated (direct) photons at relatively high 
$p_T \approx 25-30$ GeV/c. It is capable of identifying single electrons and 
pairs in intense hadron backgrounds from heavy vector mesons and W and Z decays. 
All these measurements require precise electromagnetic shower reconstruction
with high spatial resolution. This is achieved by the implementation of shower 
maximum detectors (essentially two layers of gas wire pad chambers) within the 
BEMC lead/scintillator stack. These enable measurement of shower distribution 
with high spatial resolution in two orthogonal (transverse) directions.
\begin{figure}
\begin{center}
\includegraphics[width=7.0in]{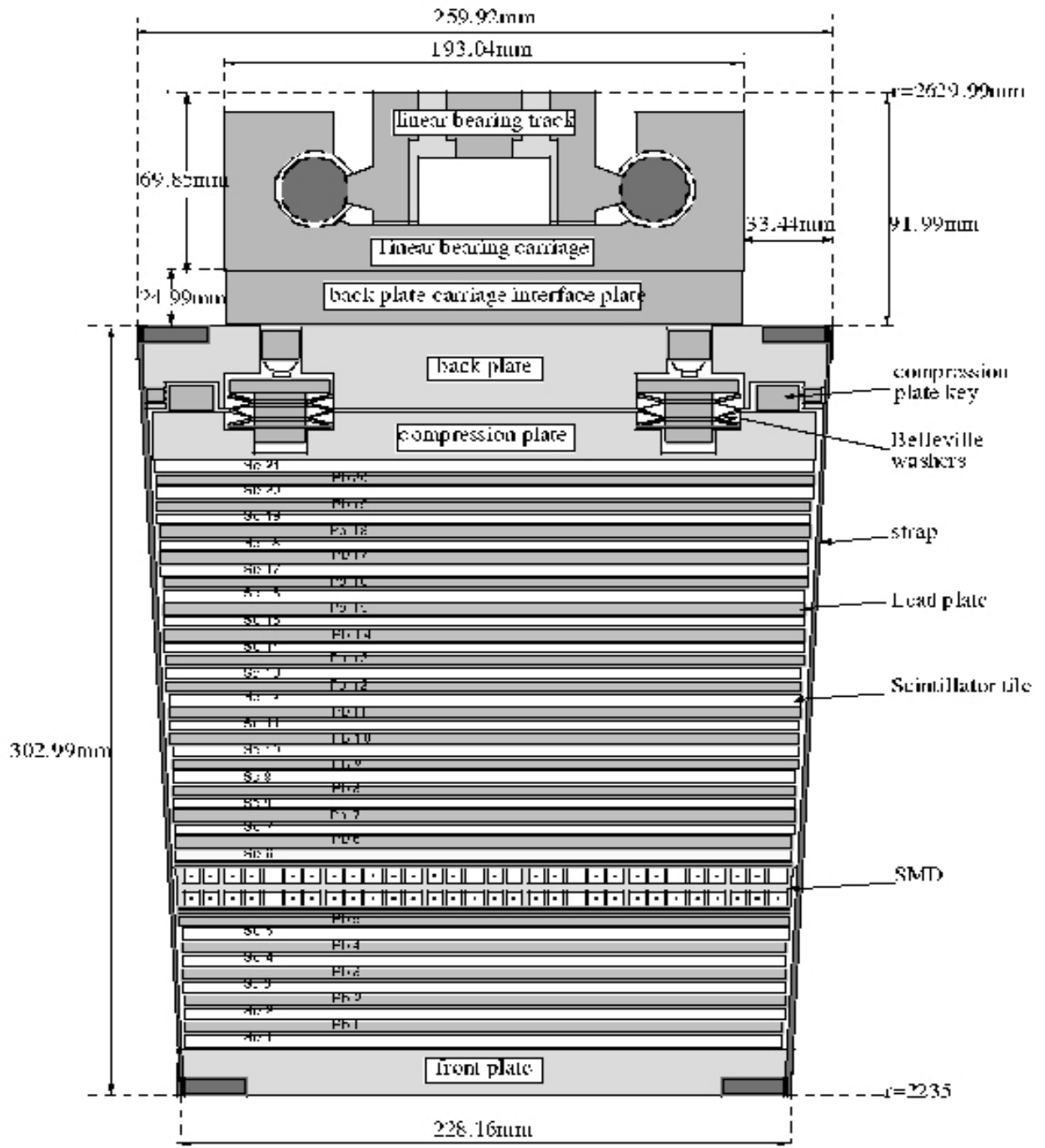}
\caption{\label{bemc4}Side view of a STAR BEMC module showing the mechanical 
assembly including the compression components and the rail mounting system. 
Also shown is the location of the two layers of SMD at a depth of approximately 
$5X_0$ from the front face at $\eta = 0$.}
\end{center}
\end{figure}

\begin{figure}
\begin{center}
\includegraphics[width=4.5in]{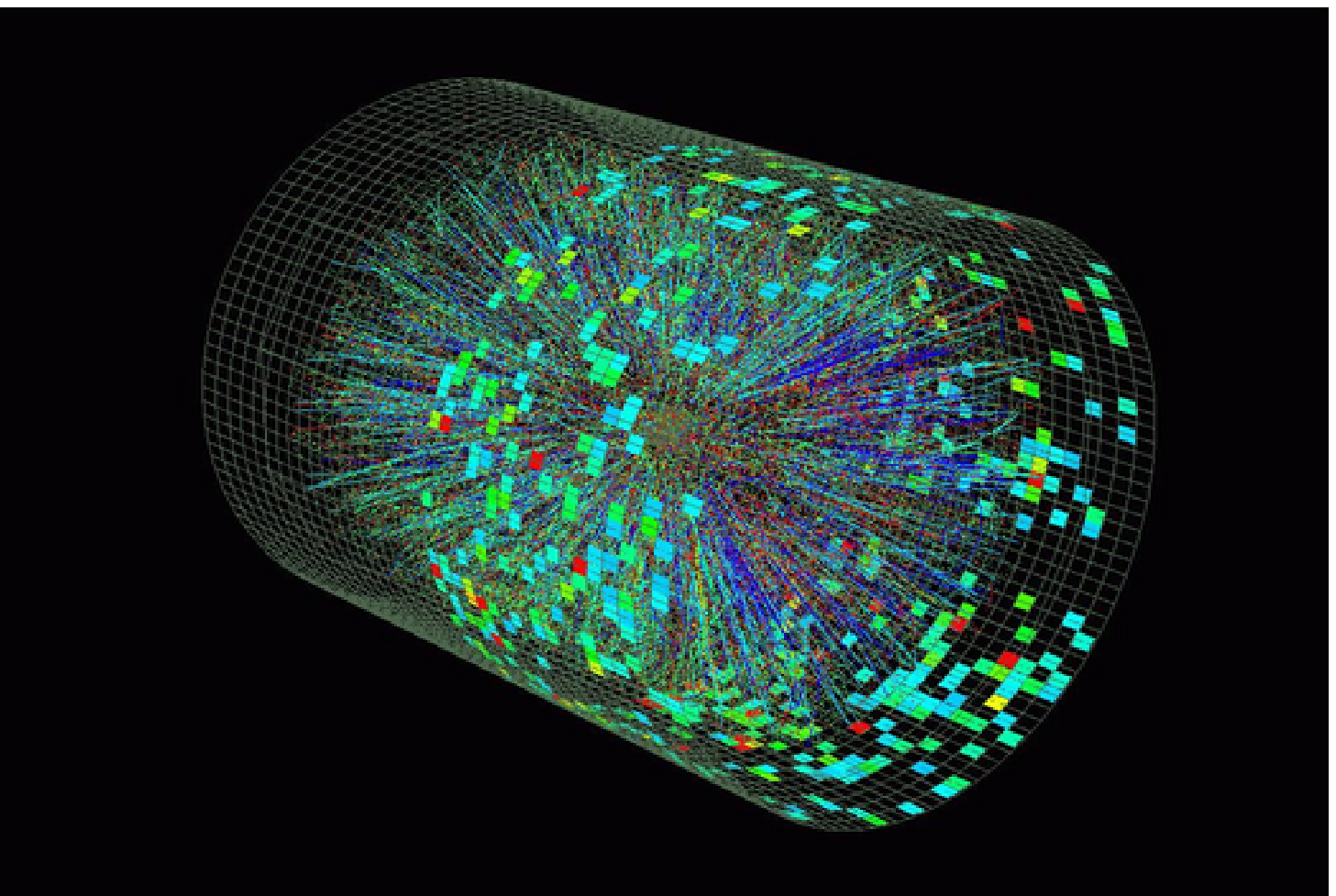}
\caption{\label{bemcEvt}The STAR BEMC event display. }
\end{center}
\end{figure}
The BEMC~\cite{bemc} is located inside the aluminum coil of STAR solenoid and 
covers a range of $|\eta| \leq 1.0$ and $2 \pi$ in azimuth. Thus it matches 
the acceptance of full TPC for tracking. This is shown schematically in 
Fig.~\ref{bemc1}. The front face of the calorimeter is at a radius of 
$\approx 220$ cm from and parallel to beam axis. The design for BEMC includes 
a total of 120 calorimeter modules, each subtending $6^0$ in $\phi ~(\sim 0.1$ 
rad) and 1.0 unit in $\eta$. These modules are mounted so that there are 60 
of them in full $\phi$ with 2 for the complete $\eta$ range for every $\phi$. 
The modules are segmented into 40 towers, 2 in $\phi$ and 20 in $\eta$, with 
$\Delta\phi, \Delta\eta$ for each tower being 0.05 and 0.05 respectively. At the
radius of the inner face of the detector, the tower size is  ~$\sim ~10 
\times 10 ~cm^2$ at $\eta = 0$ which increases towards $\eta = 1$. The 
full BEMC is thus segmented into 4800 towers, arranged in a projective geometry 
each of them pointing back to the center of the interaction diamond. 
Fig.~\ref{bemc2} shows a schematic side view of a module illustrating the 
projective nature of the towers in the $\eta$-direction while Fig.~\ref{bemc3} 
shows a photograph of the $\eta = 0$ end of a module after assembly, 
before the light tight covers are put in place. 
\begin{figure}
\begin{center}
\includegraphics[width=6.5in]{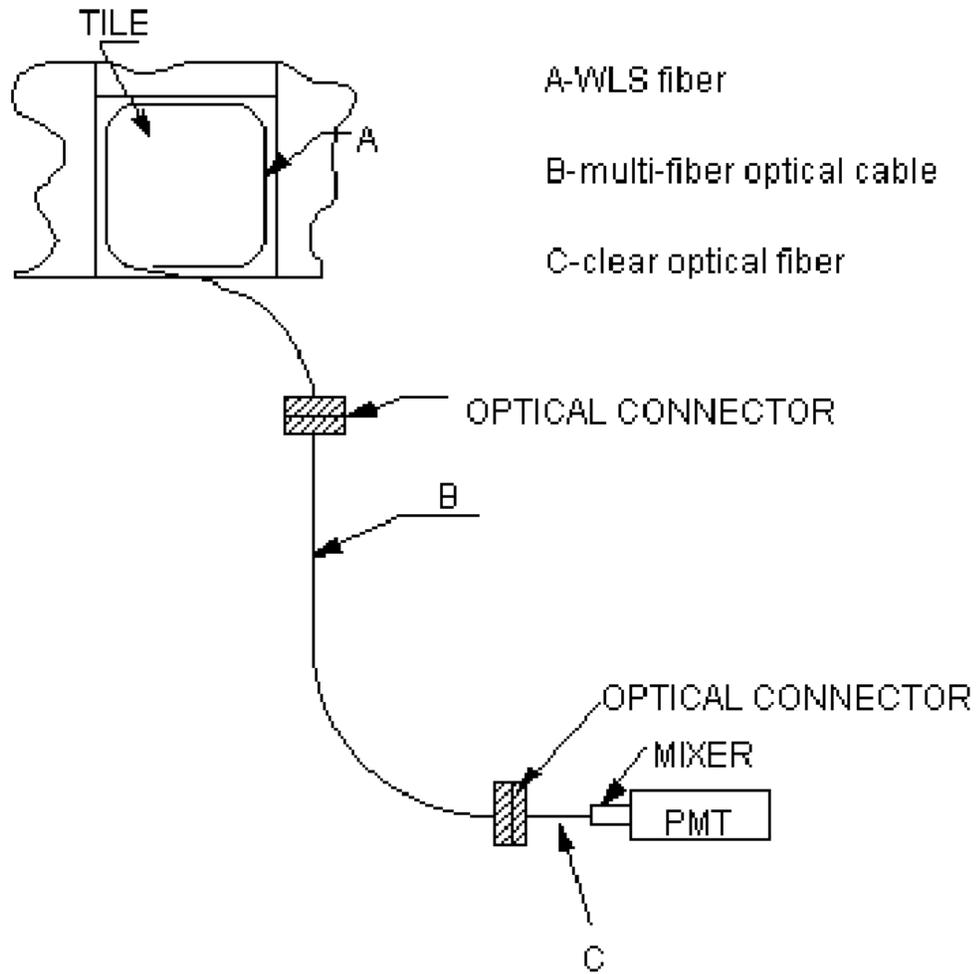}
\caption{\label{bemcOptical} Schematic diagram of the BEMC optical system 
illustrated for a single tile.}
\end{center}
\end{figure}

\begin{figure}
\begin{center}
\includegraphics[width=6.5in]{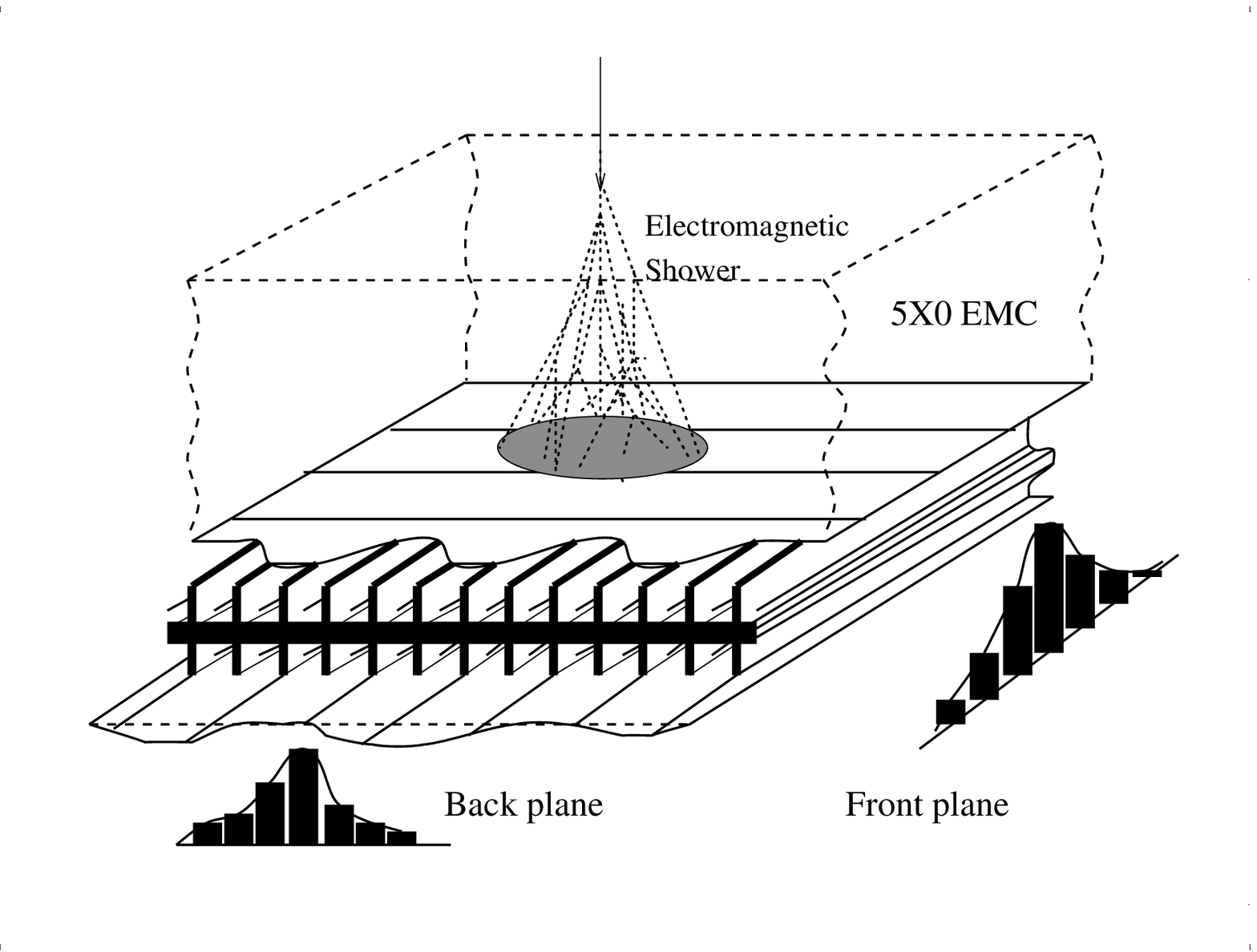}
\caption{\label{bsmd} Schematic illustration of the double layer 
STAR BEMC SMD. Two independent wire layers, separated by an aluminum extrusion, 
image electromagnetic showers in the $\eta-$ and $\phi-$ directions on 
corresponding pad layers.}
\end{center}
\end{figure}
The BEMC is a sampling calorimeter and the core of each BEMC module consists of a lead
scintillator stack and shower maximum detectors situated approximately 5 radiation 
lengths from the front of the stack. There are 20 layers of 5 mm thick lead, 19 
layers of 5 mm thick scintillator and 2 layers of 6 mm thick scintillator. The later 
ones, which are the thicker scintillator layers, are used in the pre-shower portion 
of the detector. Fig.~\ref{bemc4} shows an end view of a module showing the mounting 
system and the compression components. 

BEMC towers provide a precise energy measurement for isolated electromagnetic 
showers and the high spatial resolution provided by the SMD is essential for 
$\pi^0$ reconstruction, direct photon and electron identification. Information 
on shower position, shape, signal amplitude and the electromagnetic shower 
longitudinal development, are provided by the BEMC.

Fig.~\ref{bsmd} shows the conceptual design of the STAR BEMC SMD. 

\subsection{Endcap Electromagnetic Calorimeter (EEMC)}

The STAR endcap electromagnetic calorimeter (EEMC)~\cite{eemc} provides full 
azimuthal coverage for high-$p_T$ photons, electrons and electromagnetically 
decaying mesons over a pseudo-rapidity range $1.086 \leq \eta \leq 2.00$. It 
has a scintillating-strip shower-maximum detector (SMD) to provide $\pi^0/\gamma$ 
discrimination over an energy range of 10-40 GeV. The pre-shower and post-shower 
layers of SMD helps in distinguishing electrons and charged hadrons. The EEMC is
crucial for the STAR spin physics program for it's coverage and the jet triggering 
capabilities.

EEMC is an annular Pb/plastic scintillator sampling calorimeter which is divided into
two halves. The geometry is of alternating Pb radiator and scintillator layers. A 
scintillator strip SMD with high position resolution is located after the fifth 
radiator plate. Light from the towers and SMD  is carried on optical fibers outside 
the STAR magnet, to photomultiplier tubes mounted on the rear of the pole-tip.

\vfill 
\eject

\chapter{Transverse Energy Measurement}
\markboth{nothing}{\it Transverse Energy Measurement}

\section{Transverse Energy}

The transverse energy ($E_T$) is the energy produced transverse to the beam
direction. $E_T$ is an event-by-event variable defined as
\begin{equation}
{E_T = \sum_i E_i sin\theta_i ~~~~and~~~~ dE_T(\eta)/d\eta 
  = sin\theta(\eta)dE(\eta)/d\eta.}
\end{equation}
The sum is taken over all particles produced in an event into a fixed but
large solid angle. 

To probe the early stages of the produced fireball, it is ideal to take transverse
observables like $E_T$, $p_T$ etc. This is because, before the collision of two
nuclei, the longitudinal phase space is filled by the beam particles whereas
the transverse phase space is empty. The $E_T$ is produced due to the initial
scattering of the partonic constituents of the incoming nuclei and also by
the re-scattering among the produced partons and hadrons \cite{etGen}.
The $E_T$ production tells about the explosiveness of the interaction. 
In addition, the collision centrality can be selected using the minimum-bias 
$E_T$ distributions. 

In this thesis, the transverse energy production is studied for different
center of mass energies and with different centralities in order to estimate 
the initial energy density produced in the collision of two nuclei and to 
study the particle production mechanism. The initial energy density has been
estimated in the framework of Bjorken boost-invariant hydrodynamic model 
which has been discussed in Chapter 1. The Bjorken energy density which gives
a lower estimate of the energy density produced, has been compared with the
lattice QCD calculations in order to know if a state has been achieved where
one can search for the deconfinement transition. The details of the 
analysis procedure and results has been discussed in the subsequent chapters.

\section{Data Analysis}

We have analyzed the 62.4 GeV Au+Au data based on minimum bias Au+Au collisions
measured by the STAR detector in the 2004 RHIC run. The detectors used for
this analysis include the Time Projection Chamber (TPC) and the Barrel
Electromagnetic Calorimeter (BEMC) in the common phase space at
mid-rapidity i.e. $ 0 < \eta < 1$ and with full azimuthal coverage.

\subsection{Event Trigger} 

The trigger selection is obtained from the Zero Degree Calorimeters (ZDCs)
\cite{zdc}, the Beam Beam Counters (BBCs) and the Central Trigger Barrel
(CTB) \cite{ctb}. The ZDCs are located at $\pm 18$m from the interaction point
and measure neutron energy. The scintillator based BBCs provide the principal
luminosity measurement. The scintillator based CTB surrounds the TPC and measure 
the charged particle multiplicity at mid-rapidity within $|\eta| < 1$. The
coincidence signal from CTB, ZDCs and BBCs provides the minimum bias trigger
for Au-Au collisions. For this trigger a vertex cut of $\sim 50$ cm on BBC
(trigger Id: 35007) or on ZDC (trigger Id: 35004) and a ctbSum $> 15$ cut were
used to remove Ultra-Peripheral Collisions (UPC)-like events. The ZDC 
resolution goes bad at z-distance of $\sim 20$ cm, so a cut of $\pm 30$ cm
was used which removes a large fraction of the bias. The trigger efficiency
was found to be almost $100\%$.

\subsection{Centrality Selection} 

For any analysis it is desired to take a different detector for centrality 
selection, other than the detector used for a specific analysis to avoid 
auto-correlations. TPC and the BEMC are the detectors used for this 
analysis. Taking FTPC for centrality selection has an advantage as it 
has large separation in rapidity both from TPC and BEMC. But
unfortunately, for the data set under study, FTPC had large gain fluctuation and
different electronic losses in east and west sides, which in turn affects the 
multiplicity distribution. ZDC could not be used for centrality selection as well 
for its low resolution for high central events. Hence TPC uncorrected mid-rapidity 
multiplicity, within pseudo-rapidities $|\eta| < 0.5$ and $|V_z| < 30$ cm, was 
used for the centrality selection. However, auto-correlation is not significant 
for Au-Au collisions due to large multiplicities.

The events were divided into eight different centrality classes which 
correspond to fraction of the total geometrical cross section from central to
peripheral collisions: $0-5\%, 5-10\%, 10-20\%, 20-30\%, 30-40\%, 40-50\%,
50-60\%, 60-70\%$. Very high peripheral events are not taken for this analysis
because BEMC energy deposition corrected for hadronic contaminations,
sometimes make the tower energy negative for very peripheral events.
This is because the correction is done in a statistical fashion. 
Table~~\ref{centrality} enlists the TPC uncorrected multiplicity used for
the centrality definitions.
\begin{table}[htb]
\begin{tabular}{|l|l|l|l|l|}
\hline
 Centrality Bin   &  TPC RefMult ($\geq$)  &  $N_{part}$ & $N_{bin}$ \\
\hline
$(0-5)\%$ & 373 & 347.4 - 4.4 + 3.7 & 899.2 - 64.0 + 64.3 \\
$(5-10)\%$  & 313  & 293.3 - 7.1 + 5.5 & 709.6 - 57.7 + 59.6 \\
$(10-20)\%$ & 222 & $229.1 \pm 8.2$ & 509.5 - 50.3 + 51.9\\
$(20-30)\%$ & 154 & 162.7 - 9.7 + 9.0 & 319.9 - 39.7 + 42.0\\
$(30-40)\%$ & 102 & 112.5 - 10.5 + 8.2 & 192.4 - 29.4 + 33.0\\
$(40-50)\%$ & 65 &  & \\
$(50-60)\%$ & 38 &  & \\
$(60-70)\%$ & 38 &  & \\
%$(70-80)\%$ & 20 &  & \\
%$(80-90)\%$ & 09 &  & \\
\hline
\end{tabular}
\vspace{0.2cm}
\caption{Centrality definitions from different TPC uncorrected reference multiplicity
ranges and the corresponding $N_{part}$ and $N_{bin}$ obtained from Glauber model
calculations.}
\label{centrality}
\end{table}

\subsection{Track Selection} 

TPC is the main tracking detector used for this analysis. The TPC drift volume
was located inside the STAR magnet for particle curvature measurement which
gives the momentum information. The data were taken at magnetic field 
$|B_z| = 0.5$ Tesla, where the z-component is parallel to the beam direction.
The TPC reconstructed momentum resolution is found to be 
$\delta p_T/p_T = 0.01 + p_T/(62.4 ~ GeV/c)$ \cite{resolRef}.
For this analysis only primary tracks are taken with a cut on the distance of 
closest approach (dca): $|dca| < 3$. This is because for initial energy density 
estimation we need primary particles produced in the collision. This forbids 
taking the secondary particles which are produced due to decays and re-scatterings 
at later stages of the fireball evolution. The dca cut significantly 
reduces the pile up events. The longitudinal $z$ position of the interaction 
point is determined on-line by the measured time difference in the ZDCs. 
A cut of $30 ~cm$ was used for the $z$ position of the reconstructed primary 
vertex to ensure nearly uniform detector acceptance. Tracks can leave up to
45 hits on the TPC pad-rows. In this analysis, at least 25 hits are required for
each track to avoid track splitting effects. For track fitting, a minimum of
10 fit points has been taken to select good tracks. Tracks to escape from TPC 
inside the magnetic field of 0.5 Tesla, charged particles have to have a minimum
transverse momentum of 0.15 GeV/c. A minimum transverse momentum of $p_T > 0.15$
GeV/c is used in this analysis. 

\begin{figure}[htbp1111]
\begin{center}
\includegraphics[width=2.5in]{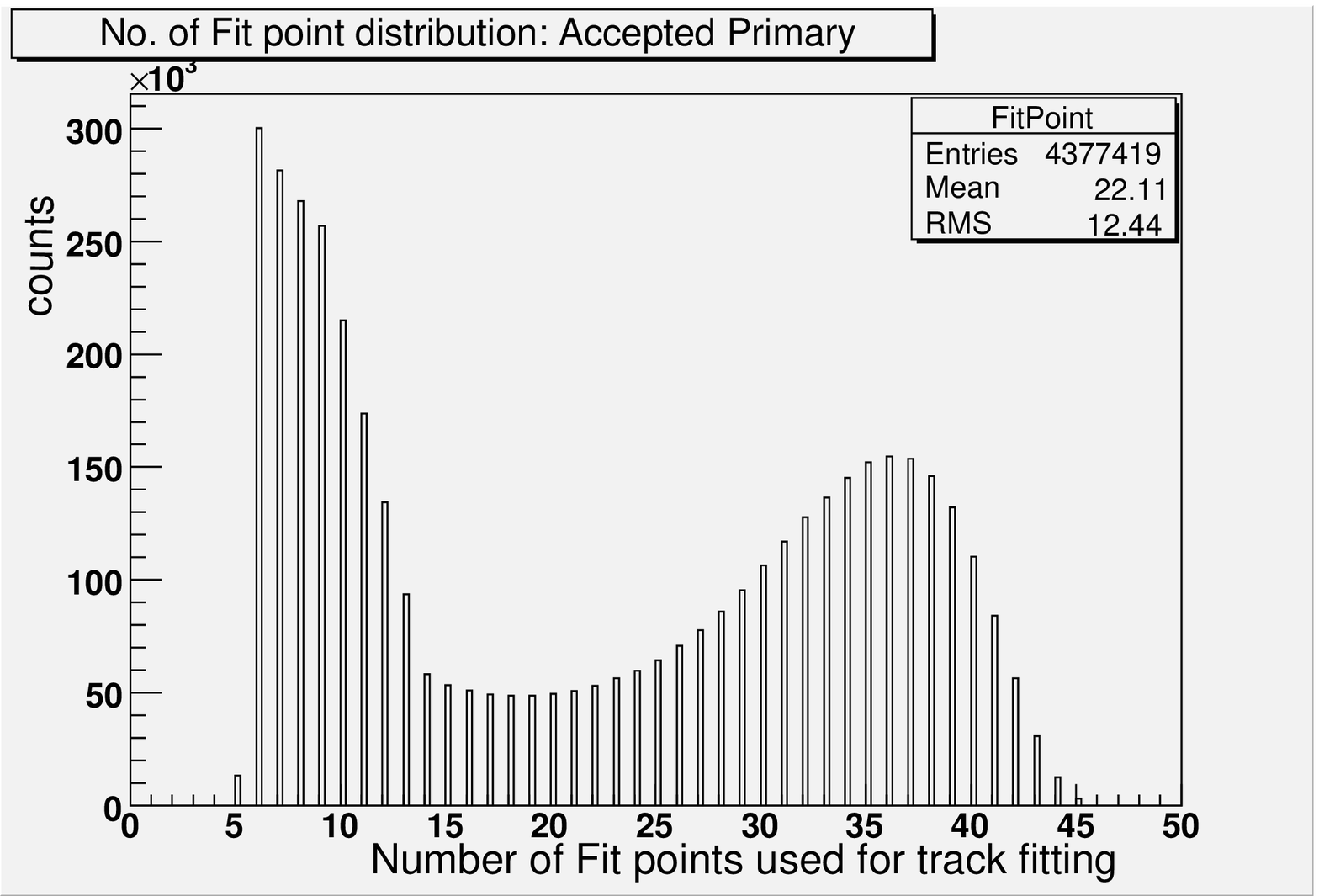}
\includegraphics[width=2.5in]{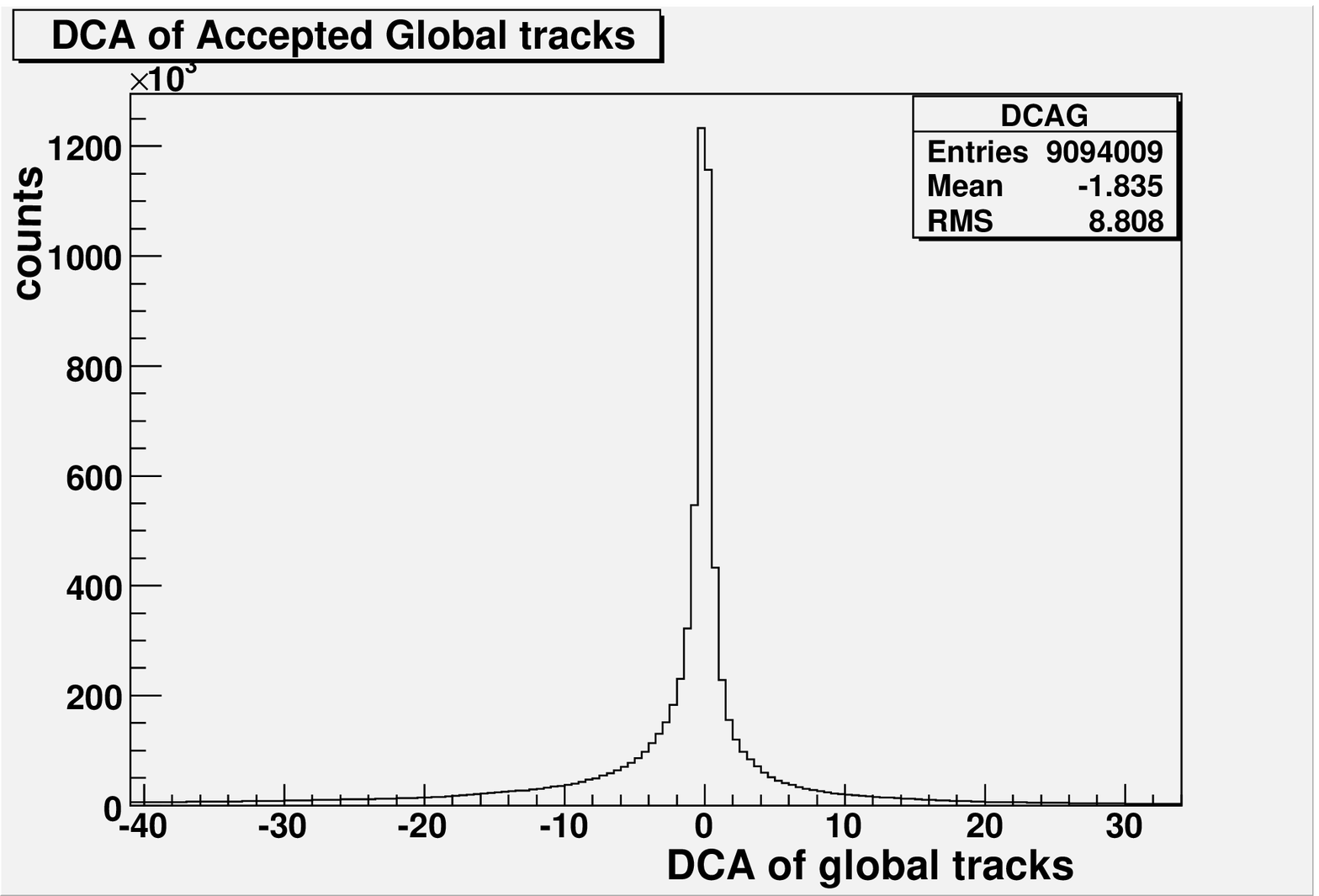}
\caption{The number of fit point distributions for primary tracks (left).  The 
distribution of track dca (distance of closest approach) (right).}  
\label{track}
\end{center}
\end{figure}

Data are analyzed in the common phase space of TPC and BEMC at mid-rapidity
with $0 < \eta < 1.0$ over the full $\phi$-range.

\section{Estimation of Transverse Energy}

The procedure used in this analysis provides an independent measurement of 
electromagnetic and hadronic transverse energy on an event by event basis
in a common phase space. The hadronic component of transverse energy is obtained
from TPC reconstructed tracks after taking into account the long-lived neutral
hadrons which couldn't be detected by TPC. The electromagnetic component 
is estimated from the energy deposited in the electromagnetic calorimeter 
towers, after correcting for hadronic contaminations by projecting TPC 
tracks onto the BEMC. In the following sections the procedure to
estimate both the components of transverse energy is discussed in details which
adds up to give the total transverse energy. This method is identical to that
used in Ref.~\cite{star200GeV}.

\subsection{Hadronic Transverse Energy ($E_T^{had}$)}
The hadronic transverse energy, $E_T^{had}$ is defined as 
\begin{equation}
{E_T^{had} = \sum_{hadrons} E^{had} sin\theta}
\label{etHad}
\end{equation}
where the sum runs over all the hadrons produced in the collision, except
$\pi^0$ and $\eta$ for their short life time. $\theta$ is the polar angle 
relative to the beam axis and the collision vertex position. $E^{had}$
is defined for nucleons as kinetic energy, for anti-nucleons as kinetic 
energy plus twice the rest mass and for all other particles as the total
energy~\cite{star200GeV,helios}. This is given in the following equations.
It takes care of the energy of the participant nucleons, thereby making 
the measured hadronic energy as the hadronic energy produced in the collision.\\
\[E^{had} = \left\{\begin{array}{lll}
\sqrt{p^2 + m^2} - m & \mbox{for nucleons}\\
\sqrt{p^2 + m^2} + m & \mbox{for anti-nucleons}\\
\sqrt{p^2 + m^2}     & \mbox{for all others}
\end{array}
\right. \]

 However, experimentally, $E_T^{had}$ is measured using charged particle
tracks in TPC by using the equation
\begin{equation}
{E_T^{had} = C_0\sum_{tracks} C_1(ID,p) E_{track}(ID,p) sin\theta}
\label{etHadExpt}
\end{equation}
The sum includes all tracks from the primary vertex in the range
$0 < \eta < 1$ and for full azimuthal coverage. $C_0$ is the correction factor
defined as
\begin{equation}
{C_0 = \frac{1}{f_{acc}}\frac{1}{f_{p_TCut}}\frac{1}{f_{neutral}}}
\label{co}
\end{equation}
which includes the effective acceptance $f_{acc}$, the correction 
$f_{neutral}$ for long-lived neutral hadrons not detected by TPC, and
$f_{p_TCut}$, for the TPC low momentum cutoff. $E_{track}(ID,p)$ is the
energy associated with a particular track as defined earlier. This is 
calculated from the measured momentum and particle identity (ID). 
The factor $C_1(ID,p)$ is defined as
\begin{equation}
{C_1(ID,p) = f_{bg}(p_T)\frac{1}{f_{notID}}\frac{1}{eff(p_T)}},
\label{c1}
\end{equation}
where\\  $f_{notID} ~~~~\equiv$ corrections for the uncertainty in the particle ID
determination\\
$eff(p_T) ~\equiv$  momentum dependent tracking efficiency\\
$f_{bg}(p_T) ~~\equiv$  momentum dependent backgrounds.\\
The estimation procedures of these correction factors are discussed below.\\
{\bf Estimation of $f_{p_TCut}$:}\\
Low momentum particles inside the magnetic field, in TPC can't come out
because of very high bending of the tracks. The TPC tracking efficiency 
drops very rapidly for $p_T < 0.15$ GeV/c. In view of the above, in this 
analysis we have accepted only particles with $p_T > 0.15$ GeV/c. 
Estimation through HIJING \cite{hijingPRD} events suggests that 
this cut excludes $5\%$ of the total $E_T^{had}$. Finally $f_{p_TCut}$ 
is estimated through the formula
\begin{equation}
{f_{p_TCut} = \frac{E_T^{had}~(particles ~ with ~ p_T > 0.15 ~GeV/c)}
  {E_T^{had}~(all ~particles)}}
\label{fPtCut}
\end{equation}
taking minimum bias events. The value of $f_{p_TCut}$ is found to be 
$0.953 \pm 0.03$. The systematic error on $f_{p_TCut}$ is estimated by taking
simulated tracks with $p_T > 0.15$ GeV/c and calculating $E_T^{had}$ assuming
pions with $p=0$ and $p= 0.15$ GeV/c. This gives a variation of $3\%$ in 
$E_T^{had}$.\\
{\bf Estimation of $f_{neutral}$:}\\
Since only the charged tracks detected by the TPC are used in this analysis,
we need to correct $E_T^{had}$ to include the contribution from the long-lived
neutral hadrons, principally $n(\bar{n}), ~K_L^0, ~K_S^0,$ and 
$\Lambda(\bar{\Lambda})$. The correction factor is defined as
\begin{equation}
{f_{neutral} = \frac{E_T^{had}~(charged)}{E_T^{had}~(charged) 
+ E_T^{had}~(neutral)}}
\label{fNeutral}
\end{equation}
 HIJING simulation shows this factor doesn't change much from 200 GeV to 
62.4 GeV Au+Au collisions and the value of $f_{neutral}$ is found to be 
$0.82 \pm 0.03$ \cite{star200GeV}.\\
{\bf Particle Identification and Estimation of $f_{notID}$:}\\
As discussed in section 2.3.1, TPC uses the ionization energy loss 
of different hadrons to separate them. The charge collected for each 
hit on a track is proportional to the energy loss of the particle. 
At low momentum, the energy loss is approximately inversely proportional
to the square of particle velocity. The particle identification procedure 
has been already discussed in section 2.3.1, which uses the measurement 
of momentum and truncated mean specific ionization. This procedure could 
only separate particles up to $p_T < 1$ GeV/c.
For $p_T < 1$ GeV/c, assignment was made to the most probable particle
type relative to the Bethe-Bloch expectation. Particles were assumed
to be pions if $<\frac{dE}{dx}>$ differed from this expectation by more
than three standard deviations, or if $p_T$ was greater than 1 GeV/c. The 
uncertainty in this procedure was gauged by calculating $E_T^{had}$ for 
$p_T < 1$ GeV/c both with correct particle assignments and with all 
particles assumed to be pions i.e.
\begin{equation}
{f_{notID} = \frac{E_T^{had}~(p_T < 1 ~GeV/c, ~with ~correct ~PID)}
  {E_T^{had}~(p_T < 1 ~GeV/c, all ~particles ~assumed ~to ~be ~pions)}}
\label{fNotId}
\end{equation}
The correction factor $f_{notID}$ found by this procedure for 62.4 GeV
Au+Au collision is $f_{notID} = 0.991 \pm 0.02$. This was estimated from data.
As this correction factor was calculated from low momentum particles and hence
it doesn't account for the centrality variations in the particle ratios with 
$p_T > 1$ GeV/c \cite{starSpectra}. On the contrary, particles with $p_T > 1$ GeV/c
only account for $20\%$ of the total number of particles. When we take care of 
the centrality dependence in the increase in particle ratios i.e. $p/\pi$, 
$K/\pi$ ratios at higher $p_T$, the estimated $E_T^{had}$ only increases 
by the order of $2\%$, which is within the systematic error of $f_{notID}$.\\
{\bf Estimation of $f_{bg}(p_T)$:}\\
The correction factor $f_{bg}(p_T)$ for background, takes care of electrons,
weak decays and secondary tracks that are mis-identified as primary. This
however depends on the type of the track and is divided into two classes. The
first is for electrons which are mis-identified as hadrons. The second is 
due to weak decays which have been included in $f_{neutral}$ and therefore 
must be excluded from the primary tracks to avoid double counting of their
contribution to hadronic energy. Assuming that this factor doesn't change
much from 200 GeV Au+Au data to 62.4 GeV, we have taken the estimation
from STAR 200 GeV transverse energy paper \cite{star200GeV}. These values are
$f_{bg}(p_T) = 0.84 \pm 0.02 ~(p_T \leq 0.25~GeV/c), ~0.94 \pm 0.02 ~
(0.25 \leq p_T \leq 1 ~ GeV/c)$.\\
{\bf Estimation of $eff(p_T)$:}\\
TPC reconstruction efficiency, $eff(p_T)$, was determined by embedding 
simulated tracks into real events and comparing the simulated input with that
of the final reconstructed event. In order to evaluate the effect of different
particle species on the reconstruction efficiency, pions, kaons and protons 
are embedded in the real events. The final charged track efficiency correction 
is the number averaged over all particle species weighted by their 
relative population. This factor depends on the transverse momentum 
of the tracks and the track density in phase space. 
For central events, the efficiency is $0.70 \pm 0.04$
for tracks with $p_T = 0.25$ GeV/c and reaches a plateau at about 0.8 for
$p_T > 0.4$ GeV/c. It is interesting to note that although the track density
changes from 200 GeV Au+Au  collisions to 62.4 GeV Au+Au  collisions
the track reconstruction efficiency is almost the same 
\cite{star200GeV, starSpectra}. This efficiency correction includes the
efficiency of track reconstruction, the probability of track splitting,
correction for ghost tracks and the dead regions of the TPC.
\begin{table}[htb1]
\begin{center}
\begin{tabular}{|l|l|l|}
\hline
 Correction Factor & Correction \\
\hline
$f_{p_TCut}$ & $0.953 \pm 0.03$ \\
$f_{neutral}$ & $0.820 \pm 0.03$ \\
$f_{notID}$ &  $0.991 \pm 0.02$\\
$f_{bg}(pT)$ & $0.84 \pm 0.02 ~( p_T \leq 0.25~GeV/c)$ \\
& $0.94 \pm 0.02 ~( 0.25 \leq p_T \leq 1 ~ GeV/c)$\\
$eff(p_T)$ & $0.70 \pm 0.04~(p_T \leq 0.25~GeV/c)$\\
& $0.80 \pm 0.04~(0.25 \leq p_T \leq 1.0~ GeV/c)$\\
\hline
\end{tabular}
\vspace{0.2cm}
\caption{Correction factors and their estimated values with uncertainties for
  $E_T^{had}$ for the $5\%$ most central collisions. }
\label{hadCorr}
\end{center}
\end{table}
 The resulting systematic uncertainty in $E_T^{had}$, taking into account 
all corrections added in quadrature is obtained to be $8.5\%$. 
Table~~\ref{hadCorr} enlists all individual corrections and the uncertainties 
in their measurements. The systematic uncertainties from the dynamic cuts used 
in the analysis, when added in quadrature give an error of $5.8\%$ in $E_T^{had}$.
This is discussed separately in subsequent sections. The final systematic 
uncertainty in the estimation of $E_T^{had}$ is found to be $10.3\%$.

The final $E_T^{had}$ is corrected for vertex reconstruction efficiency which
depends on the number of tracks measured in TPC. This varies from peripheral 
to central events. For central events, the vertex reconstruction efficiency
is $94.5\%$. For top $5\%$ central events, the event-by-event resolution 
in $E_T^{had}$ is found to be $9\%$. The estimated values of $dE_T^{had}/d\eta$ 
for different centralities in Au+Au collisions at $\sqrt{s_{NN}}=$ 62.4 GeV  are 
given in Table-~\ref{valuesT}.

\subsection{Electromagnetic Transverse Energy ($E_T^{em}$)}
The electromagnetic transverse energy is the sum of transverse energy of 
electrons, positrons and photons. The largest fraction in this sector comes
from $\pi^0$ decays. Electrons and positrons are included because most of
them are produced in the conversion of photons in the detector materials.
Photons and electrons deposit their full energy in the BEMC. Charged and
neutral hadrons can also deposit significant fraction of their energy in the
BEMC. Hence, this contribution must be subtracted to permit a measurement
of $E_T^{em}$. In order to remove the hadronic contribution from the $E_T^{em}$
measurement, the full spatial profile of the energy deposition of identified
hadrons in EMC has been studied. An extensive experimental database of hadronic
shower clusters in the calorimeter has been obtained, which, with the help of 
TPC tracking, allow a correction of the hadronic background in the calorimeter
\cite{star200GeV}.

\subsubsection{The BEMC Calibration}
The BEMC has a total of 4800 towers. The gain of each tower can have different
values due to i) the variation in their electronic gains, ii) the variation
in operating voltage or iii) could be due to some other intrinsic factors. Hence,
there is a requirement of normalizing the gain of each tower before the real
data taking. This is called gain calibration of the BEMC. The methods of 
calibration of BEMC is discussed in brief in the following paragraphs.

The hadronic particles produced in the heavy ion collisions, when interact 
with BEMC, deposit a widely fluctuating fraction of their energy through 
hadronic showers. In addition, $\sim~ 30-40\%$ of relativistic charged
hadrons penetrate the entire depth of the BEMC without hadronic interaction.
However, when such a non-showering charged hadron has sufficient momentum, it
will behave like a minimum ionizing particle (MIP). When these MIPs traverse 
through the scintillator layers, deposit their total energy uniformly. This
energy deposition (MIP peak) is nearly independent of the incident momentum 
and the particle species. It varies linearly with the total path length 
covered in the scintillator. Due to the projective nature of the BEMC towers, 
the total length of the scintillator increases with $\eta$ and hence the MIP 
peak. The absolute energy of the MIP peak and it's $\eta$ dependence is 
determined from the cosmic tests and from the test beam measurements 
\cite{bemcCalib}.
 
\begin{figure}[htbp11]
\begin{center}
\includegraphics[width=2.5in]{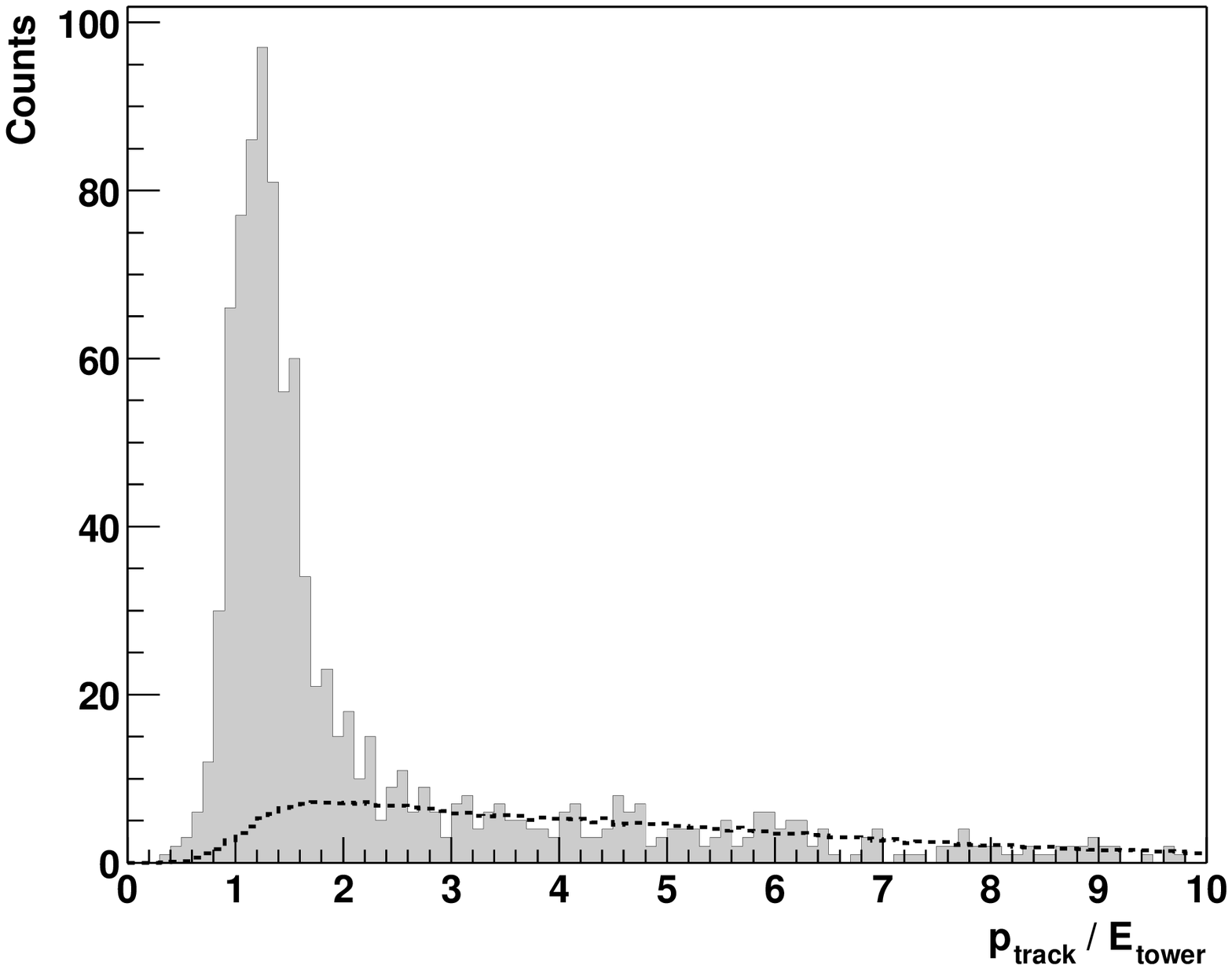}
\includegraphics[width=2.5in]{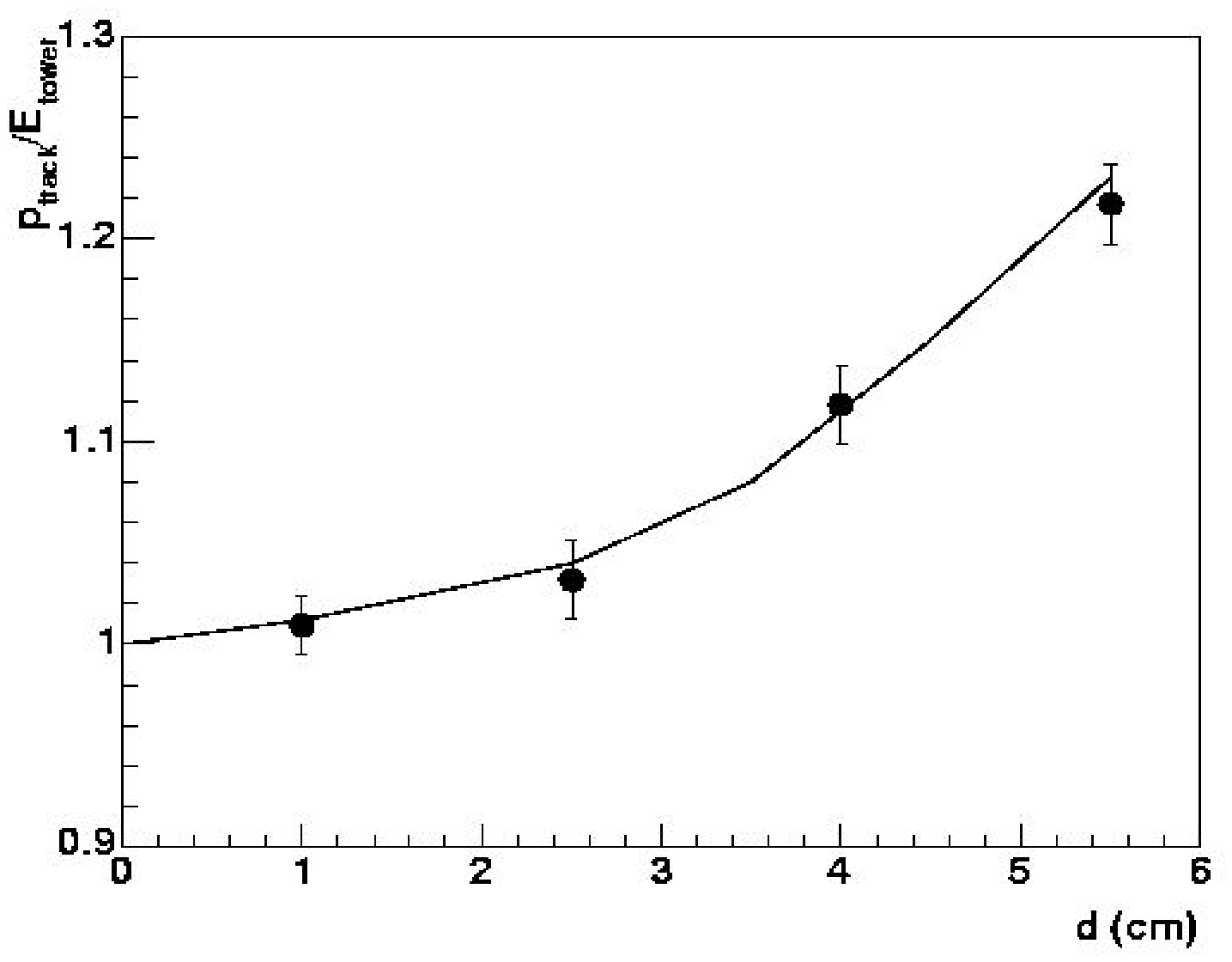}
\caption{a) A typical $p/E_{tower}$ spectrum for electron candidates, selected 
through $dE/dx$ from the TPC, with $1.5 < p < 5.0$ GeV/c. A well-defined electron
peak is observed and the dashed line corresponds to the hadronic background
in the $dE/dx$-identified electron sample. b) Data point are the measured 
$p/E_{tower}$ electron peak positions as a function of the distance to the 
center of the tower. The solid line is a calculation based on full GEANT 
simulation of the detector response to electrons. Fig. Ref.~\cite{star200GeV}}
\label{electron}
\end{center}
\end{figure}
The above factors along with the high yield of charged hadrons, makes it 
feasible to use high energy MIPs for calibration and continuous monitoring 
of variations in the BEMC tower gains. 

The method of MIP-calibration requires that the initial gains should be set
so that MIP peaks from penetrating charged particles, should sit somewhere
within digitizers' (ADC) ranges, reasonably far from their lower and upper
limits. The calibration using MIP-hits has two stages. In the first stage,
a sample of BEMC modules is exposed to an external beam (in STAR, it is the
AGS beam at BNL). The external beam should consists of hadrons, electrons and 
muons in a selected momentum range. The ratios of each tower responses to MIP's
and electron hits i.e. MIP/e-ratios are measured simultaneously. Here ``MIP'' 
is the MIP-peak position and the notation ``e'' is used for the ratio 
$S_e(E_e)/E_e$, where $S_e$ is the mean BEMC signal from electrons of energy $E_e$.
Thus, the MIP/e-ratio represents the energy of electrons which would generate
in the BEMC, the same mean signal as MIPs. In the second stage, after the
BEMC modules are installed in STAR, samples of MIP-hits as close as possible to
those in the test beam are accumulated for each tower and the positions of the
resulting MIP peaks are measured from real events from RHIC runs. This step
essentially completes the procedure of transferring beam-test results to STAR.
For those towers already exposed to test beam, their responses to electron
hits (momenta used in the test beam-stage) can immediately be predicted, using
the known MIP/e-ratios that have been measured from the test beam. For all other
modules, these ratios are expected to be close to those of the tested ones,
provided that the key design tolerances are kept at the module manufacturing stage.

 MIPs of $p_T~\sim$ 1 GeV/c are used for the calibration purposes to make an
compromise between the particle yield $p_T$ and their utility. This is because,
we can't choose very low $p_T$ MIPs, as they hit BEMC at very large angles
for their deflection in high STAR magnetic field. On the other hand, when the
$p_T$-threshold is chosen very high, the useful event rate would be very low
because of the steep drop of particle yield with high $p_T$. 

The absolute calibration of BEMC is also done by selecting identified electron
tracks in the TPC, in a wider momentum range ($1.5 < p < 5.0$ GeV/c). The 
selection of electron candidates in TPC is done by the $dE/dx$ measurements. 
The purity of electron candidates in this momentum range is poorer than it is 
for the low momentum ones. However, the hadron rejection factor obtained from
the TPC $dE/dx$ provides a clear electron signal in the calorimeter. The 
Bethe-Bloch predictions for $dE/dx$ of electrons and heavy particles show that
the main background in this momentum range comes from deuterons and heavier
particles as well as the tails in the distributions of protons and lower mass
particles. Tracks with number of space points greater than 25 are used in this
procedure to reduce systematic uncertainties. In addition, such long tracks
show better $dE/dx$-resolution. It is also required that the tracks should be
isolated in a $3 \times 3$ tower patch in the calorimeter.
%\begin{figure}[htbp12]
%\begin{center}
%\includegraphics[width=3.5in]{geant.eps}
%\caption{}
%\label{geant}
%\end{center}
%\end{figure}

\begin{figure}[htbp13]
\begin{center}
\includegraphics[width=3.5in]{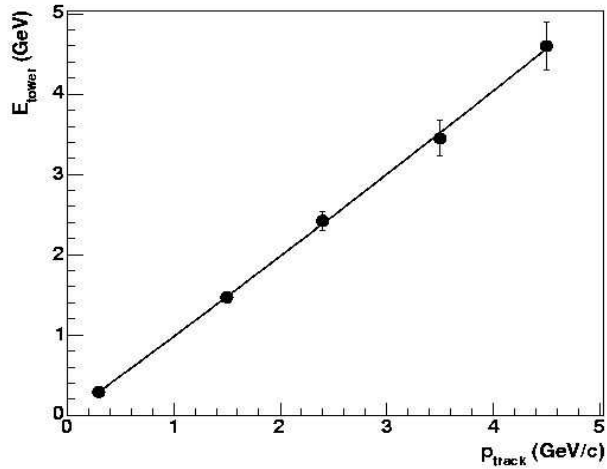}
\caption{The data points show measured energy deposited by electrons in 
the tower as a function of the momentum for distances to the center of the
tower smaller than 2.0 cm. The first point is the electron equivalent energy
of the MIPs. The solid line is a second order polynomial fit of the data 
\cite{star200GeV}.}
\label{ep}
\end{center}
\end{figure}
For a better accuracy of electron identification, finally the energy deposition
by a track in the BEMC tower ($E_{tower}$) is compared to the momentum of the
track ($p$) in the momentum range $1.5 < p < 5.0$ GeV/c. Fig.~\ref{electron} (a)
shows the $p/E_{tower}$ spectrum for the electron candidates, where a well-defined
electron peak is observed. The residual hadronic background in this figure is
evaluated by shifting the $dE/dx$ selection window toward the pion region.
The obtained hadronic background is shown in the figure as the dashed line.
The peak position is not centered at 1, after the hadronic background subtraction.
This shows the leakage of energy to the neighboring towers which are not taken
into account in this procedure. The amount of leakage depend on the distance to
the center of the tower hit by the electron and will shift the peak position
to higher values as this distance increases. This effect is well-reproduced
by the GEANT simulations of the detector response when it is hit by electrons
in the momentum range used in this calibration procedure. This is shown by
Fig.~\ref{electron} (b).  This figure also shows the position of the electron 
$p/E_{tower}$ peak as a function of this distance. The solid line is a prediction 
from GEANT simulations.  Fig.~\ref{ep} shows the energy deposited in the 
calorimeter tower as a function of its momentum for electrons in the case where 
the distance to the center of the tower is less than 2.0 cm. (A distance of 5.0 
cm and 7.5 cm correspond to the border of the tower at $\eta =  0$ and $\eta = 1$ 
respectively.) The first point is the electron equivalent energy of the MIPs. 
A fit to the data with a second order polynomial i.e. $f(x)=a_0+a_1x+a_2x^2$ is 
represented by the solid line. 

The combination of the MIP calibration and the electron calibration of the BEMC,
gives an overall systematic uncertainty  of less than $2\%$ on the total energy
measured by the calorimeter. The stability of the tower response was evaluated by
monitoring the time dependence of the shape of the raw ADC spectra for each tower.
This is the tower response to all particles that reach the calorimeter. The overall
gain variation of the detector was less than  $5\%$ for the entire RHIC run.

\subsubsection{The Hadronic Contaminations in BEMC Energy}
As discussed earlier, hadrons produced in the collisions traverse through the TPC
and then hit the BEMC, as both the detectors share a common phase space and deposit
part of their energy in BEMC. In order to measure the electromagnetic transverse
energy it is essential to subtract the hadronic energy deposited in the calorimeter.
For charged hadrons, the hit locations on the calorimeter are well determined and
in case of an isolated hadron, a cluster of energy is identified easily. However,
in the high density environment of Au+Au collisions, it is difficult to identify
a hadron track. In this case, an average energy deposition based on the measured 
momentum of the incident track is subtracted out. As one is interested in the 
cumulative distribution averaged over many events, where each event has many
tracks, this averaged correction results in a negligible contribution to the
uncertainty in the measured electromagnetic energy. The following is the 
procedure which explains, how the hadronic contaminations in BEMC is taken
care of \cite{star200GeV}. 

The spatial and energy distributions of hadronic showers in the calorimeter 
is studied both from the real data and from the GEANT simulations. A library 
of separate profiles for pions, kaons, protons and anti-protons was obtained 
from GEANT simulations of the detector response in the STAR environment (GSTAR). 
Events were chosen with a uniform momentum distribution in the range $0 < p < 10$ 
GeV/c and with vertex positions limited by $|z_{vertex}| < 20$ cm. The constraint 
on the longitudinal co-ordinate of the vertex ensures that the trajectory of a
particle would extrapolate through only one tower of BEMC. Because of the fact 
that BEMC has a projective geometry, this constraint on the extrapolated track 
is strongly related to the vertex constraint. The simulated tracks were projected 
on BEMC using a helix model for the particle trajectory in the magnetic field. 
The energy distributions and the corresponding mean values are then obtained 
as a function of the momentum, $p$, the pseudo-rapidity of the BEMC towers, $\eta$
and the distance of the incident hit point to the center of the tower, $d$. The 
distributions were then binned in the intervals of $\Delta \eta = 0.2$. For all
particle species, the total mean deposited hadronic energy in a particular
tower has been found to increase approximately linearly with the momentum. 
This shows very little dependence on $\eta$ and decreases with increasing $d$.  
From the real data, the hadronic shower profiles are obtained from the  
minimum-bias events, by projecting tracks on the BEMC. This is achieved by 
accepting tracks which are isolated in a $5 \times 5$ tower patch to ensure that 
the energy in the towers are only from one particle and calculating the energy 
distributions and the mean values. Profiles for all particles excluding electrons 
and positrons, for both positive and negative tracks were recorded with good 
statistics up to $p = 2.0$ GeV/c.
\begin{figure}[htbp13]
\begin{center}
\includegraphics[width=2.5in]{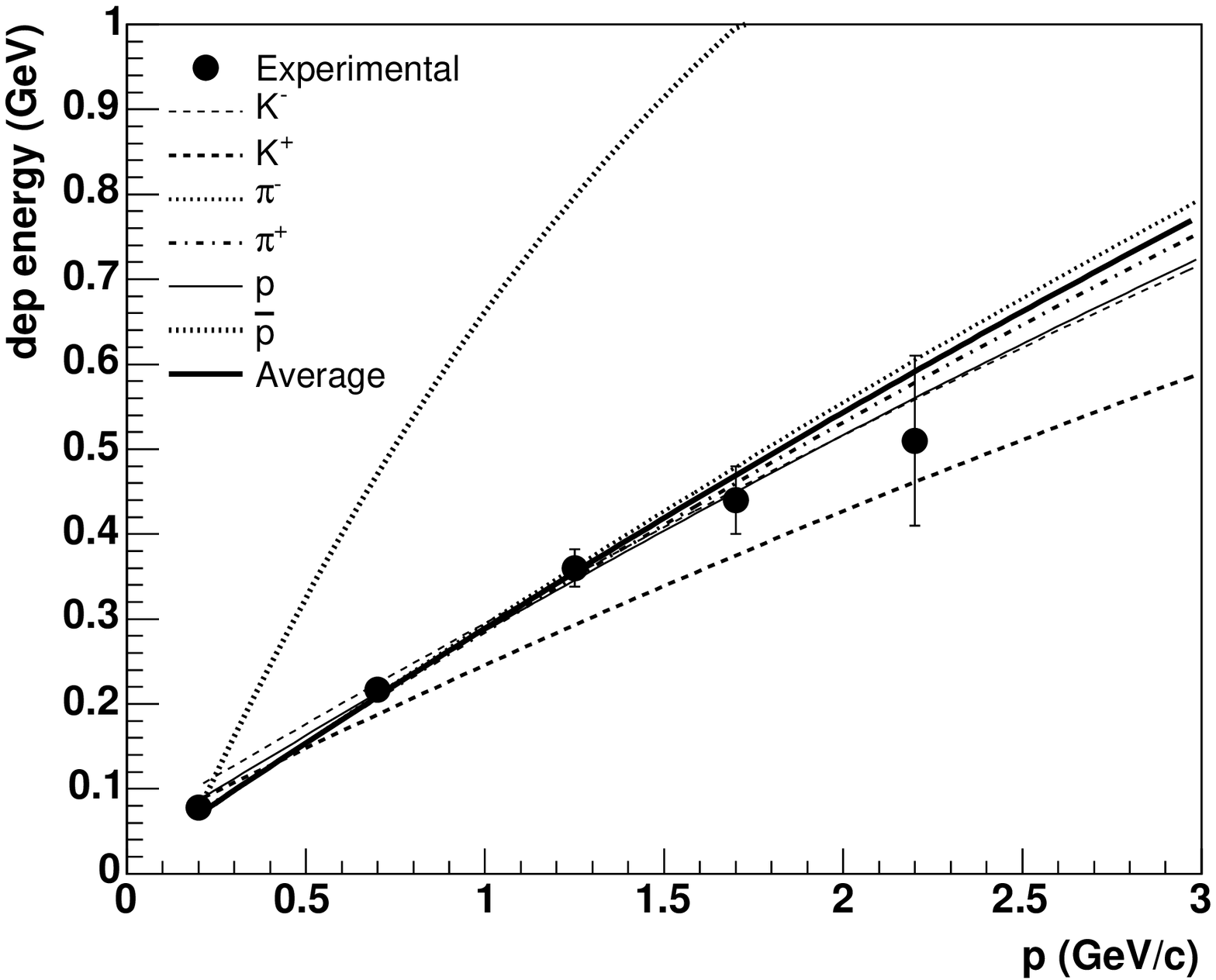}
\includegraphics[width=2.5in]{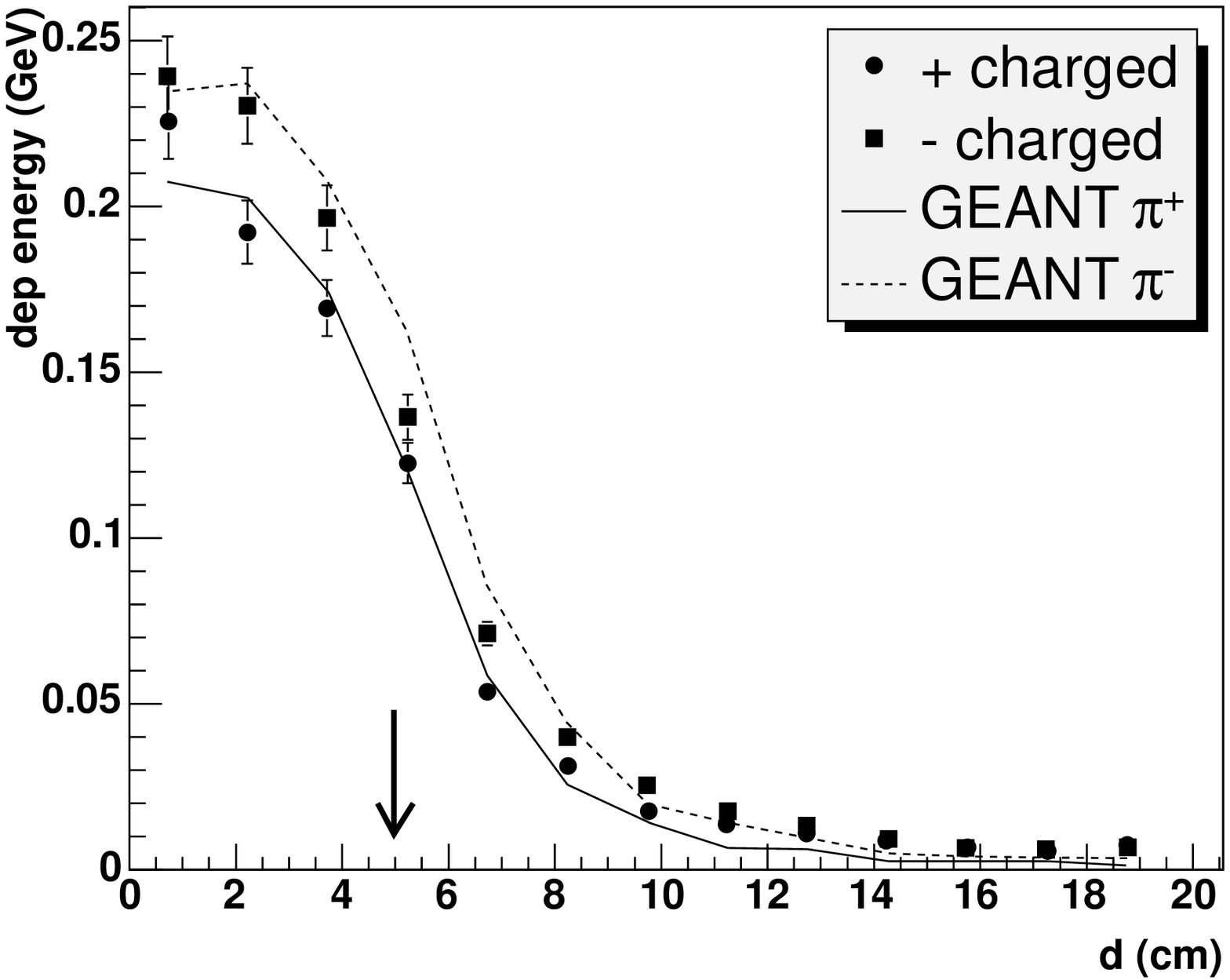}
\caption{a) The mean values of the energy deposited in the BEMC by various hadronic
species as a function of momentum. This is from GEANT simulations. b) The spatial 
profiles of the energy deposition in the BEMC as a function of the distance 
($d$) from the hit point to the center of the tower for $\pi^-$,
$\pi^+$ from simulations and for positive and negative hadrons from data. The 
arrow indicates the distance corresponding to the border of a tower in 
$0 < \eta < 0.2$.}
\label{eHadEMCFig}
\end{center}
\end{figure}

%\begin{figure}[htbp14]
%\begin{center}
%\includegraphics[width=3.5in]{spatial.eps}
%\caption{}
%\label{spatial}
%\end{center}
%\end{figure}

\begin{figure}[htbp141]
\begin{center}
\includegraphics[width=3.5in]{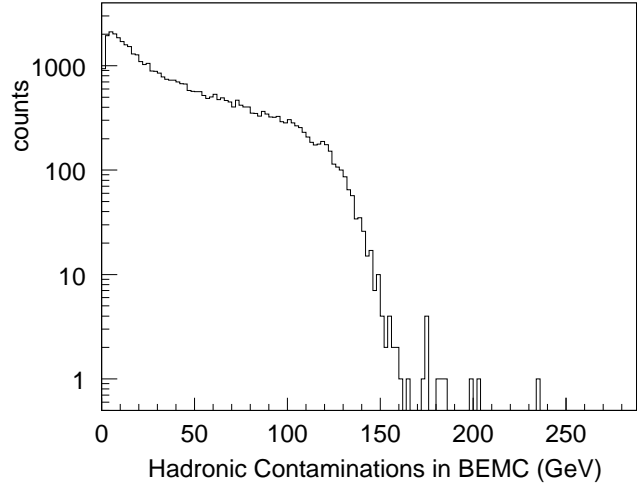}
\caption{The event-by event distribution of the hadronic energy deposited 
in the BEMC for 62.4 GeV Au+Au minimum-bias collisions.}
\label{hadContam}
\end{center}
\end{figure}

Fig.~\ref{eHadEMCFig} (a) shows the mean energy deposited for different particles 
from GEANT simulations, as a function of $p$, for a fixed $\eta$ and $d$. An
average curve, based on the relative yield of different particles is also
presented. Small differences are observed for most of the particles except
the anti-proton, for which additional annihilation energy is expected.
The solid points are the deposited energy obtained from experimental data
for charged hadrons. The experimental profiles for charged hadrons agree
quite well with the averaged profile. Fig.~\ref{eHadEMCFig} (b) shows the simulated
profiles for $\pi^-$, $\pi^+$ and the experimental profiles for all positive
and negatively charged tracks in the momentum range $0.5 < p < 1.0$ GeV/c, as
a function of $d$. Experimental profiles are well described by simulations,
except for a normalization factor of the order of $20\%$ for $0 < p < 0.5$ GeV/c
and $5\%$ for $p > 0.5$ GeV/c. After renormalization, all experimental profiles
up to $p = 2.0$ GeV/c are seen to be in good agreement with simulation. 
Hence the renormalized simulated profiles are used to allow smooth interpolation 
in the data analysis and for extrapolation to allow corrections for higher 
momentum tracks. Since the interval $p < 2.0$ GeV/c contains $98\%$ of the 
tracks, the magnitude of this extrapolation is small for the $E_T$ measurement.

Fig.~\ref{hadContam} shows the distribution of the hadronic energy deposited in
the BEMC for minimum-bias events in 62.4 GeV Au+Au collisions. This is the
hadronic contamination in the electromagnetic energy measured in BEMC.

\subsubsection{Measurement of ($E_T^{em}$)}
The electromagnetic transverse energy, $E_T^{em}$ is defined as 
\begin{equation}
{E_T^{em} = \sum_{towers} E_{tower}^{em}~ sin(\theta_{tower})},
\label{etEm}
\end{equation}
where, $E_{tower}^{em}$ is the electromagnetic energy measured in an BEMC tower
and $\theta_{tower}$ is the polar angle of the center of the tower relative to
the beam axis and the collision vertex position. However, experimentally 
$E_T^{em}$ is given by
\begin{equation}
{E_T^{em} = \frac{1}{f_{acc}}\sum_{towers} ~(E_{tower} - \Delta E_{tower}^{had})
~ sin(\theta_{tower})},
\label{etEmExpt}
\end{equation}
where, the sum over BEMC towers corresponds to $0 < \eta < 1$ with full azimuthal
coverage i.e. the common phase space of TPC and BEMC.\\ 
$f_{acc} ~~~~~~\equiv$  the EMC acceptance correction factor,\\ 
$E_{tower} ~~~\equiv$  the energy measured by an BEMC tower and\\
$\Delta E_{tower}^{had} \equiv$ total correction for each tower to exclude the
hadronic contribution to tower energy.\\
The  $\Delta E_{tower}^{had}$ is given by
\begin{equation}
{\Delta E_{tower}^{had} = \frac{1}{f_{neutral}}~\sum_{tracks} \frac{f_{elec}(p_T)}
{eff(p_T)}~\Delta E(p,\eta,d)}~,
\label{towerHad}
\end{equation}
where, $\Delta E(p,\eta,d)$ is the energy deposited by a track projected on an
BEMC tower as a function of its momentum $p$, pseudo-rapidity $\eta$ and distance
$d$ to the center of the tower from the track hit point. This track projection
takes into account the magnetic field within which the particles are traversing.
$f_{elec}(p_T)$ is a correction factor to exclude the electrons that are 
misidentified as hadrons in the particle identification procedure through
$dE/dx$ method and therefore should not be added to $\Delta E_{tower}^{had}$.
The procedure of estimation of this correction factor is the same as described
in the previous section to exclude the real electrons from the $E_T^{had}$ 
measurement. $eff(p_T)$ is the TPC track reconstruction efficiency as discussed
previously. The factor $f_{neutral}$ is the correction factor to exclude the
long-lived neutral hadron contributions. This is given by
\begin{equation}
{f_{neutral} = \frac{\Delta E_{tower}^{charged}}{\Delta E_{tower}^{charged}+ 
\Delta E_{tower}^{neutral}}}
\label{fNeutralEm}
\end{equation}
Here $\Delta E_{tower}^{neutral}$ is defined as the energy deposited by all
long-lived neutral hadrons. Through HIJING simulation we have observed that 
this factor doesn't change from 200 GeV Au+Au collision to 62.4 GeV Au+Au 
collisions. Therefore for this analysis, we have used the value of $f_{neutral}$ 
used in 200 GeV Au+Au analysis \cite {star200GeV}. The value of this correction 
factor is $0.86 \pm 0.03$.

The estimation of hadronic correction for charged tracks, $\Delta E(p,\eta,d)$,
is based primarily on measured hadronic shower profiles with GEANT simulations
used for interpolation between measurements and extrapolation beyond $p = 2$
GeV/c. The systematic uncertainty for this correction to $E_T^{em}$ is 
estimated from the observed uncertainties in the calculation of hadronic
profile at points in the shower library where full measurements were made.
After normalization, a $5\%$ systematic uncertainty is found to be consistent 
with the comparison of measured and calculated shower profiles. Unlike the 
correction for $p_T$ cut-off for $E_T^{had}$, there is no correction for 
$p_T$ cut-off in the hadronic background subtraction in the electromagnetic energy.
Low $p_T$ tracks will not reach the calorimeter because of the strength 
of the magnetic field, and therefore will not deposit any energy in the
calorimeter.

 The EMC acceptance correction is given by 
\begin{equation}
{f_{acc} = \frac{Number ~of ~working ~EMC ~towers}
{Total ~number ~of ~EMC ~towers}}
\label{fccEm}
\end{equation}
The number of working towers, which affects the acceptance factor, 
can fluctuate in case of electronic readout failures.  
The value of $f_{acc}$ is estimated to be $0.87 \pm 0.02$. The systematic
error in $f_{acc}$ is estimated by measuring it's variation over several
runs. The systematic uncertainty due to the calibration of the calorimeter is 
of the order of $2\%$ and this uncertainty contributes to the uncertainty
in $E_T^{em}$. The systematic uncertainty due to the electron background
track correction is negligible $(< 0.5\%)$ \cite{star200GeV}.

Assuming that all the sources of uncertainties are uncorrelated, when added in 
quadrature the overall systematic uncertainty estimate for $E_T^{em}$ is $3.0\%$.
All the corrections and corresponding systematic uncertainties are summarized
in Table \ref{emCorr}. The final $E_T^{em}$ is then corrected for the $94.5\%$
vertex reconstruction efficiency. For the $5\%$ most central collisions, the
 event-by-event resolution in $E_T^{em}$ is found to be $15.6\%$. The measured 
values of $dE_T^{em}/d\eta$ for different centralities in Au+Au collisions 
at $\sqrt{s_{NN}}=$ 62.4 GeV are given in Table-~\ref{valuesT}.
\begin{table}[htb1]
\begin{center}
\begin{tabular}{|l|l|l|}
\hline
 Correction Factor & Correction \\
\hline
$f_{acc}$ & $0.87 \pm 0.02$\\
$f_{neutral}$ & $0.86 \pm 0.03$ \\
$f_{elec}(p_T)$ &  $0.96 \pm < 0.005~(p_T \leq 0.25~GeV/c)$\\
&$1.00 \pm < 0.005~(0.25 \leq p_T \leq 1~GeV/c)$\\
$eff(p_T)$ & $0.70 \pm 0.04~(p_T \leq 0.25~GeV/c)$\\
& $0.80 \pm 0.04~(0.25 \leq p_T \leq 1.0~ GeV/c)$\\
\hline
\end{tabular}
\vspace{0.2cm}
\caption{Correction factors and their estimated values with uncertainties for
  $E_T^{em}$ for the $5\%$ most central collisions. }
\label{emCorr}
\end{center}
\end{table}
In order to evaluate the hadronic background subtraction procedure and estimate
the event-by-event resolution of the reconstructed electromagnetic energy, 
simulations have been performed where we compare the reconstructed $E_T^{em}$ 
energy and the input HIJING. The event-by-event resolution improves when the 
event becomes more central. The main factors that determine this resolution 
are the hadronic energy subtraction, the corrections for tracking efficiency 
and the corrections for long-lived neutral hadrons \cite{star200GeV}.

\subsection{Total Transverse Energy ($E_T$)}
\begin{figure}[htbp]
\begin{center}
\includegraphics[width=2.8in]{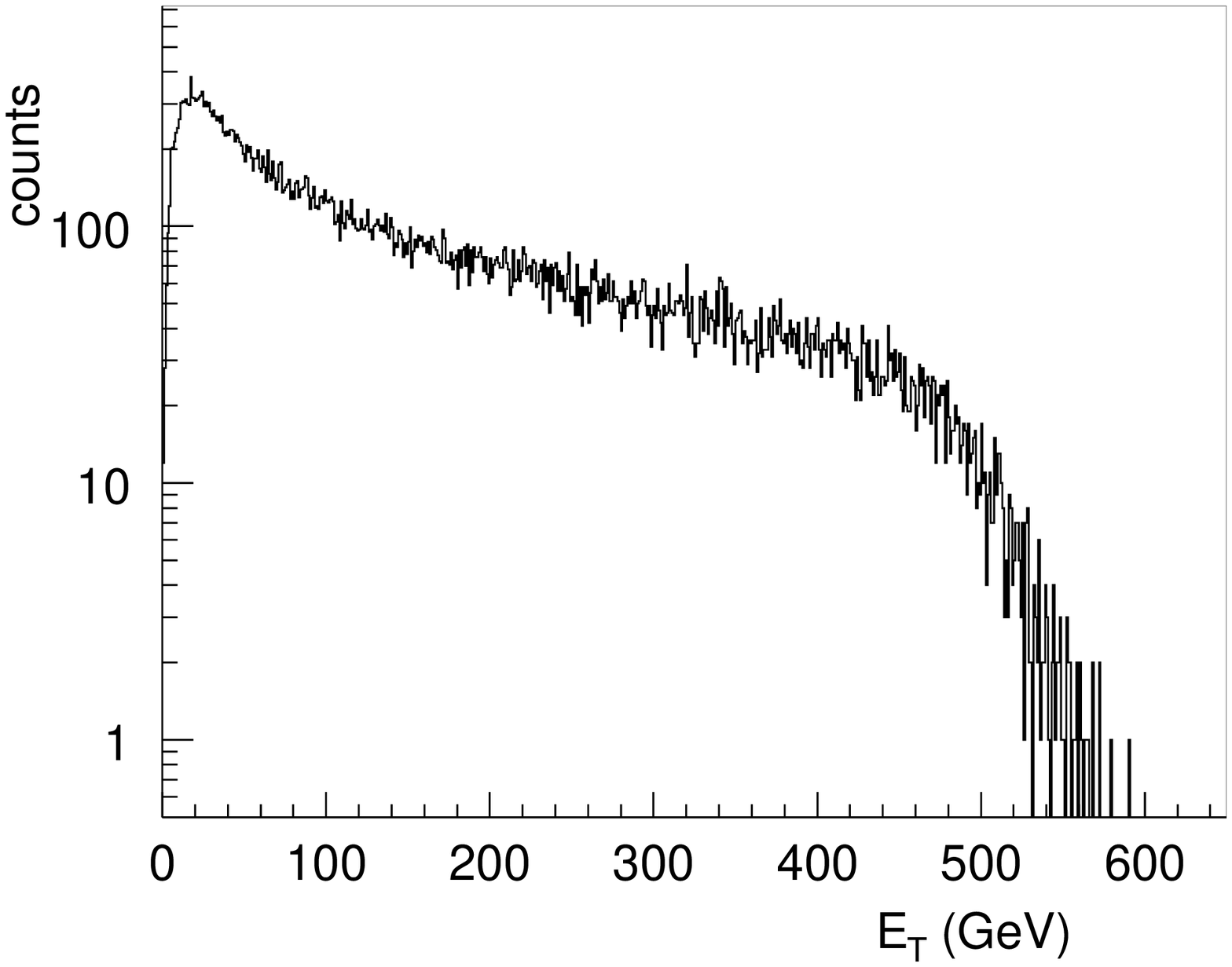}
\includegraphics[width=2.8in]{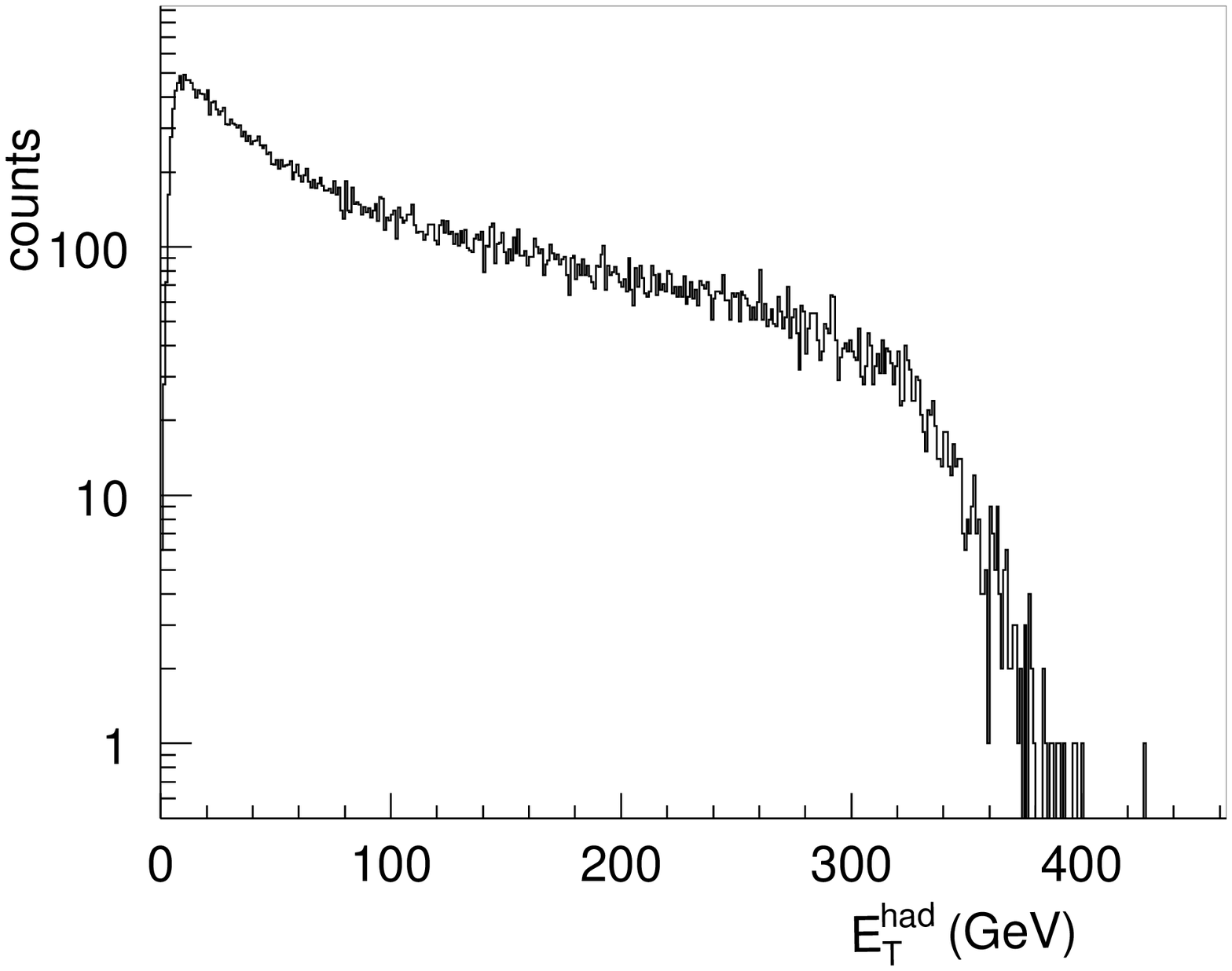}
\includegraphics[width=2.8in]{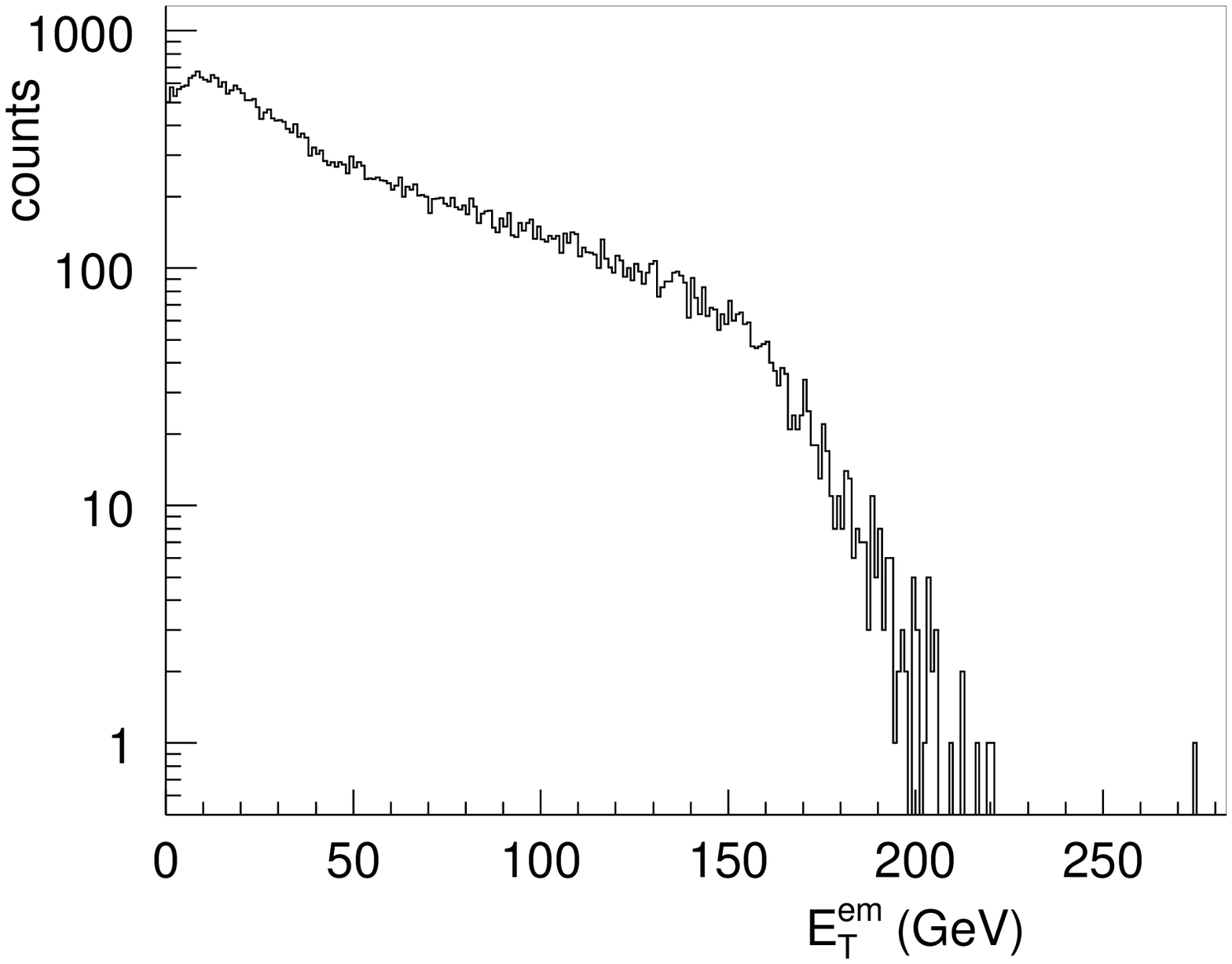}
\includegraphics[width=2.8in]{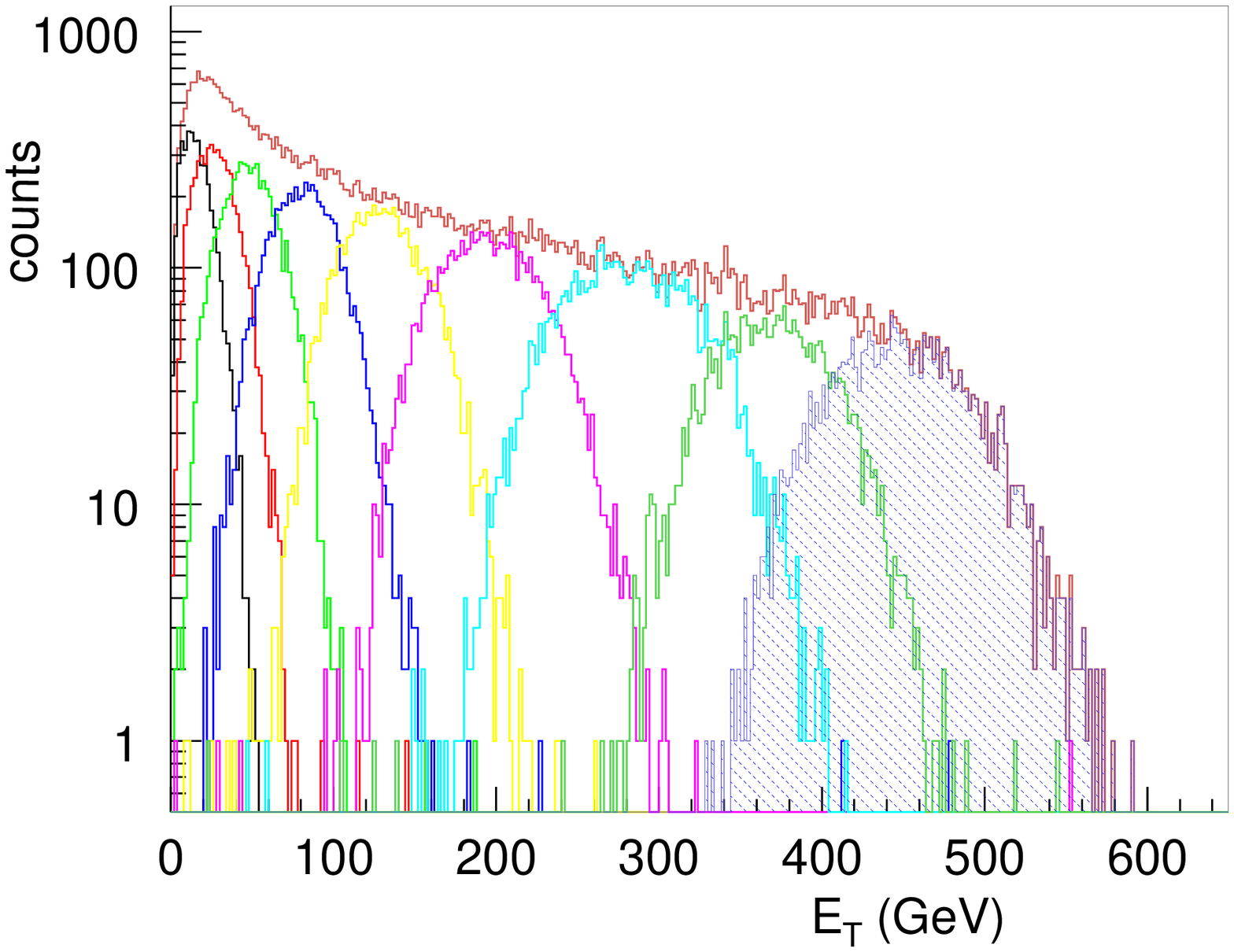}
\caption{Upper panel: Minimum bias distributions of total transverse energy 
(left side) and hadronic transverse energy (right side), Lower panel:
 Minimum bias distributions of electromagnetic transverse energy (left side)
and total transverse energy for all centrality classes (right side) 
for $\sqrt{s_{NN}} = 62.4$ GeV Au+Au collisions. 
The shaded area corresponds to the top most central bin.}
\label{minBias}
\end{center}
\end{figure}

The sum of $E_T^{had}$ and $E_T^{em}$ is the total transverse energy, $E_T$.
The transverse energy distributions (Fig.~\ref{minBias})show a peak and a sharp 
drop off at the lower energy edge corresponding to the most peripheral 
collisions with grazing impact. It reaches a broad, gently sloping plateau 
at the middle which corresponds to the mid-central collisions 
(mid-range of impact parameters). This is dominated by the nuclear geometry. 
For higher values of $E_T$, which corresponds to the most central collisions 
(where the colliding nuclei fully overlap), the shape of the distribution 
has a ``knee'' leading to a falloff which is very steep for larger acceptances 
and less steep for smaller acceptance. This shape is mostly determined by 
statistical fluctuations and depends on the experimental acceptance \cite{abbott}. 
Figs.~\ref{minBias} shows the event-by-event distribution of 
transverse energy and its components. Fig.~\ref{minBias} (lower panel and right
figure) shows $E_T$ distributions for different centrality classes defined by 
the percentage of the total cross-section as discussed earlier 
(Table.~\ref{centrality}). The shaded area corresponds to the most 
$5\%$ central collision.
\begin{figure}[htbp1000]
\begin{center}
\includegraphics[width=3.5in]{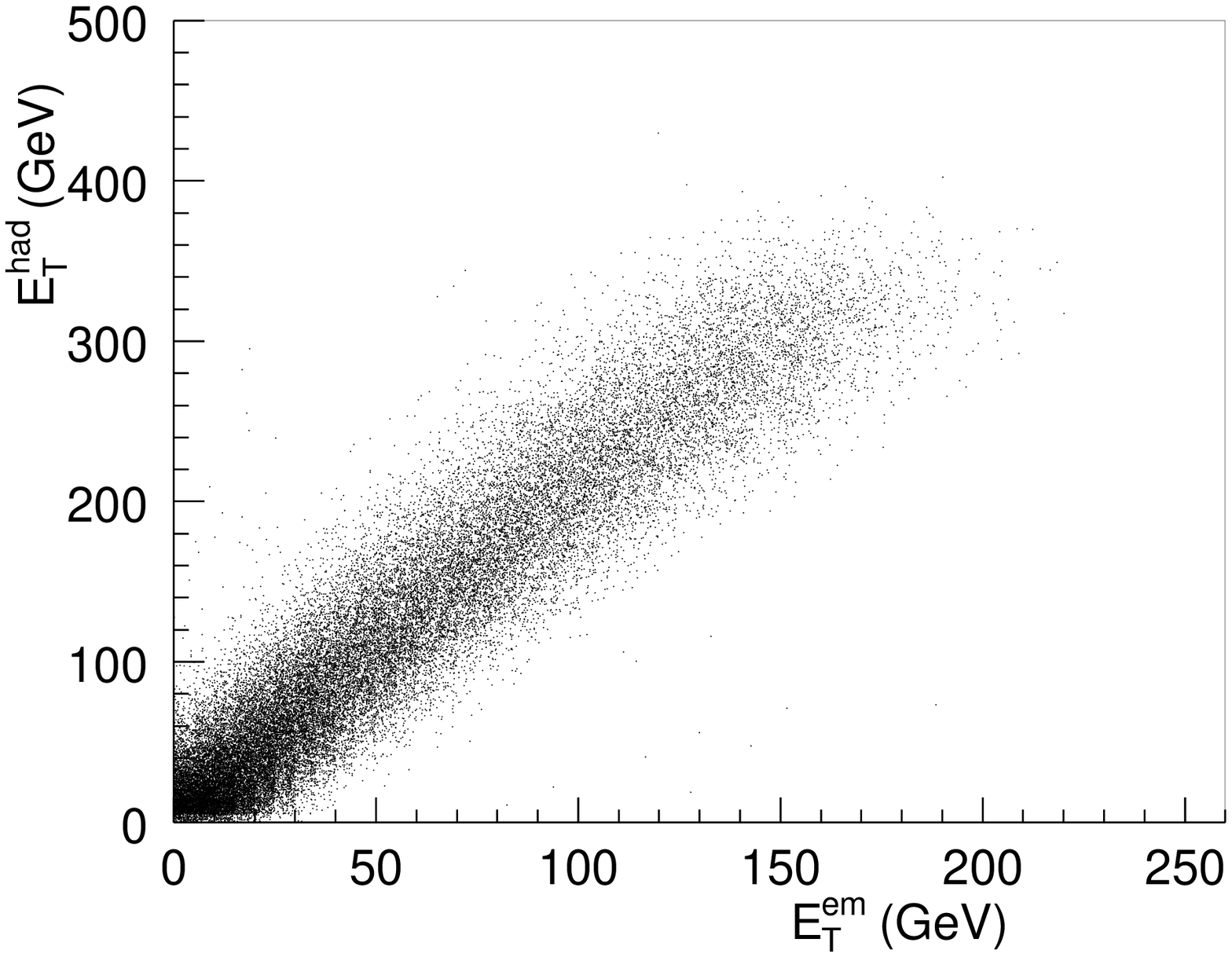}
\includegraphics[width=3.5in]{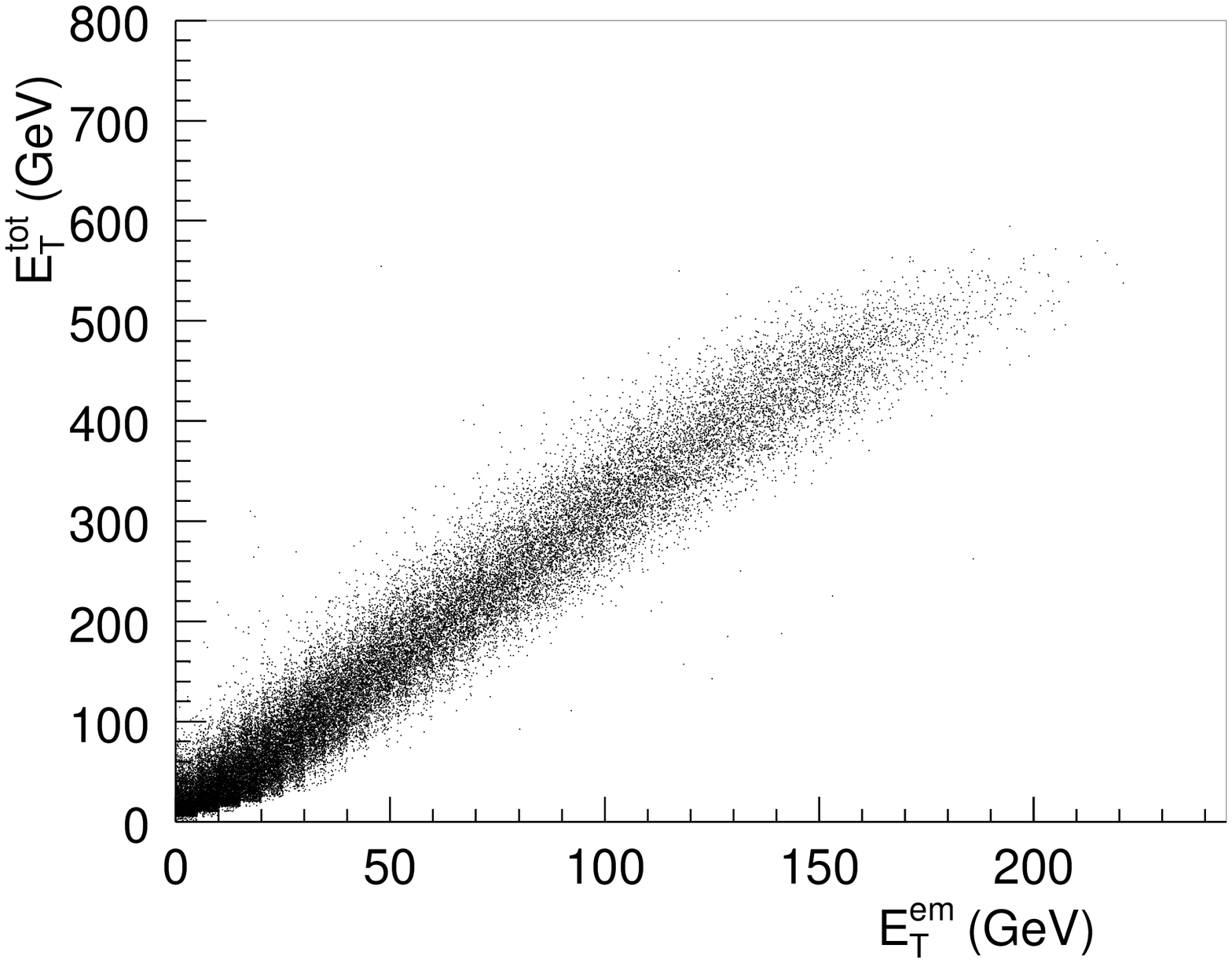}
\includegraphics[width=3.5in]{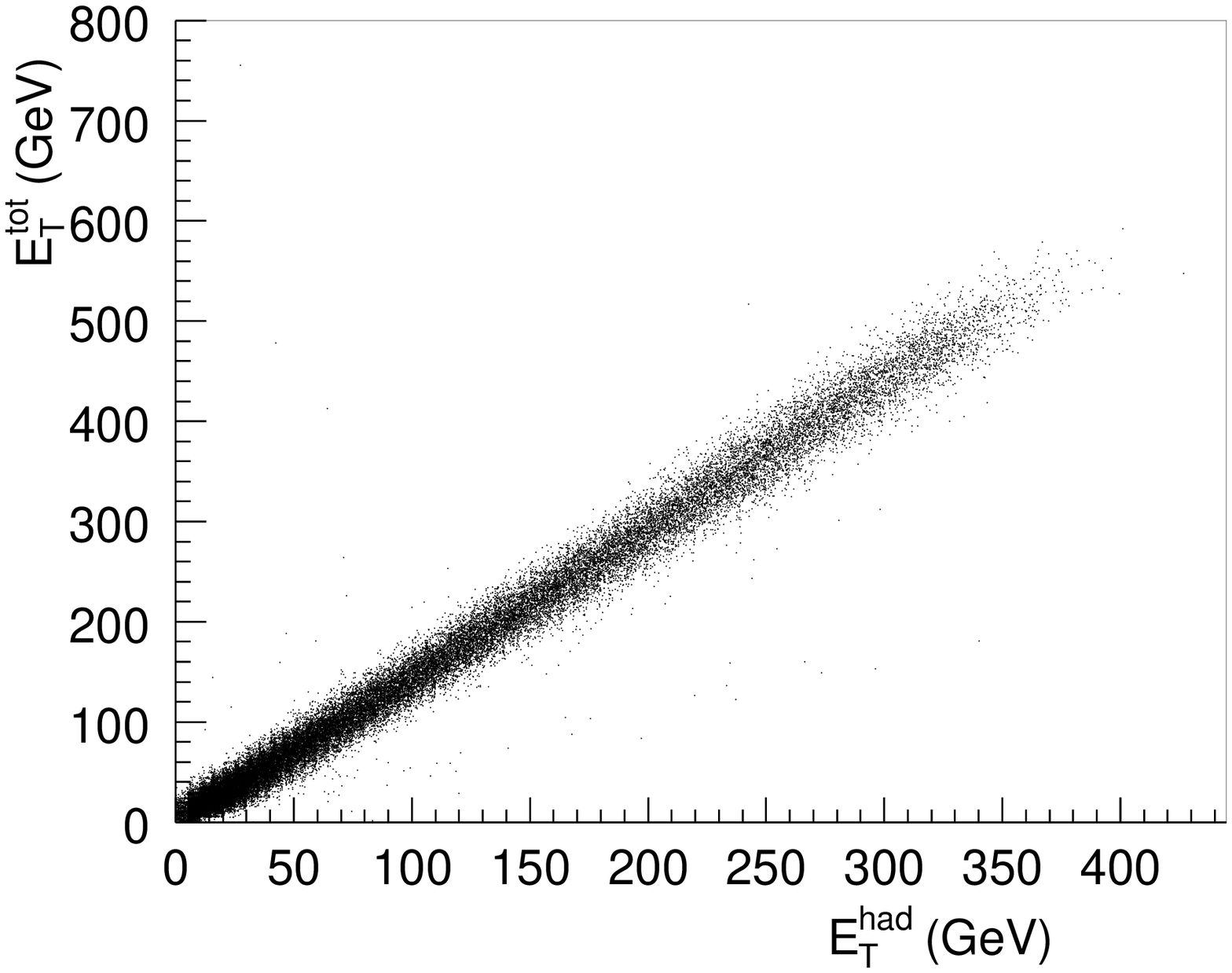}
\caption{The correlation plots for $E_T^{em}$ and $E_T^{had}$ (top), 
 $E_T^{em}$ and $E_T$ (middle) and  $E_T^{had}$ and $E_T$ (bottom); 
for $\sqrt{s_{NN}} = 62.4$ GeV Au+Au collisions.}
\label{correl}
\end{center}
\end{figure}

Fig.~\ref{correl} shows the correlation plots for both the components of 
transverse energy and the total transverse energy for $\sqrt{s_{NN}} = 62.4$ 
GeV Au+Au collisions.

Combining both the components of transverse energy and after properly 
taking into account all the systematic uncertainties, we estimate a combined
systematic uncertainty of $11\%$ in $E_T$. For the top $5\%$ central 
collisions, $<dE_T/d\eta|_{\eta=0.5}> ~= ~<E_T>_{5\%}~ 
= ~474 \pm 51 ~(syst) \pm 1~(stat)$ GeV. The measured values of 
$dE_T/d\eta$ for different centralities in Au+Au collisions at 
$\sqrt{s_{NN}}=$ 62.4 GeV are given in Table-~\ref{valuesT}. 
For the $5\%$ most central collisions, the event-by-event resolution 
in $E_T$ is found to be $8.9\%$.

\section{Results and Discussions}

\subsection{Variation of $<dE_T/d\eta|_{\eta = 0.5}>$ per $N_{part}$ pair
 with Centrality of the Collision}

\begin{figure}[htbp011]
\begin{center}
\includegraphics[width=4.0in]{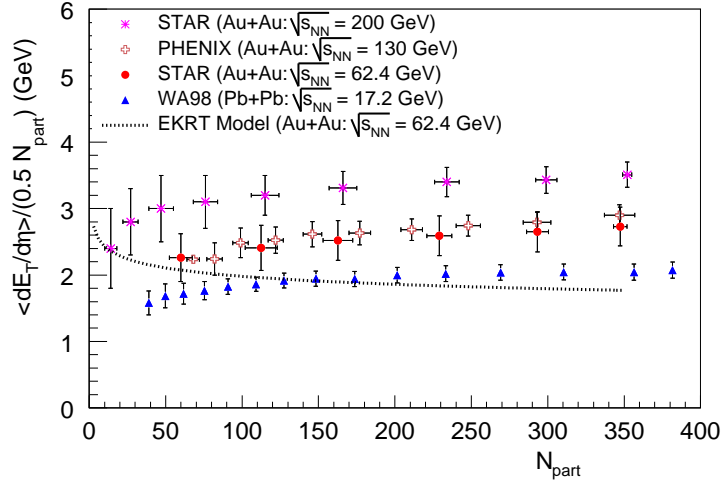}
\caption{$<dE_T/d\eta|_{\eta = 0.5}>$ per $N_{part}$ pair vs $N_{part}$. 
The dotted line is the estimation from EKRT model (Eqn.~\ref{ekrtEq}), for 62.4 
GeV Au+Au collisions.}
\label{fig8F}
\end{center}
\end{figure}

The variation of $<dE_T/d\eta|_{\eta = 0.5}>$ per $N_{part}$ pair as a function 
of $N_{part}$ (obtained from Monte Carlo Glauber calculations) is shown in 
Fig.~\ref{fig8F}. Here, the result for $\sqrt{s_{NN}}=$ 62.4 GeV Au+Au 
collisions is shown together with similar measurements from Pb+Pb collisions 
at $\sqrt{s_{NN}}=$ 17.2 GeV from WA98 \cite{wa98Et}, Au+Au collisions at 
$\sqrt{s_{NN}}=$ 130 GeV from PHENIX \cite{phenixEt} and Au+Au collisions 
at $\sqrt{s_{NN}}=$ 200 GeV from STAR \cite{star200GeV}. The STAR 200 GeV and
62.4 GeV measurements are at $0 < \eta < 1$ and measurements by WA98 and PHENIX
are at $\eta = 0$.  The dotted line is the EKRT model \cite{ekrt} estimation 
for 62.4 GeV Au+Au collisions. The measured values of 
$<dE_T/d\eta|_{\eta = 0.5}>$ per $N_{part}$ pair for different centralities
in Au+Au collisions at $\sqrt{s_{NN}}=$ 62.4 GeV are given in 
Table-~\ref{valuesT1}.

The EKRT model, which is based on final state gluon saturation, 
predicts that for more central collisions, for both charge particle 
multiplicity per participant pair and $E_T$ density in pseudo-rapidity 
per participant pair will decrease. The hydrodynamic work plays a role during 
the expansion of the fireball by decreasing the observed $E_T$ relative to 
the initially generated $E_T$ \cite{ekrt}. However, the transverse expansion 
(radial flow) at later times compensates for this effect \cite{ekrt,hydro}.
According to the EKRT model, the observed $E_T$ in terms of the center of
mass energy ($\sqrt{s_{NN}}$) and the system size ($A$) is given by
\begin{equation}
{E_T^{b=0} = 0.46~A^{0.92}(\sqrt{s})^{0.40}(1-0.012~lnA~+~0.061~ ln \sqrt{s})}
\label{ekrtEq}
\end{equation}
The centrality dependence of the above equation can be approximated by replacing
$A$ by $N_{part}/2$ \cite{kharzeevNpart}. The EKRT model is seen not to agree
with the data. It shows a significantly different behavior in the centrality 
dependence. A more precise comparison of the system size dependence of $E_T$ 
predicted by the EKRT model, requires either a further refinement of the 
Glauber calculations or measurements for central collisions with varying mass $A$.

\subsection{Why to Compare the Data with EKRT Model?}

The bulk hadron multiplicities measured at mid-rapidity in central Au+Au 
collisions at $\sqrt{s_{NN}} = 200$ GeV is comparatively lower than the
expectations of models with ``mini-jet'' dominated scenarios, soft Regge 
models (without accounting for strong shadowing effects) or extrapolation 
from an incoherent sum of proton-proton collisions. These models fail to 
explain the data for bulk hadron multiplicities measured at mid-rapidity 
in central Au+Au collisions. The data show a lower value as expected from 
these models \cite{enterria}.

  Whereas, models like EKRT, based on gluon saturation which takes into account 
a reduced initial number of scattering centers in the nuclear parton distribution 
functions agree well with the experimental data.

\subsection{Variation of $<dE_T/dy>$ per $N_{part}$ pair with 
Center of mass Energy}

\begin{figure}[htbp0]
\begin{center}
\includegraphics[width=4.0in]{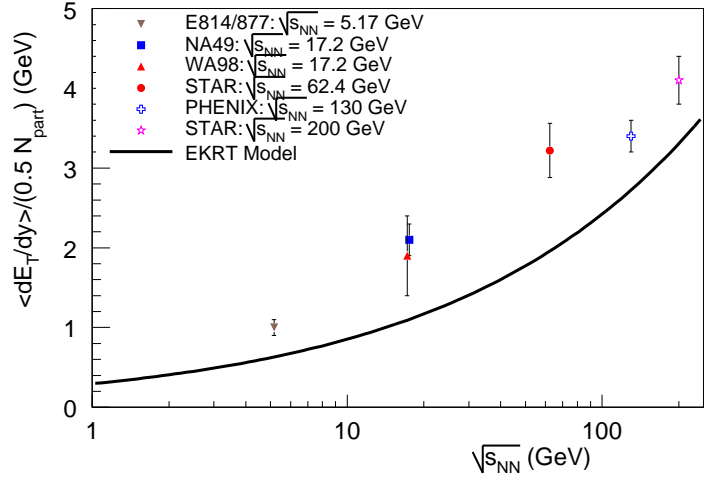}
\caption{$<dE_T/dy>$ per $N_{part}$ pair vs $\sqrt{s_{NN}}$ for central
events. The solid line is the EKRT \cite{ekrt} model prediction.}
\label{fig9F}
\end{center}
\end{figure}

\begin{figure}[htbp1]
\begin{center}
\includegraphics[width=4.5in]{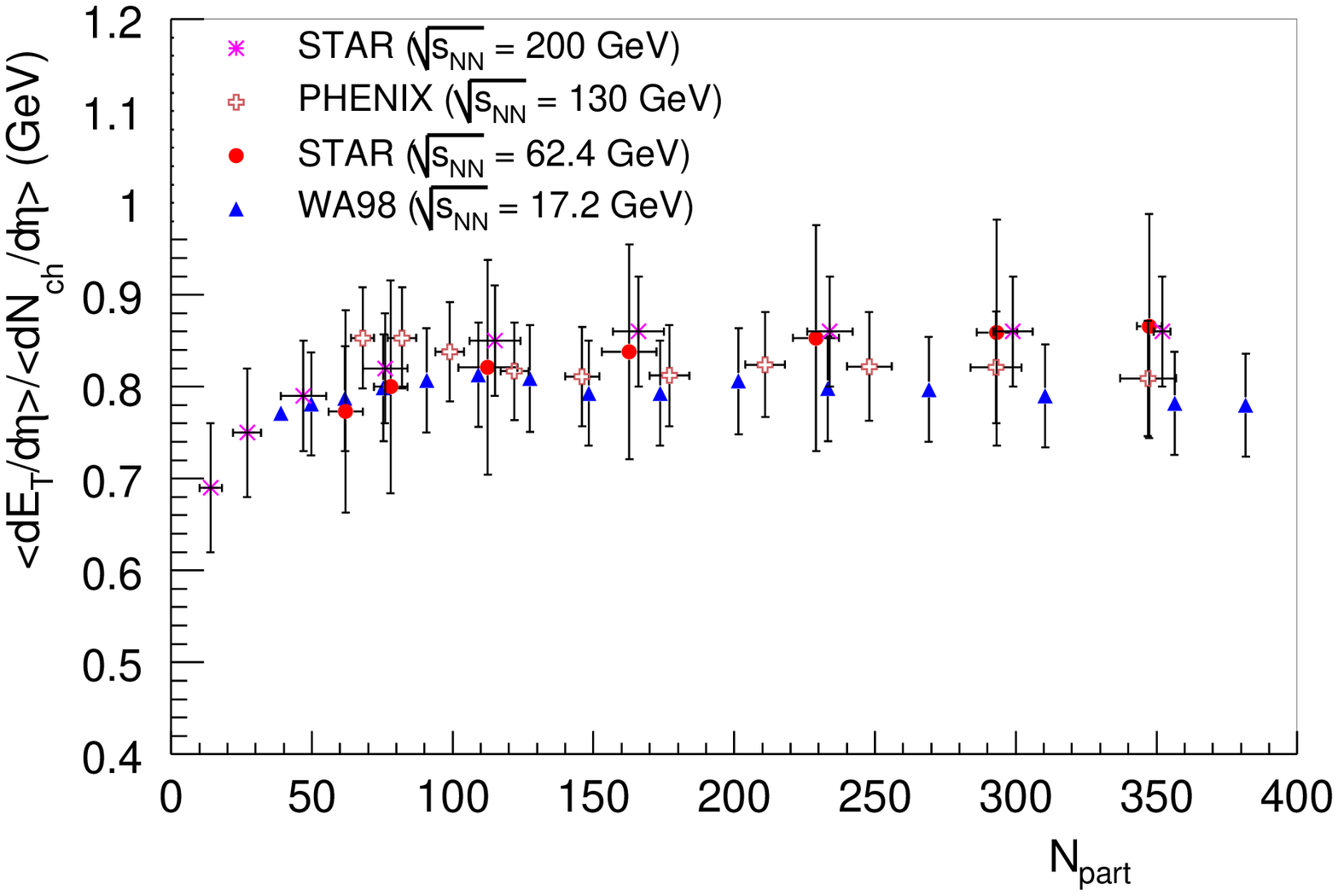}
\includegraphics[width=4.5in]{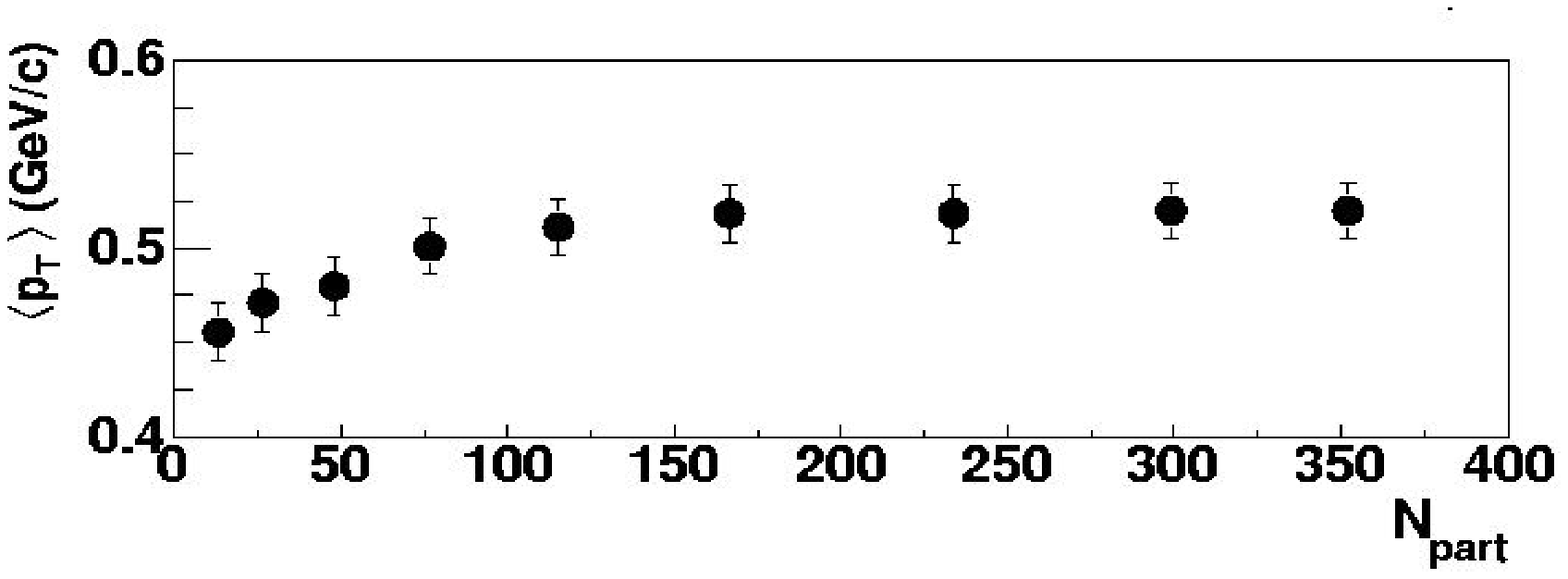}
\caption{$<dE_T/d\eta>/<dN_{ch}/d\eta>$ Vs $N_{part}$ (top). 
Results from WA98 \cite{wa98Et} and PHENIX \cite{phenixEt}
are shown with STAR measurements. The mean transverse momentum of charged 
hadrons as a function of $N_{part}$ \cite{starPRL} (bottom).}
\label{fig10F}
\end{center}
\end{figure}
The variation of $<dE_T/dy>$ per participant pair with the center of mass energy
is shown in Fig.~ \ref{fig9F}. For the top $5\%$ central events in 62.4 GeV Au+Au
collisions, the value of $<dE_T/d\eta>$ is estimated and $<dE_T/dy>$ is measured 
using a factor of 1.18 for the conversion of pseudo-rapidity phase space to 
rapidity phase space, using HIJING simulations. The value of $<dE_T/dy>$ per 
participant pair for the top $5\%$ central Au+Au collisions at
$\sqrt{s_{NN}}$ = 62.4 GeV is estimated to be $3.22 \pm 0.34$ GeV. 
This result is compared with results from other experiments from AGS to 
RHIC \cite{Barrettee, Alber, wa98Et, phenixEt} for the most central collisions. 
The result is consistent with the 
fact that $<dE_T/dy>/(0.5~N_{part})$ increases logarithmically with 
$\sqrt{s_{NN}}$. The solid line is the EKRT model prediction for central 
Au+Au collisions. It is seen that the EKRT model underestimates the final 
transverse energy by $\sim 24\%$ for 200 GeV Au+Au collisions and $\sim 64\%$  
for 62.4 GeV Au+Au collisions.

\subsection{Variation of $<dE_T/d\eta>/<dN_{ch}/d\eta>$ with 
Collision Centrality}

In order to understand the systematic growth in transverse energy with 
collision energy shown in Fig. \ref{fig9F}, the centrality dependence of
$<dE_T/d\eta>/<dN_{ch}/d\eta>$ i.e. the scaling of transverse energy relative
to the number of charged particles produced in the collision is studied.
The centrality dependence of the ratio may indicate the effects of 
hydrodynamic flow \cite{hydro}. In this scenario if we assume the expansion
of the produced fireball is isentropic (entropy is conserved), then 
$<dN_{ch}/d\eta>$ will remain constant while $<dE_T/d\eta>$ will decrease
due to performance of longitudinal work. This is clearly reflected in
the peripheral events as a dip (see Fig.~ \ref{fig10F}) where we do expect 
hydrodynamic flow.

Fig~\ref{fig10F} (top) shows the centrality dependence of 
$<dE_T/d\eta>/<dN_{ch}/d\eta>$ from STAR measurements at $\sqrt{s_{NN}}$
= 200 GeV and 62.4 GeV, compared to similar measurements at 17.3 and 130
GeV. For the top $5\%$ central Au+Au collisions at $\sqrt{s_{NN}}$ = 62.4
GeV, the value of $<dE_T/d\eta>/<dN_{ch}/d\eta>$ is found to be $0.866 \pm 0.122$ 
GeV. The measured values for different centralities in Au+Au collisions 
at $\sqrt{s_{NN}}=$ 62.4 GeV are given in Table-~\ref{valuesT}.

Data at all energies starting from lower SPS to RHIC, within uncertainties, 
are seen to fall on a common curve. It shows a modest increase from the most 
peripheral collisions at $N_{part} < 50$, reaching a roughly constant value 
of the ratio at $N_{part} = 100$. Fig~\ref{fig10F} (bottom) shows the $<p_T>$ for 
200 GeV Au+Au collisions measured by STAR \cite{starPRL}. 
This shows a similar centrality dependence  as that of the transverse energy
per charge particle: modest increase with $N_{part}$ for $N_{part} < 100 $,
with constant value for more central collisions. $E_T$, multiplicity and $<p_T>$ 
all show a similar behavior. This indicates that the
increase of $E_T$ is due to increased particle production. However, the
quantitative comparison of theoretical models of particle production with
the measured centrality dependences of $<dE_T/d\eta>/<dN_{ch}/d\eta>$ and
$<p_T>$ of charged particles will constrain the profile of initial energy
deposition and the role of hydrodynamic work during the expansion.

\subsection{The Global Barometric Observable: $E_T/N_{ch}$}
Average transverse energy per charge particle is an important global
barometric measure of the internal pressure in the ultra-dense matter
produced in heavy ion collisions. 
\begin{figure}[htbp2]
\begin{center}
\includegraphics[width=4.3in]{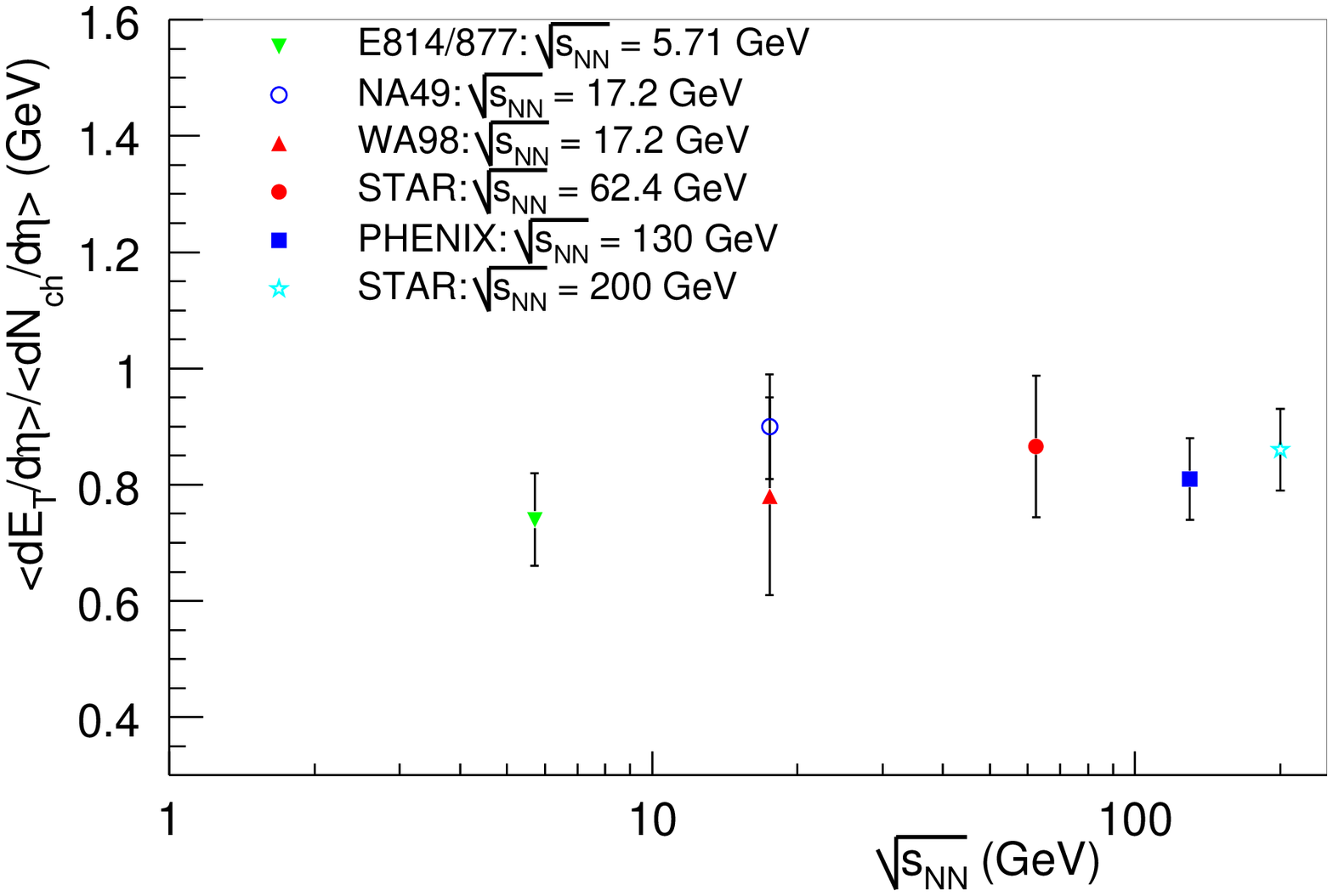}
\caption{$<dE_T/d\eta>/<dN_{ch}/d\eta>$ Vs $\sqrt{s_{NN}}$ for central
events.} 
\label{fig11F}
\end{center}
\end{figure}
Fig.~\ref{fig10F} and Fig.~\ref{fig11F} show the most amazing fact that
a constant transverse energy per charge particle ($E_T/N_{ch}$) $\sim 800$
MeV has been produced which is almost independent of collision centrality
and the collision energy. There is a slow logarithmic increase in $E_T/N_{ch}$
amounting to $< 10\%$, which characterizes all the measurements within errors, 
over a range in which $E_T$ per participant grows by a factor of 4. 
This has been observed from lower energy AGS measurements to the highest 
RHIC energies. This is shown in Ref.~\cite{phenixSyst}. HIJING predicts that 
$E_T$ per charged particle should increase from 0.8
(SPS) to 0.9 (RHIC) due to the enhanced mini-jet activity at RHIC. The
EKRT initial state gluon saturation model predicts a growth of this quantity
in the initial state by about a factor of 3. The reason that EKRT model 
remains viable after these data is that the assumed entropy conservation
implies that a large amount of $pdV$ work due to longitudinal expansion
is performed by the plasma. In 1+1D hydrodynamics the energy per particle,
$\epsilon/\rho$ ($~\approx 2.7~T$) decrease as the system expands and cools 
($T ~\sim 1/\tau^{1/3}$). If the freeze-out is assumed to occur at all
energies and impact parameters in A+A collisions, on a fixed decoupling 
isotherm, then the energy per particle will always be the same.

 However, there are theoretical problems in justifying hydrodynamics
and the freeze-out prescription. The observed NULL effect in $E_T/N_{ch}$
is very interesting because it is so difficult to obtain in any transport
theory with finite pQCD relaxation rates \cite{lnp}.

\subsection{Estimation of Energy Density}

The estimation of initial energy density of the produced fireball in  heavy ion
collisions has been discussed in details in Chapter 1. This has been
estimated in a boost invariant Bjorken hydrodynamic model. The Bjorken
energy density obtained in this framework is given by 
\begin{equation}
{\epsilon_{Bj} = \frac{dE_{T}}{dy} ~\frac{1}{\tau_0 \pi R^2}
~ \simeq ~ <m_T> \frac{3}{2} \frac{dN_{ch}}{dy} ~\frac{1}{\tau_0 \pi R^2}}
\label{bjEqn}
\end{equation}
where, $\tau_0$ is the formation time, usually assumed to be $1~ fm/c$ and $\pi R^2$ is
the transverse overlap area of the colliding nuclei. The formation time is usually 
estimated from model calculations and has been a matter of debate. There are 
different ways to estimate the transverse overlap area. It goes like $N_{part}^{2/3}$, 
in an approach which accounts for only the common area of colliding nucleons but not 
the nuclei (chosen by STAR). In this approach, the transverse overlap area 
$F = \pi R^2$, where $R =R_0 A^{1/3}$. When we replace $A$ with the number of 
participants by, $A=N_{part}/2$ \cite{kharzeevNpart}, $F$ becomes,
\begin{equation}
{F = \pi R_0^2 ~(\frac{N_{part}}{2})^{2/3} }
\label{FEqn}
\end{equation}
\begin{figure}[htbp26]
\begin{center}
\includegraphics[width=3.5in]{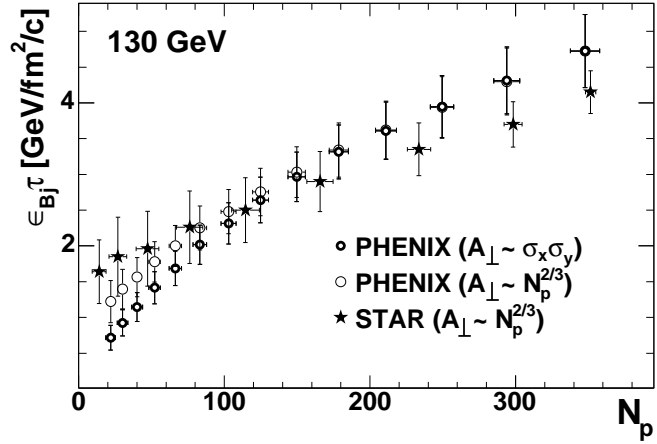}
\caption{The Bjorken energy density vs $N_{part}$ using different estimates of
the transverse overlap area at $\sqrt{s_{NN}} = 130$ GeV. Ref.~\cite{phenixSyst}.} 
\label{phenixTransA}
\end{center}
\end{figure}
In the other approach (adopted by PHENIX) \cite{phenixSyst}, the transverse overlap 
area of the colliding species, $F$, is estimated in the following way.
The Woods-Saxon parametrization for the nuclear density profile is given by
\begin{equation}
{\rho(r) = \frac{1}{(1+e^{(r-r_n)/d})}},
\label{woods}
\end{equation}
where, $\rho(r)$ is the nuclear density profile, $r_n$ is the nuclear radius and
$d$ is a diffuseness parameter. Based on the measurements of electron scattering 
from Au nuclei \cite{eAuScatt}, $r_n$ is set to $(6.38 \pm 0.27)$ fm and $d$ to 
$(0.54 \pm 0.01)$ fm. A Monte Carlo-Glauber model with $F\sim \sigma_x \sigma_y$, 
(where $\sigma_x$ and $\sigma_y$ are the widths of $x$ and $y$ position 
distributions of the participating nucleons in the transverse plane) is used to 
estimate the transverse overlap area of two colliding nuclei. In this approach, 
$F$ is the transverse overlap area of two colliding nuclei not the 
participating nucleons. The normalization to $\pi R^2$, where $R$ is the sum of 
$r_n$ and $d$ parameters in the Woods-Saxon parametrization (given by 
Eqn.~\ref{woods}), is done for most central collisions at the impact parameter $b=0$.
\begin{figure}[htbp22]
\begin{center}
\includegraphics[width=5.5in]{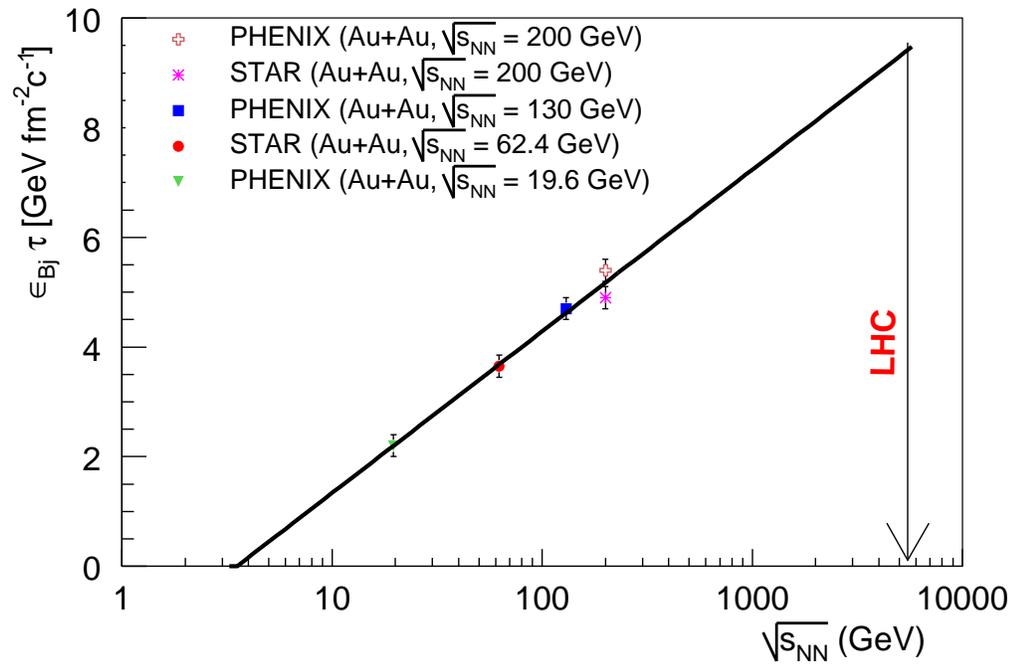}
\caption{The excitation function of $\epsilon_{Bj} \tau~ [GeV~fm^{-2}~c^{-1}]$.
A prediction for LHC is indicated by an arrow, on the assumption of the 
validity of Bjorken hydrodynamic scenario.} 
\label{bjLHC}
\end{center}
\end{figure}

The results obtained in these two methods, as shown in Fig.~\ref{phenixTransA}, 
are different only in the peripheral bins. The results obtained by STAR agree 
with PHENIX results, within systematic errors. However, STAR data show a 
a smaller rate of increase of the energy density with $N_p$. As can be seen from 
the figure, the results agree rather well for central collisions, where we expect
a deconfinement of quarks and gluons to take place.

Using the procedure outlined above (for STAR), the $\epsilon_{Bj}$ for the $5\%$ 
most central Au+Au collisions at $\sqrt{s_{NN}}=$ 62.4 GeV is found to be 
$3.65 \pm 0.39 ~GeV/fm^3$. The estimated values of $\epsilon_{Bj} \tau$ for 
different centralities in Au+Au collisions at $\sqrt{s_{NN}}=$ 62.4 GeV are 
given in Table-~\ref{valuesT1}. In these estimations we have used a factor of
1.18 for $\eta \rightarrow y$-phase space conversion, as compared to 1.25 used by
PHENIX \cite{phenixEt, phenixSyst}. The value of $\epsilon_{Bj}$ for Au+Au collisions
at $\sqrt{s_{NN}} =$ 19.6, 130 \cite{phenixEt,phenixSyst} and 200 GeV \cite{star200GeV}
are $2.2 \pm 0.2, ~4.7 \pm 0.5$ and $4.9 \pm 0.3~~ GeV/fm^3$ 
($5.4 \pm 0.6 ~ GeV/fm^3$, PHENIX) respectively. Compared to this, $\epsilon_{Bj}$
at SPS for Pb+Pb collisions at $\sqrt{s_{NN}} =$ 17.2 GeV is found to be 
$3.2 ~GeV/fm^3$ \cite{Alber}. This value of $\epsilon_{Bj}$ is much higher than the
same for Au+Au collisions at the SPS-like energy i.e $\sqrt{s_{NN}} =$ 19.6 GeV
at RHIC. As all these estimations assume the same formation time of 1 fm/c, there 
is an over estimation of $\epsilon_{Bj}$ at SPS. In any case these energy densities 
are significantly larger than the energy density ($\sim ~ 1~ GeV/fm^3$) predicted 
by lattice QCD calculations \cite{latticeQCD} for a transition to a deconfined 
quark gluon plasma phase. Following the deconfinement transition, there is a
hydrodynamic expansion. Subsequently local equilibrium is achieved at 
$\tau_0 \sim 1$ fm/c. This picture is indeed valid, if we compare the RHIC data
for elliptic flow to the hydrodynamic calculations \cite{hydro1, hydro2, hydro3}.
\begin{figure}[htbp222]
\begin{center}
\includegraphics[width=4.5in]{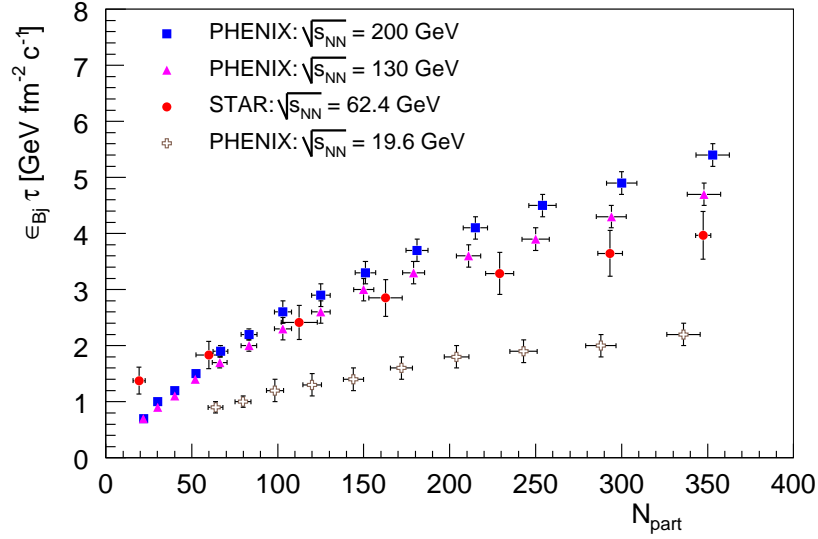}
\caption{The $N_{part}$ dependence of the product of the Bjorken energy density
and the formation time ($\epsilon_{Bj}. \tau$) for Au+Au system at different 
energies at RHIC.}
\label{bjNpart}
\end{center}
\end{figure}
Taking all $\epsilon_{Bj}$s measured, for similar type of colliding species 
i.e. Au, at RHIC energies and assuming that the Bjorken hydrodynamic model 
works fine at energies higher than RHIC energy, we have made a prediction for 
the $\epsilon_{Bj} \tau$ for LHC. This is done using the Eqn.~\ref{bjEqn}.
This is shown in Fig.~ \ref{bjLHC}. The solid line is a logarithmic fit. The 
extrapolated value of $\epsilon_{Bj} \tau$  at LHC is $9.42 \pm 0.55~GeV~fm^{-2}~c^{-1}$. 
However, as the formation time at LHC will be much less than $1~ GeV/fm^3$, 
the above value sets a lower bound to the initial energy density to be formed 
at LHC.

Fig.~ \ref{bjNpart} shows the estimate of the product of the Bjorken energy
density and the formation time ($\epsilon_{Bj}. \tau$) as a function of the 
centrality of the collision in terms of $N_{part}$. The STAR estimation of
 $\epsilon_{Bj}. \tau$ at $\sqrt{s_{NN}}$ = 62.4 GeV Au+Au collisions for different
centrality classes, has been compared with similar data from PHENIX experiment, 
for similar collision species at different energies. 
As expected there is an increase in $\epsilon_{Bj}. \tau$ with increasing 
centrality of the collision. The STAR data for $\epsilon_{Bj}. \tau$ for Au+Au 
collisions at $\sqrt{s_{NN}}$ = 62.4 GeV show a cross over with PHENIX data for 
the peripheral collisions. This is because of different ways of estimation of 
the transverse overlap area by STAR and PHENIX (as explained earlier, 
see Fig.~ \ref{phenixTransA}).

While comparing the results from different experiments, related to the initial 
energy density, one need to take care of the following factors: (i) value of the 
formation time taken into the calculations, (ii) the procedure of estimation of 
the transverse overlap area and (iii) the value of the Jaccobian used to transform 
$\eta$ to $y$ phase space.

\subsection{Variation of $dE_T/d\eta$ and $dN_{ch}/d\eta$ per $N_{part}$ pair 
with $\sqrt{s_{NN}}$}

\begin{figure}[htbp23]
\begin{center}
\includegraphics[width=5.5in]{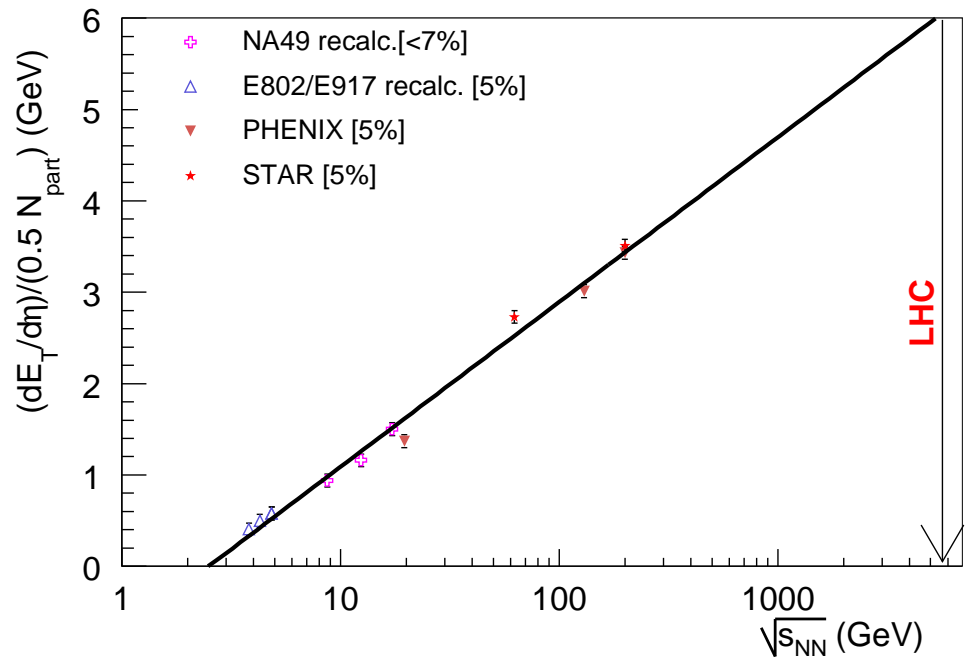}
\caption{$dE_T/d\eta$ per participant pairs measured in most central collisions
(value given in brackets) as a function of incident nucleon energy. The line 
is a logarithmic fit.}
\label{etCoM}
\end{center}
\end{figure}
\begin{figure}[htbp22]
\begin{center}
\includegraphics[width=5.5in]{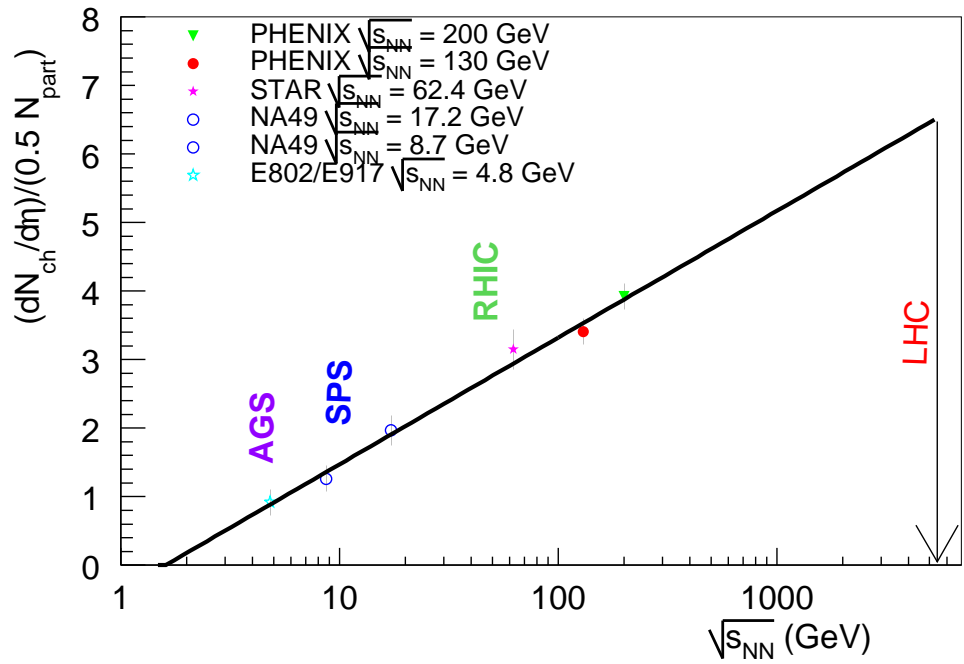}
\caption{$dN_{ch}/d\eta$ per participant pairs measured in most central collisions
as a function of incident nucleon energy. The line is a logarithmic fit. A prediction
to LHC energy is shown by an arrow.}
\label{nChCoM}
\end{center}
\end{figure}

To study the variation of $dE_T/d\eta$ and $dN_{ch}/d\eta$ per $N_{part}$ pair 
with center of mass energy we have taken the most central collisions data form 
different experiments at AGS \cite{ags1,ags2,ags3,ags4}, SPS \cite{Alber, 
na49-1, na49-2, na49-3}  and RHIC \cite{star200GeV,phenixEt,phenixSyst}.This is 
shown in Fig.~\ref{etCoM} for $E_T$ and in Fig.~\ref{nChCoM} $N_{ch}$. The values
of $dE_T/d\eta$ and $dN_{ch}/d\eta$ per participant pair for the top $5\%$ central
Au+Au collisions at $\sqrt{s_{NN}}$ = 62.4 GeV are found to be $2.73 \pm 0.29$ 
and $3.15 \pm 0.29$ respectively. We have tried to fit a phenomenological 
logarithmic function to both the figures motivated by the trend of the data 
in the range of available measurements. The fitted function is
\begin{equation}
{(dX/d\eta)/(0.5 ~N_{part}) = A~ ln (\sqrt{s_{NN}}/\sqrt{s_{NN}^0})},
\label{nChFit}
\end{equation}
where $X ~=~ E_T$ or $N_{ch}$. One can see the agreement of the fits in both 
the figures is very good.  The results of the fits are:\\
for $E_T$, $\sqrt{s_{NN}^0}~ = ~ 2.49 \pm 0.12$ GeV and $A~=~ 0.78 \pm 0.014$ GeV,\\
for $N_{ch}$, $\sqrt{s_{NN}^0}~ = ~ 1.60 \pm 0.38$ GeV and $A~=~ 0.80 \pm 0.05$,\\

The parameter $\sqrt{s_{NN}^0} = 2.49$ obtained from the $E_T$  fit is slightly 
greater than the minimum possible value of the center of mass energy i.e. 
$\sqrt{s_{NN}^{min}}= 2~amu$ (1.86) GeV. The FOPI experiment 
\cite{fopi1, fopi2, fopi3} has carried out a measurement of $dE_T/d\eta$ at 
$\sqrt{s_{NN}} = 2.05$ which is close to $\sqrt{s_{NN}^0} = 2.49$. For most central
collisions corresponding to $N_{part} = 359$, it has got $dE_T/d\eta$ = 5.0 GeV
\cite{phenixSyst}. The extrapolation of the fit to the lower energy suggests
that the logarithmic parametrization requires higher order terms to
describe how the $E_T$ production starts at very low $\sqrt{s_{NN}}$. This is
because, at very lower energies, this logarithmic parametrization gives a value
of $(dE_T/d\eta)/(0.5 ~N_{part})$ which is negative. But energy conservation
demands that this value must approach zero smoothly at very low center of mass 
energies. An estimation of the value of $dE_T/d\eta$ per participant pair at
 the LHC energy is found to be around $6.03 \pm 0.11$ GeV, based on the 
extrapolation of the fit. From top SPS energy ($\sqrt{s_{NN}}$ = 17.2 GeV) 
to the RHIC top energy 
($\sqrt{s_{NN}}$ = 200 GeV), there is an $131\%$ increase in $dE_T/d\eta$ per 
participant pair. Whereas, there is an increase of $73\%$ in $dE_T/d\eta$ per 
participant pair from RHIC top energy to LHC energy ($\sqrt{s_{NN}}$ = 5.5 TeV).
Fig.~\ref{nChCoM} shows the earlier mentioned logarithmic fit to $dN_{ch}/d\eta$ 
per participant pairs for $N_{part} = 350$. Unlike for $E_T$, the fit parameter 
$\sqrt{s_{NN}^0}$ for $N_{ch}$ is $1.60 \pm 0.38$ GeV. This is lower than the 
minimum center of mass energy as given by $\sqrt{s_{NN}^{min}} = 1.86$ GeV. It 
suggests that above $2~amu$, the $N_{ch}$ production as a function of center 
of mass energy should undergo a threshold-like behavior, unlike the $E_T$ production,  
which must approach zero smoothly because of energy conservation. 

However, the FOPI measurements at $\sqrt{s_{NN}}$ = 1.94 and 2.05 GeV agree with
the extrapolation of the fit to lower $\sqrt{s_{NN}}$ close to $2~amu$. It is an
interesting result that colliding nuclei with kinetic energies of 0.037 and 0.095 
GeV per nucleon in center of mass follow the same particle production trend, as is 
seen at AGS, SPS and RHIC energies.

The $dN_{ch}/d\eta$ per participant pairs for the most central events, shows about
$11.4\%$ increase from the highest AGS energy ($\sqrt{s_{NN}}$ = 4.8 GeV) to the 
top SPS energy. From the highest SPS energy to the highest RHIC energy, there is a
$100\%$ increase in $dN_{ch}/d\eta/(0.5~N_{part})$. Assuming that the same logarithmic
behavior extends to the LHC energy, the predicted value of $dN_{ch}/d\eta$ 
= $(6.53 \pm 0.45) \times (0.5 N_{part})$. This suggests that the increase in 
charge particle production from the highest RHIC energy to LHC, will be $\sim 65\%$ 
for the most central events.
\begin{figure}[htbp244]
\begin{center}
\includegraphics[width=3.5in]{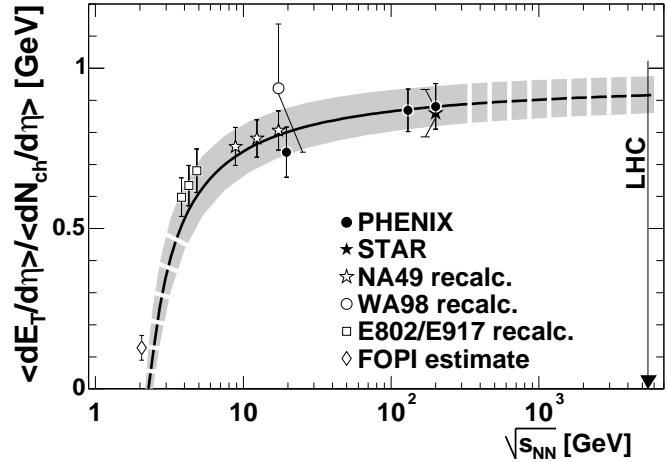}
\caption{The ratio of $E_T$ over $N_{ch}$ for the most central events as a function of
$\sqrt{s_{NN}}$. The line is the ratio of two fits shown in Fig.~\ref{etCoM} and 
Fig.~\ref{nChCoM}. The band corresponds to one standard deviation of the 
combined error.} 
\label{phenixfig14}
\end{center}
\end{figure}
PHENIX \cite{phenixSyst} explains the above behavior of  $E_T$ and $N_{ch}$
in the following way. In Fig.~\ref{phenixfig14} we have shown the $E_T/N_{ch}$ ratio 
for the most central collisions as a function of $\sqrt{s_{NN}}$. The solid line 
shown in the figure is the ratio of the fits used in Fig.~\ref{etCoM} (for $E_T$)
and Fig.~\ref{nChCoM} (for $N_{ch}$). This calculation agrees with the data very well. 
One can see there are two regions in the figure which could be clearly separated. 
The region from the lowest allowed energy to SPS energy is
characterized by a steep increase of the $E_T/N_{ch}$ ratio with $\sqrt{s_{NN}}$.
In this region, the increase in the incident energy causes an increase in the
$<m_T>$ of the produced particles \cite{na49-1,phenixPRC}. The second region starts 
from the SPS energies and continues above. In this region, $E_T/N_{ch}$ is very 
weakly dependent on $\sqrt{s_{NN}}$. The incident energy goes for particle 
production, rather increasing the energy per particle.

The shape of the $E_T/N_{ch}$ curve in the first region is governed by the difference 
in the $\sqrt{s_{NN}^0}$ parameter between $E_T$ and $N_{ch}$. However, in the second
region it is dominated by the ratio of the $A$ parameters in the fits, which is close 
to  1 GeV. The extrapolation of the curve to LHC energy gives a value of $E_T/N_{ch}$ 
equal to $(0.92 \pm 0.06)$ GeV.

\subsection{The Electromagnetic Component of Total Transverse energy}

\begin{figure}[htbp24]
\begin{center}
\includegraphics[width=5.5in]{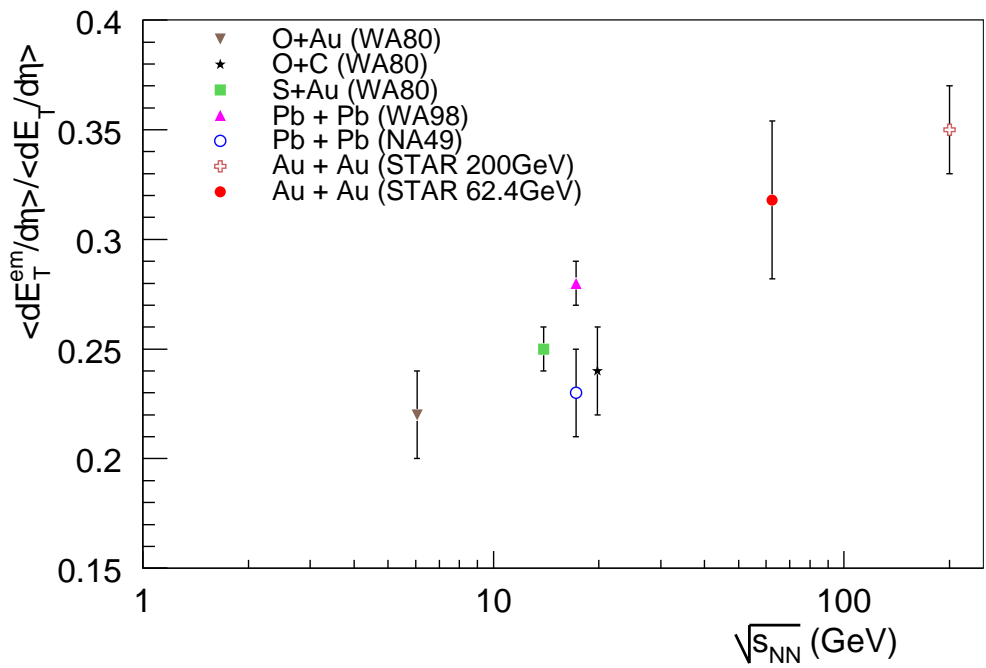}
\caption{The center of mass energy dependence of electromagnetic fraction of
total transverse energy for a number of systems from SPS to RHIC energy for
central events.} 
\label{fig12F}
\end{center}
\end{figure}
The electromagnetic fraction of total transverse energy ($E_T^{em}/E_T$) has 
contributions predominantly from photons emitted at all stages of evolution
of the fireball. This is because, the production cross-section for electrons
and positrons, which come in electromagnetic sector, is very less. Hence, 
this ratio can roughly give information on the integrated photon energy produced
in the collision process. As photons are emitted at all stages of the fireball
evolution, we have direct photons coming from the plasma and also thermal or
decayed photons (from electromagnetic and weak decays) coming at later stages. 
The main sources of decayed photon are $\pi^0, \omega^0$ and $\rho$. In the 
hadronic sector the production cross-section of pions is the highest. If we 
assume all the isospin partners of pions are equally probable to be produced
in the collision, then the decay of each neutral pion into two photons would
result in an $E_T^{em}/E_T$ ratio of 0.33. If we take other decay sources of 
photons, then this number should be slightly larger than 0.33. For a long-lived 
plasma we would expect a very high photon yield \cite{halzenF}. Hence, if we 
observe an increase in this ratio significantly higher than this value, then 
we can say something on the formation of a long-lived deconfined plasma phase. 
In addition, the electromagnetic fraction of total energy will strongly be influenced
by the meson to baryon ratio. This is because, the main sources of photons are 
mesons. At lower energies due to nuclear stopping, the baryon content in the produced
matter will be very high. At very large energies where the system is almost
baryon free, we would expect virtually all the $E_T$ to be carried by 
mesons. Hence the electromagnetic fraction of total energy should increase 
with center of mass energy.

Taking top $5\%$ central events, we have estimated the electromagnetic fraction
of the total transverse energy for 62.4 GeV Au+Au collisions. This result is shown 
in Fig.~\ref{fig12F}, as a function of the center of mass energy, along 
with similar data from other measurements ranging from lower SPS energies 
\cite{wa80prc, wa98Syst} to the top RHIC energy. The values of 
$<dE_T^{em}/d\eta>/<dE_T/d\eta>$ ratio are $0.35 \pm 0.02$ and $0.318 \pm 0.03$, 
for $\sqrt{s_{NN}} =$ 200 GeV and 62.4 GeV respectively.  The estimated values of
$<dE_T^{em}/d\eta>/<dE_T/d\eta>$ for different centralities in Au+Au collisions
at $\sqrt{s_{NN}} =$ 62.4 GeV are given in Table-~\ref{valuesT1}.
At SPS, for Pb+Pb collisions at $\sqrt{s_{NN}} =$ 17.2 GeV, this ratio has a value of 
$0.23 \pm 0.02 $ \cite{Alber}. There is a $52\%$ increase of the electromagnetic 
fraction of total energy from SPS to RHIC top energy. However, from the values of 
these ratios, observed at RHIC energies, it is not conclusive to say anything about 
the formation of a long-lived deconfined phase of quarks and gluons.
\begin{figure}[htbp25]
\begin{center}
\includegraphics[width=5.5in]{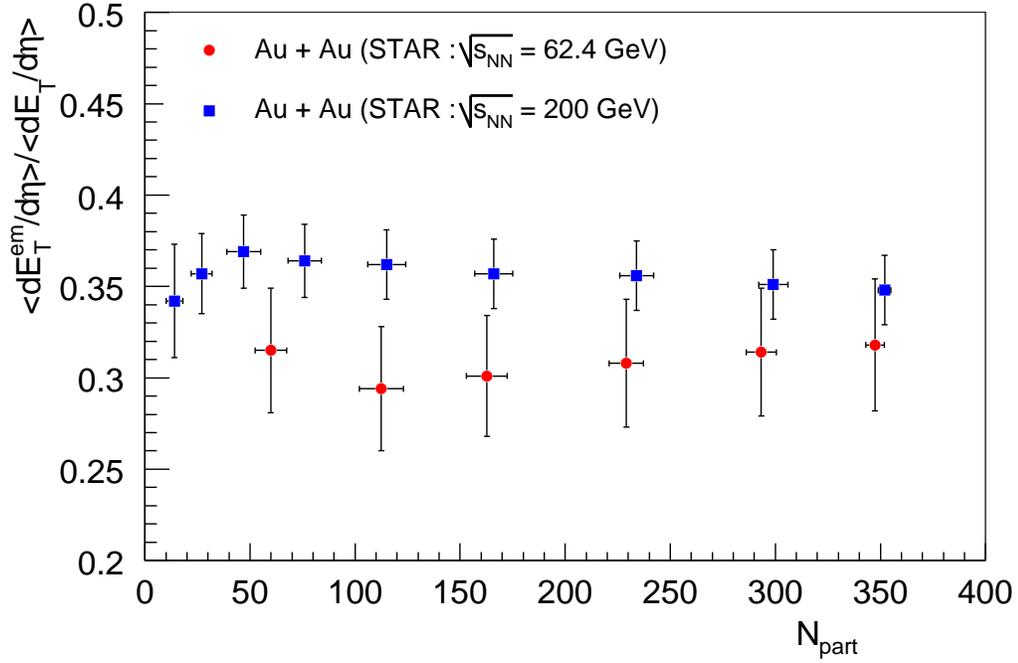}
\caption{The centrality dependence of electromagnetic fraction of
total transverse energy.} 
\label{fig13F}
\end{center}
\end{figure}
The energy dependence seen in Fig.~\ref{fig12F} is presumably dominated by
the total meson content of the final state. The centrality dependence of this 
ratio may provide additional information about the reaction mechanisms. 
This is shown in Fig.~\ref{fig13F} for the Au+Au collisions at $\sqrt{s_{NN}}$
 = 62.4 GeV and 200 GeV. However, no significant centrality dependence of the 
electromagnetic fraction has been observed. This is similar to what has been 
found in Au+Au collisions at $\sqrt{s_{NN}}$ = 200 GeV \cite{star200GeV}. 

\subsection{The Ratios of $E_T^{had}$ and $E_T^{em}$}

\begin{figure}[htbp26]
\begin{center}
\includegraphics[width=2.6in]{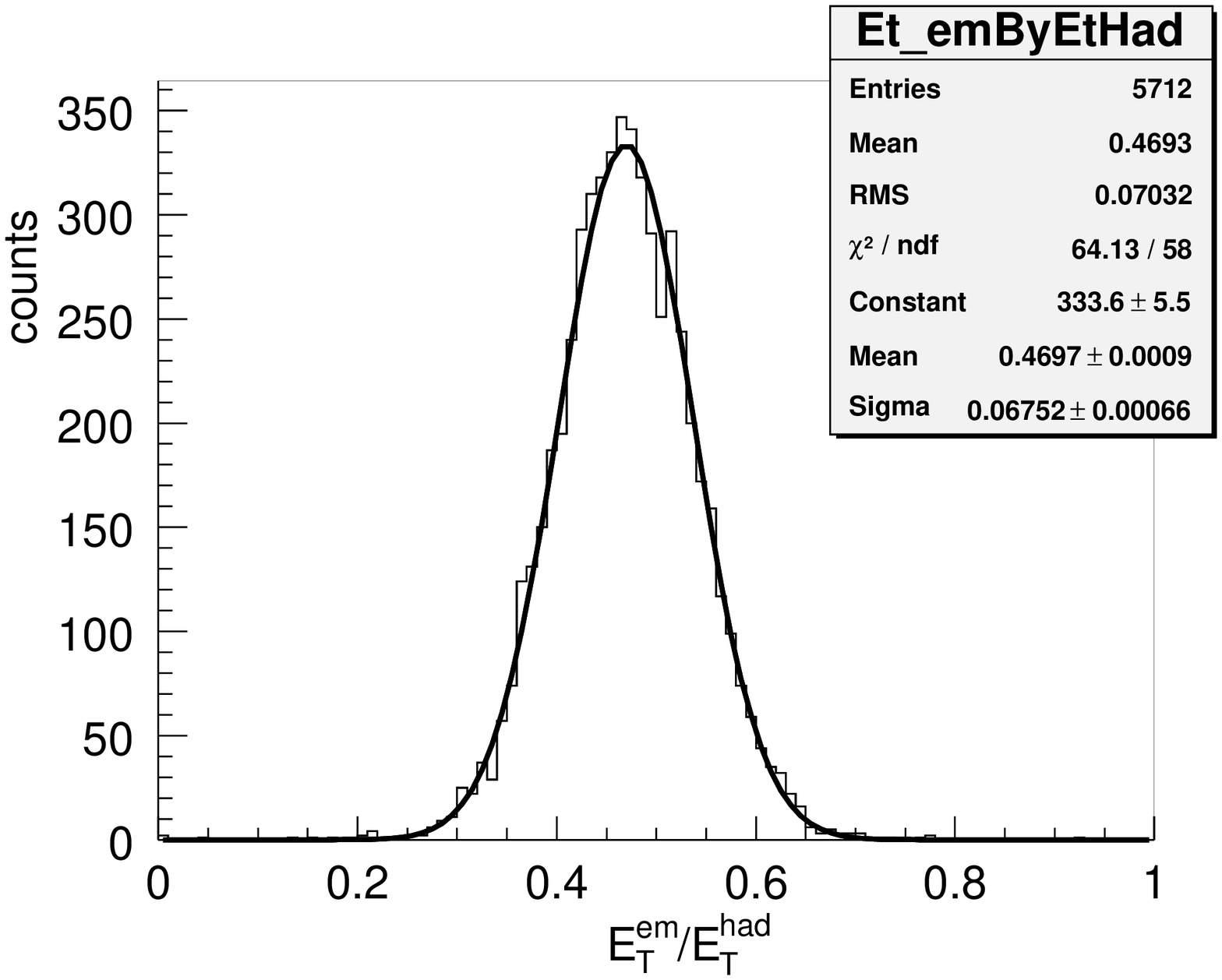}
\includegraphics[width=2.6in]{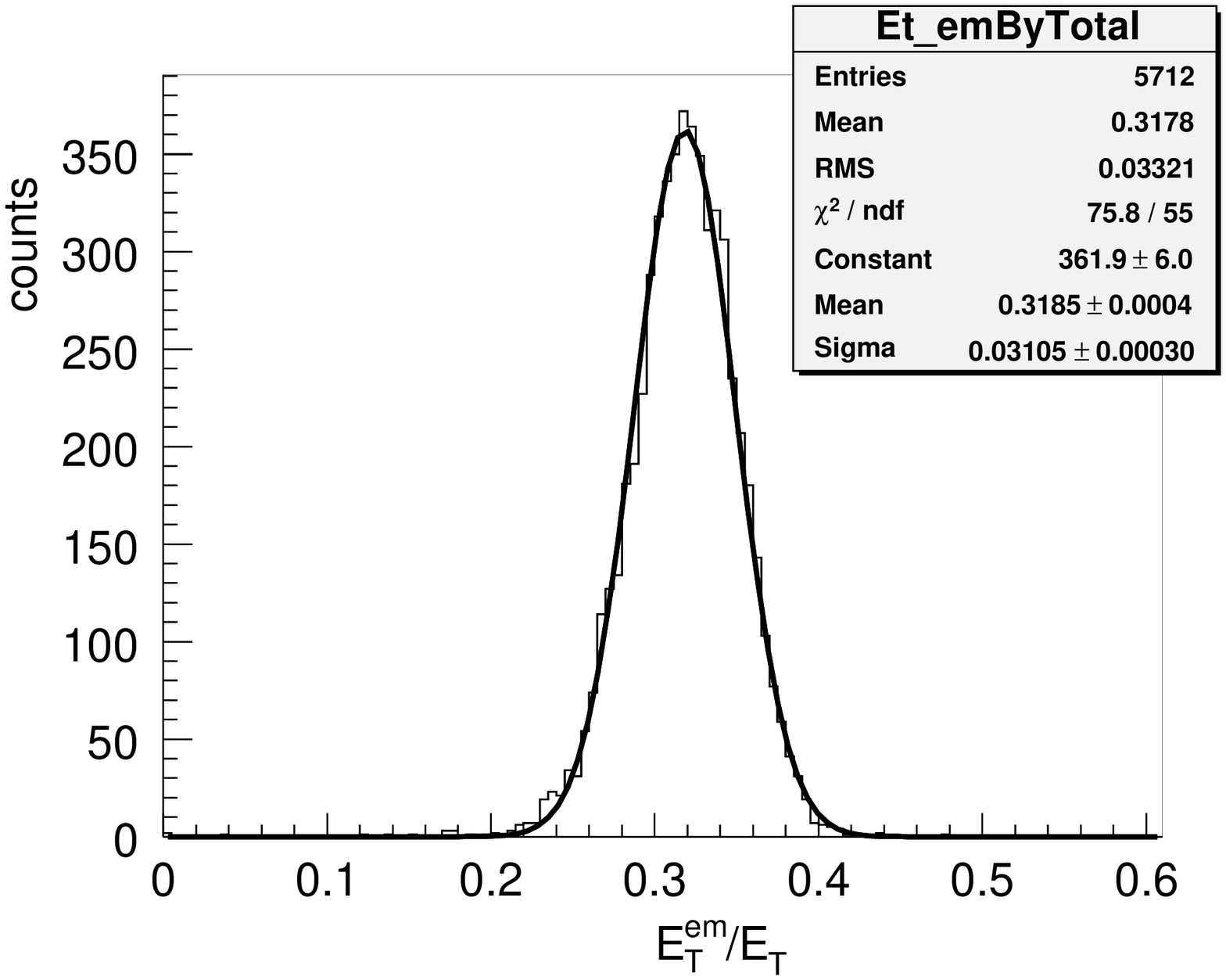}
\caption{Event-by-event distributions of a) $E_T^{em}/E_T^{had}$ and b) $E_T^{em}/E_T$ 
for the top $5\%$ central events in $\sqrt{s_{NN}}$ = 62.4 GeV Au+Au collisions. A 
gaussian function is fitted to the distributions.}
\label{ratios}
\end{center}
\end{figure}
\begin{figure}[htbp27]
\begin{center}
\includegraphics[width=4.5in]{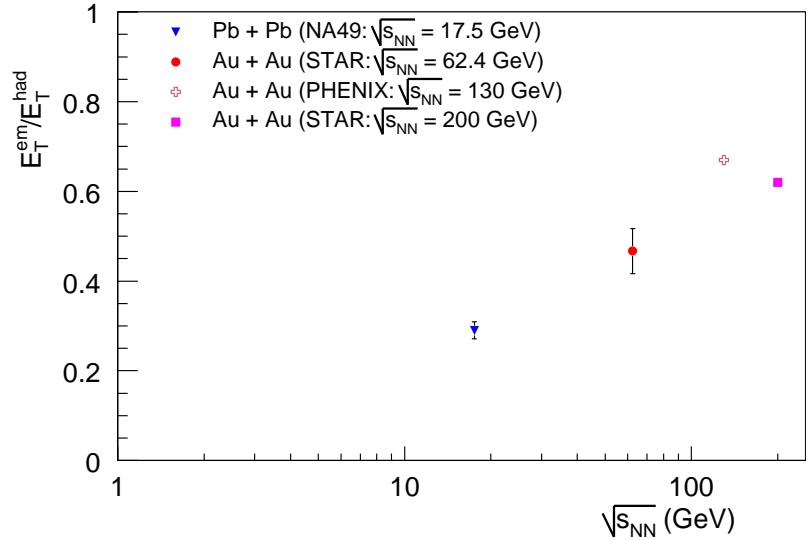}
\caption{$E_T^{em}/E_T^{had}$ is shown as a function of center of mass energy.
Data points are from NA49 \cite{Alber}, PHENIX \cite{phenixEt} and 
STAR \cite{star200GeV}}
\label{emByHadVsCM}
\end{center}
\end{figure}

Taking top $5\%$ central events in 62.4 GeV Au+Au collisions, we have plotted
the ratio of $E_T^{em}$ to $E_T^{had}$  and the ratio of $E_T^{em}$ to $E_T$
in Fig.~\ref{ratios}. The mean value of $E_T^{em}/E_T^{had}$ and of$E_T^{em}/E_T$
are found to be $0.47 $ and $0.32$ respectively. These have been obtained from
gaussian fittings to the distributions. The gaussian behavior of these distributions
indicate the possibility of carrying out fluctuation studies in these ratios.
We have estimated the ratio of $E_T^{em}/E_T^{had}$ for top $5\%$ central events
in $\sqrt{s_{NN}}$ = 62.4 GeV Au+Au collisions. The value of this ratio is
$0.47 \pm 0.05$. This ratio is shown in Fig.~\ref{emByHadVsCM} as a function of
center of mass energy, along with similar data from other measurements (NA49 at SPS,
PHENIX and STAR at RHIC). This ratio is found to increase with center of mass energy. 
\section{Estimation of Systematic Uncertainties}
\begin{table}[htb]
\begin{center}
\begin{tabular}{|l|l|l|l|l|}
\hline
 Sources of Syst. Errors  & $E_T^{had}$  & $N_{ch}$ & $E_T^{em}$ \\
\hline
$|V_z| < 50~cm$   & -0.46 \%& -0.62\% & - \\
$|dac| < 4.5~cm$  & -0.26 \%& -0.27\% & - \\
$nFitPoints > 5$  & +3.3 \% & +3.3 \% & - \\
$nFitPoints > 15$ & -5.8 \% & -6.6 \% & - \\
correction factors & $\pm 8.5 \%$ & $\pm 6.4 \%$ & $\pm 2.2 \%$\\
calibration & - & - & $\pm 2. \%$ \\
\hline
Total & $10.3 \%$ & $9.1\%$ & $3.\%$\\
\hline
\end{tabular}
\vspace{0.1cm}
\caption{The systematic uncertainties in percentage for top $5\%$ central Au+Au 
collisions at $\sqrt{s_{NN}}$ = 62.4 GeV. }
\label{systT}
\end{center}
\end{table}

\begin{table}[htb111]
\begin{center}
\begin{tabular}{llllll}
\hline
\hline
 Centrality (\%) & $N_{part}$ & $E_T$ (GeV)  & $E_T^{had}$ (GeV) & $E_T^{em}$ (GeV) & 
$E_T/N_{ch}$ (GeV)\\ 
&&$\pm 11\%$&$\pm 10.3\%$&$\pm 3\%$ & $\pm 14\%$ \\ \hline
0-5   &$347.4 \pm 4.4$&$474 \pm 51$ & $322 \pm 33$&$151 \pm 5$    & $0.87 \pm 0.12$ \\
5-10  &$293.3 \pm 7.1$&$389 \pm 43$ & $267 \pm 27$&$122 \pm 3$    & $0.86 \pm 0.12$ \\
10-20 &$229.1 \pm 8.2$&$297 \pm 33$ & $206 \pm 21$&$92 \pm 3$     & $0.85 \pm 0.12$ \\
20-30 &$162.7 \pm 9.7$&$205 \pm 22$ & $143 \pm 15$&$62 \pm 2$     & $0.84 \pm 0.12$ \\
30-40 &$112.5 \pm 10.5$&$136 \pm 15$ &$96 \pm 10$ &$40 \pm 1$     & $0.82 \pm 0.12$ \\
40-60 &$59.9 \pm 7.5$&$68 \pm 7$   & $49 \pm 5$&$21.4 \pm 0.6$ & $0.80 \pm 0.12$ \\
60-80 &$19.3 \pm 3.5$&$24 \pm 3$   & $16 \pm 2$&$11.2 \pm 0.3$ & $0.77 \pm 0.11$ \\
\hline
\hline
\end{tabular}
\vspace{0.1cm}
\caption{$E_T$, $E_T^{had}$, $E_T^{em}$ and $E_T/N_{ch}$ for different centralities
in Au+Au collisions at $\sqrt{s_{NN}}$ = 62.4 GeV. All uncertainties are systematic 
and statistical errors are negligible.}
\label{valuesT}
\end{center}
\end{table}

\begin{table}[htb1111]
\begin{center}
\begin{tabular}{lllll}
\hline
\hline
 Centrality (\%) &$N_{part}$ & $E_T/(0.5N_{part})$ (GeV) & $E_T^{em}/E_T$ & 
$\epsilon_{Bj} \tau ~(GeV fm^{-2}c^{-1})$ \\ 
&&$\pm 10.6\%$&$\pm 11.3\%$&$\pm 10.8\%$ \\ \hline
0-5   &$347.4 \pm 4.4$ & $2.73 \pm 0.29$  &$0.318 \pm 0.036$    & $ 3.97\pm 0.43$ \\
5-10  &$293.3 \pm 7.1$ &$2.65 \pm 0.3$  &$0.314 \pm 0.035$    & $3.64 \pm 0.41$ \\
10-20 &$229.1 \pm 8.2$ &$2.59 \pm 0.3$  &$0.308 \pm 0.035$     & $3.29 \pm 0.37$ \\
20-30 &$162.7 \pm 9.7$ &$2.52 \pm 0.3$  &$0.301 \pm 0.033$     & $2.85 \pm 0.32$ \\
30-40 &$112.5 \pm 10.5$ &$2.41 \pm 0.34$  &$0.294 \pm 0.034$     & $2.41 \pm 0.30$ \\
40-60 &$59.9 \pm 7.5$  &$2.26 \pm 0.36$    &$0.315 \pm 0.034$ & $1.83 \pm 0.24$ \\
60-80 &$19.3 \pm 3.5$ &$2.47 \pm 0.54$    &$0.468 \pm 0.054$ & $1.37 \pm 0.24$ \\
\hline
\hline
\end{tabular}
\vspace{0.1cm}
\caption{ $E_T/(0.5N_{part})$, $E_T^{em}/E_T$  and 
$\epsilon_{Bj} \tau $ for different centrality classes in Au+Au collision at 
$\sqrt{s_{NN}}$ = 62.4 GeV. All uncertainties are systematic and statistical 
errors are negligible.}
\label{valuesT1}
\end{center}
\end{table}

There are various sources of systematic uncertainties in the estimation of
both the components of $E_T$. We have divided these sources into two classes.
One class consists of uncertainties in the estimation of various correction
factors used in the measurement. This has been discussed previously in this
chapter. The other class is having contributions from the dynamical cuts
used in the analysis. The final systematic uncertainty is the quadratic
sum of the contributions from both the classes. Here, we discuss the estimation
procedure of the uncertainties coming from the dynamic cuts used. The dynamic
cuts used in this analysis include a DCA cut ($|dca| < 3~cm$) to select primary 
tracks, a cut on z-position of the vertex ($|V_z| < 30 ~ cm$) and the number of
fit points cut on TPC tracks ($nFitPoints > 10$). For the estimation of $E_T^{had}$
and the charge particle multiplicity ($N_{ch}$), we have varied the DCA-cut to 
$|dca| < 4.5~cm$, the z-position of the vertex to $|V_z| < 50 ~ cm$ and the number 
of fit points cut to $nFitPoints > 5$ in one case and $nFitPoints > 15$ 
in another case.  While varying individual cuts, we keep other baseline 
cuts unchanged and find out the percentage of variation in $E_T^{had}$ and 
$N_{ch}$. For the estimation of 
final systematic uncertainties in $E_T^{em}$, we have taken the quadratic sum of the
uncertainties due to correction factors and that due to the calibration procedure.
The positive and negative errors are added up separately in quadrature rule. The larger
of the two is the final systematic error. Table.~\ref{systT} enlists the sources of
systematic errors and their estimated values for $E_T^{had}$, $E_T^{em}$ and $N_{ch}$.
The uncertainty in $E_T$ is the quadratic sum of the same estimated in $E_T^{had}$
and $E_T^{em}$. We have estimated the systematic uncertainties in the top $5\%$ central
Au+Au collisions and have applied the same percentage to the systematic uncertainties
in other centrality classes. 

Taking the systematic uncertainties in $E_T$, $E_T^{had}$, $E_T^{em}$ and $N_{ch}$
we have estimated the uncertainties in various observables such as 
$E_T/(0.5N_{part})$, $E_T^{em}/E_T$, $\epsilon_{Bj} \tau$ etc., using 
the usual error propagation formula (see Appendix-1). The estimated values of  
these observables and the coresponding systematic errors are tabulated in 
Table.~\ref{valuesT} and Table.~\ref{valuesT1}. The method and
formulae used to estimate the systematic uncertainties in different observables
are given in Appendix-1.

\vfill 
\eject

\chapter{Transverse Energy Fluctuations}
\markboth{nothing}{\it Transverse Energy Fluctuations}

\section{Introduction}

Fluctuations in physical observables in heavy ion collisions have
been a topic of interest since decades. They provide important
signals regarding the formation of QGP and also help to address
important questions related to thermalization \cite{fluctTherm}.
The study of fluctuations on an event-by-event basis has become a 
reality with the advent of modern accelerators where a large number
of particles are produced in heavy ion collisions. There have been
several new methods for the study of event-by-event fluctuations in
global observables to probe the nature of the QCD phase transition
\cite{QCDPhase,baymHH,asakawaMuller,jeonKoch,fluctMethods}. In a 
thermodynamic picture of a strongly interacting system produced in the
collision, the fluctuations in particle multiplicities, mean transverse
momentum ($<p_T>$), transverse energy ($E_T$), net charge and other 
global observables, are related to the fundamental properties of the 
system, such as the chemical potential, specific heat and the matter
compressibility. These help in the understanding of the critical 
fluctuations near the QCD phase boundary. The existence of a tricritical
point at the QCD phase boundary, has been predicted to be associated with 
large event-by-event fluctuations in the above observables \cite{QCDPhase}.

It is believed that in the first order phase transition scenario, the
supercooling might lead to density fluctuations which result in droplet
and hot spot formations \cite{droplet}. These might lead to fluctuations
in rapidity in the form of spikes and gaps. The event-by-event fluctuations
of the number of photons to charged particles has been proposed to be 
a probe for the formation of the disoriented chiral condensates (DCC)
\cite{dcc1, dcc2}. The study of fluctuations in particle ratios, e.g.
$K/\pi, \pi^+/\pi^-, \pi^0/\pi^\pm$ etc. are interesting, as many systematic
errors get nullified. There are also theoretical predictions on strangeness 
fluctuations \cite{strange}.

$E_T$ is an extensive global variable which provides a direct measure
of the violence of an interaction. It is also an indicator of the energy
density produced in the collision. Since energy density is directly 
associated with the quark-hadron phase transition, it is extremely
important to study $E_T$ and fluctuations in $E_T$. In  addition, it is
interesting to compare the fluctuations in $E_T$ to that observed in the
particle multiplicities. This is because, $E_T$ has a very good 
correlation with the charge particle multiplicity and the center of mass 
energy. In addition, $E_T$ fluctuation has been predicted to be leading 
to $J/\Psi$ suppression \cite{etJPsi}. It is also interesting to 
study the fluctuations in both the components of $E_T$, as we have an 
independent event-by-event measurement of both the electromagnetic and
hadronic components of transverse energy.

There have been a lot of interest to study event-by-event fluctuations 
which is motivated by the experimental observation of near perfect 
Gaussian distributions of $<p_T>$, particle multiplicity and particle ratios 
at various centralities \cite{gaussian,phenixEtFluct}. The variance or 
width of these Gaussian distributions contain information about the 
reaction mechanism as well as the nuclear geometry 
\cite{QCDPhase,baymFluct,rAlbrecht,na34Fluct}.

\section{Fluctuation Measures}
There are various different methods used to quantify fluctuations in 
different observables. This also helps in comparing the measurements
carried out by different experiments which use different methods for 
fluctuation studies.  

\subsection{The Variance and the $\Phi $ Measure}
The relative fluctuation ($\omega_A$), in an observable $A$, could be 
expressed as
\begin{equation}
{\omega_A = \frac{\sigma_A}{<A>}~= \frac{(<A^2> - <A>^2)^{1/2}}{<A>}}
\label{var}
\end{equation}
where, $\sigma_A$ is the standard deviation of the distribution and $<A>$ is 
the mean value. The variance which we get from the experimental data
has contributions both from the statistical and dynamical sources.
One need to know the statistical and other sources of fluctuations
in order to extract new physics from the observed dynamical fluctuations.
The known sources of fluctuations contributing to the $\omega_A$ include,
the finite particle multiplicity, impact parameter fluctuations in the real
experiment, effect of limited detector acceptance, fluctuations in the
number of primary collisions, resonance decays, Bose-Einstein correlations,
fluctuations in the transverse flow velocity and the effects of 
re-scattering of the secondary particles \cite{QCDPhase}. 
$\frac{\sigma_A^2}{<A>}$ is also used as a measure of fluctaution 
\cite{wa98_1}.

To get rid of the statistical fluctuations in $\omega_A$, it is calculated
independently from data and from the baseline i.e. mixed event distributions.
The difference in fluctuation from a random baseline is defined as
\begin{equation}
{d = \omega_{(A,data)} ~-~ \omega_{(A,baseline)}}
\label{d}
\end{equation}
The sign of $d$ is positive if the data distribution contains a correlation.
The fraction of the fluctuations that deviate from the random baseline
expectation is
\begin{eqnarray}
F_A = \frac{\omega_{(A,data)} ~-~ \omega_{(A,baseline)}}
{\omega_{(A,baseline)}}
=\frac{\sigma_{(A,data)} ~-~ \sigma_{(A,baseline)}}
{\sigma_{(A,baseline)}}
\label{fa}
\end{eqnarray}
where $\sigma_{(A,data)}$ refers to the standard deviation of the 
event-by-event data distribution and $\sigma_{(A,baseline)}$ refers
to the standard deviation of the mixed event distribution.
The most commonly used quantity $\Phi$ \cite{phiMeasure} is given by
\begin{equation}
{\Phi_A =[\sigma_{(A,data)} ~-~ \sigma_{(A,baseline)}]\sqrt{<N_A>}
=d<A>\sqrt{<N_A>}}
\label{phim}
\end{equation}
where $N_A$ is the number of particles or tracks or clusters depending on
specific cases. $\Phi_A$ is related to $F_A$ by
\begin{equation}
{\Phi_A = F_A ~ \sigma_{(A,baseline)}~\sqrt{<N_A>}.}
\label{phiF}
\end{equation}
All the above variables are used to quantify the fluctuations and they are
all correlated through the above equations.

\subsection{The $\nu_{dyn}$}
The variable $\nu_{dyn}$, as a measure of fluctuations was first suggested
in Ref.\cite{fluctMethods}. The basis of this variable is the single- and
two-particle distribution functions. The $\nu_{dyn}$ used for the net-charge
fluctuation is given by
\begin{equation}
{\nu_{+-,dyn} = \frac{<N_+(N_+-1)>}{<N_+>^2}+ \frac{<N_-(N_--1)>}{<N_->^2}
-2\frac{<N_+N_->}{<N_+><N_->}}
\label{nuDynamic}
\end{equation}
This is based on two-particle correlation functions. The $\nu_{dyn}$ is very robust, 
as it is independent of the detector efficiency and acceptance.

\subsection{The Balance Function}
The balance function, as a technique to study the dynamics of hadronization
in relativistic heavy ion collisions, was first proposed by Bass {\it et al.}
\cite{balanceFn}. The idea behind balance function is that the rapidity 
range of the correlations is changed when a collision forms QGP. To be
specific, charged hadrons form late in the reaction, after hadronization,
resulting in shorter-ranged correlations in rapidity space for 
charge/anti-charge pairs than expected in the absence of the plasma.
The balance function is defined as
\begin{equation}
{B(p_2|p_1) \equiv \frac{1}{2}[\rho(b,p_2|a,p_1)-\rho(b,p_2|b,p_1)
+\rho(a,p_2|b,p_1) - \rho(a,p_2|a,p_1)],}
\label{balance}
\end{equation}
where $\rho(b,p_2|a,p_1)$ is the conditional probability of observing a 
particle of type $b$ in bin $p_2$ given the existence of a particle of type
$a$ in bin $p_1$. The conditional probability $\rho(b,p_2|a,p_1)$ is generated
by first counting the number $N(b,p_2|a,p_1)$ of pairs that satisfy both
criteria and then dividing by the number $N(a,p_1)$ of particles of type
$a$ that satisfy the first criteria.
\begin{equation}
{\rho(b,p_2|a,p_1) = \frac{N(b,p_2|a,p_1)}{N(a,p_1)}}
\label{condProb}
\end{equation}
The balance function is also related to the $\nu_{dyn}$ \cite{fluctMethods}.

\begin{figure}[htbp41]
\begin{center}
\includegraphics[width=3.5in]{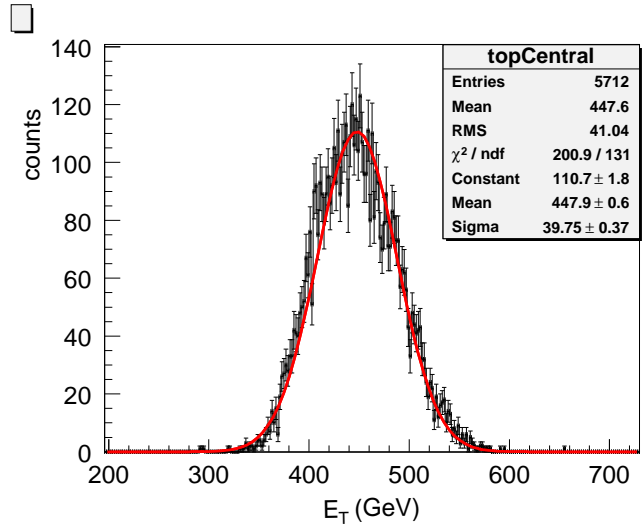}
\caption{The $E_T$ distribution for the top $5\%$ central events for $\sqrt{s_{NN}}$ = 62.4 GeV
Au+Au collisions, is shown to fit to a Gaussian distribution function.}
\label{topcentral}
\end{center}
\end{figure}

\section{The $E_T$ Fluctuation}
\begin{figure}[htbp41]
\begin{center}
\includegraphics[width=3.5in]{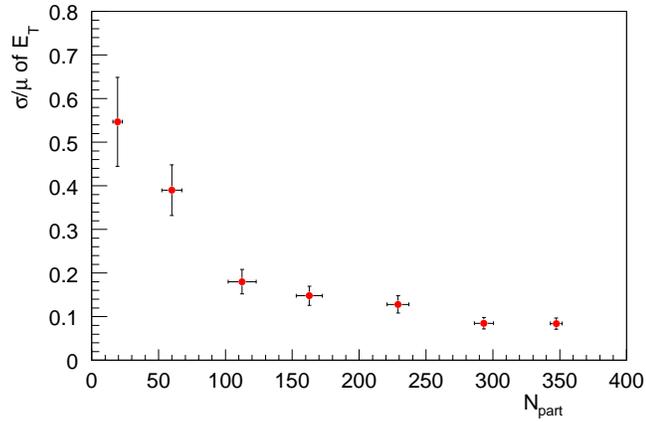}
\caption{The $\sigma/\mu$ of $E_T$ for $\sqrt{s_{NN}}$ = 62.4 GeV Au+Au 
collisions is shown as a function of the centrality of the collision.}
\label{sigmaBymean}
\end{center}
\end{figure}

As discussed in Chapter 3, for all centrality classes, the $E_T$ distributions
are Gaussians. Hence, we can study fluctuations in $E_T$. 
Fig.~\ref{topcentral} shows the $E_T$ distribution for top $5\%$ central
events in $\sqrt{s_{NN}}$ = 62.4 GeV Au+Au collisions. We have estimated
the $\sigma/<E_T>$ for all centrality classes. This is shown in 
Fig.~\ref{sigmaBymean} as a function of centrality. When we go from the central 
to peripheral collisions, the fluctuation in $E_T$ increases. However, there is 
a necessity of taking narrower centrality bins and more efficient fluctuation 
variable to extract physics from dynamical fluctuations in $E_T$. For the top 
$5\%$ central events in $\sqrt{s_{NN}}$ = 62.4 GeV Au+Au collisions, the 
$\sigma/<E_T>$ is found to be $8.4\%$.

For the top $5\%$ central events, we have already shown the event-by-event
distribution of $E_T^{em}/E_T^{had}$ and $E_T^{em}/E_T$ in Chapter 3. They fit 
very well to Gaussian functions.  The fluctuations in these observables in terms 
of their $\sigma/\mu$ for top $5\%$ central events in $\sqrt{s_{NN}}$ = 62.4 GeV 
Au+Au collisions are $14.4\%$ and $9.74\%$ respectively.

\vfill 
\eject

\chapter{Summary and Conclusion}
\markboth{nothing}{\it Summary and Conclusion}
In thesis, we have reported results on the transverse energy measurement 
and fluctuation studies in  Au+Au collisions at $\sqrt{s_{NN}}$ = 62.4 GeV.
This data set was taken by the STAR experiment at RHIC, BNL. We have used 
the STAR TPC and the BEMC detectors for this data analysis. In the common 
phase space of both the detectors ($0 < \eta < 1$ and full $\phi$), we 
have obtained the hadronic transverse energy from the TPC reconstructed 
tracks and the electromagnetic transverse energy from the BEMC tower 
hits after correcting for the hadronic contaminations in the calorimeter. 
This method provides an independent measurement of both the components 
of transverse energy on an event-by-event basis. The measurement 
of $E_T$ is very important as it helps in the estimation of the initial 
energy density of the produced fireball. The initial energy density 
tells about the QCD phase transition which is related to the formation 
of QGP. The centrality dependence of $E_T$ production is crucial in 
understanding the particle production mechanism. The fluctuations in 
$E_T$ can provide important information related to the QCD phase transition.

With the above motivations, we have measured the $E_T$ from both of it's 
components and have made a detailed study with center of mass
energy and centrality of the collisions.  For the top $5\%$ central collisions
we have obtained $<E_T>_{5\%} = 474 \pm 51 ~(syst)~\pm 1 ~(stat)$ GeV. 
We have also investigated
the scaling of transverse energy with the number of participating nucleons
and with the number of charged particles produced in the collision. For the
$5\%$ most central collisions we have obtained, $<dE_T/d\eta>/(0.5 N_{part})
= 2.73 \pm 0.29$ GeV and $<dE_T/d\eta>/<dN_{ch}/d\eta> = 866 \pm 122$ 
 MeV respectively. The most striking feature is the observation of a
nearly constant value of $<dE_T/d\eta>/<dN_{ch}/d\eta>$ from AGS, SPS to RHIC.
This also shows a very weak centrality dependence. It is found that the 
increase in $E_T$ production from lower AGS to higher RHIC energies 
comes mostly from the increase in particle multiplicity. 
A final state gluon saturation model (EKRT) is compared with the data. 
A different centrality behavior has been predicted
by this model. However, the uncertainty in the $N_{part}$ estimation rules
out discarding this model. $<dE_T/dy>/(0.5 N_{part})$ is seen to increase
logarithmically with center of mass energy: from AGS, SPS to RHIC. 

Theoretical calculations along with the measurements at RHIC suggest that
a dense and equilibrated system has been formed in the collision. The
expansion of the system also respects the hydrodynamics of an ideal 
fluid. There is a very good agreement between the hydrodynamic calculations
and the results from identified spectra and elliptic flow \cite{elliptic},
on the onset of hydrodynamic evolution at a time $\tau_0 < 1$ fm/c after
the collision \cite{agreement}. In addition, the suppression phenomenon
for high-$p_T$ hadrons observed at RHIC \cite{supression}, suggests that
the system at it's early times is very dense. Based on the boost-invariant 
Bjorken hydrodynamic model, the estimation of the initial density of the 
produced system is about $3.65 \pm 0.39~GeV/fm^3$. This has been calculated from the
measured $E_T$ for the top central collisions at $\sqrt{s_{NN}}$ = 62.4 GeV.
This energy density is almost 50-100 times the cold nuclear matter density and
is a lower bound to the estimated energy density. This is because, there
is a strong reduction in observed $E_T$ relative to that produced initially,
due to the longitudinal hydrodynamic work done during the expansion 
of the system. All these different approaches agree roughly for the 
estimated energy density, with a value which is well in excess 
of that predicted by lattice QCD for the
deconfinement phase transition. Taking similar colliding species i.e. Au,
from all the similar measurements (top $5\%$ central collisions) of the 
initial energy density at RHIC energies and assuming the validity of Bjorken 
hydrodynamic model at energies higher to RHIC, we have predicted the 
value of $\epsilon_{Bj} \tau \sim 9.42 \pm 0.55 ~GeV/fm^{-2}c^{-1}$ for LHC. 
In this approach the advantage of taking similar colliding species is
that, the estimation will be independent of the number of participant nucleons.

We have also made predictions for the $E_T$ and charge particle multiplicity
for the LHC energies, based on the measurements of $dE_T/d\eta$ per participant
pair and $dN_{ch}/d\eta$ per participant pair at different energies from
AGS to RHIC. These quantities show a logarithmic increase with the 
center of mass energy. Based on the extrapolation of the above logarithmic
behavior, we have given prediction for the LHC energy. The values of
 $dN_{ch}/d\eta$ and $dE_T/d\eta$ at LHC, are  $\sim (6.53 \pm 0.45) \times 
(0.5 N_{part})$ and $\sim (6.03 \pm 0.11) \times (0.5 N_{part})$ respectively.

 The electromagnetic fraction of the total transverse energy for top $5\%$
central events obtained in this work is $0.318 \pm 0.036$, consistent with a final
state dominated by mesons. There are theoretical predictions regarding the
formation of a long-lived deconfined plasma state in central events, which
increase the yield of direct photons. This, however should increase the 
electromagnetic fraction of the transverse energy. Our observations at RHIC,
suggest the electromagnetic fraction is almost independent of collision
centrality. Hence, from this behavior and the values of the electromagnetic 
fraction of transverse energy obtained at RHIC, it is not possible to 
conclude regarding the formation of a long-lived plasma phase. 
This however, doesn't discard the possibility of formation of a plasma 
for a very short period of time. In addition, the ratio of 
$E_T^{em}$ to $E_T^{had}$ increases with center of mass energy. The meson 
dominance at higher energies is reflected from this behavior.

Furthermore, we have studied the event-by event fluctuations in $E_T$ and
in the ratio of it's components. We have taken the $\sigma/\mu$ as the
fluctuation variable. For $5\%$ most central collisions, the fluctuation
in $E_T$ is found to be $8.4\%$. The fluctuation in $E_T$ has been found to 
increase when we go from central to peripheral collisions. The fluctuations
in $E_T^{em}/E_T^{had}$ and $E_T^{em}/E_T$ are found to be $14.4\%$
and $9.74\%$ respectively, for the $5\%$ most central collisions at
$\sqrt{s_{NN}}$ = 62.4 GeV.

\vfill 
\eject

\appendix
\chapter{Estimation of Systematic Errors}
\markboth{nothing}{\it Estimation of Systematic Errors}

\section{Estimation of Systematic Errors}
If we measure $x_1,~x_2,~.......,x_n$ with uncertainties
$\delta x_1,~\delta x_2,.......,\delta x_n$ and these measured values are
used to compute the function $q = f(x_1,....,x_n)$, then the uncertainty in 
$q$ is given by
\begin{equation}
{\delta q = \sqrt{\sum_i^n ({\frac{\partial q}{\partial x_i}})^2 (\delta x_i)^{2}}}
\label{errorP}
\end{equation}
This is valid if the uncertainties in $x_1,~x_2,~.......,x_n$ are independent 
and random in nature. This is the general formula for propagation of errors.
We have used this formula for the estimation of the uncertainties in the
measurements.

\section{Estimation of Systematic Errors in $E_T$}

For the estimation of $E_T^{had}$, the experimental formula is
\begin{equation}
{E_T^{had} = C_0\sum_{tracks} C_1(ID,p) E_{track}(ID,p) sin\theta}
\label{etHadExpt1}
\end{equation}
The sum includes all tracks from the primary vertex in the range
$0 < \eta < 1$ and for full azimuthal coverage. $C_0$ is the correction factor
defined as
\begin{equation}
{C_0 = \frac{1}{f_{acc}}\frac{1}{f_{p_TCut}}\frac{1}{f_{neutral}}.}
\label{co1}
\end{equation}
The factor $C_1(ID,p)$ is defined as
\begin{equation}
{C_1(ID,p) = f_{bg}(p_T)\frac{1}{f_{notID}}\frac{1}{eff(p_T)}},
\label{c11}
\end{equation}
The factors used in the estimation of $C_0$ and $C_1$ are already discussed
in Chapter 3. Using Eqn.~\ref{errorP} and the correction factors,
we have first estimated the systematic errors in the factors $C_0$ and $C_1$.
The uncertainties in the estimation of $C_0$ and $C_1$ are given by the following
formulae.

\begin{equation}
{\delta C_0 = \sqrt{(\frac{1}{f_{neutral}~f_{p_TCut}^2})^2 (\delta f_{pTCut})^2 + 
(\frac{1}{f_{p_TCut}f_{neutral}^2})^2 (\delta f_{neutral})^2}.}
\label{co1err}
\end{equation}
Here, $f_{acc}$ = 1.

\begin{equation}
{\delta C_1 = \sqrt{(\frac{1}{f_{notID}~eff})^2 (\delta f_{bg})^2 
+(\frac{f_{bg}}{eff ~f_{notID}^2})^2 (\delta f_{notID})^2
+(\frac{f_{bg}}{(eff)^2 f_{notID}})^2 (\delta eff)^2}}.
\label{c11err}
\end{equation}

The uncertainties in $C_0$ and $C_1$ are then used to estimate the error in 
$E_T^{had}$ due to the uncertainties in correction factors. Finally, the 
uncertainty in $E_T^{had}$ is measured by adding in quadrature, the 
uncertainties due to the correction factors and that due to the dynamic cuts.

In a similar fashion, the systematic errors on $E_T^{em}$ and $N_{ch}$ have been
estimated using their experimental formulae. The systematic error in $E_T$ is 
given by
\begin{equation}
{\delta E_T = \sqrt{(\delta E_T^{had})^2 + (\delta E_T^{em})^2}}
\label{errET}
\end{equation}

\section{Estimation of Systematic Errors in Ratios}
We have used observables like $(dE_T^{em}/d\eta)/(dE_T/d\eta)$ and 
$(dE_T/d\eta)/(dN_{ch}/d\eta)$ to study $E_T$ production. The systematic
uncertainties in these ratios have been estimated using the following
formulae.

\begin{equation}
{\delta (E_T^{em}/E_T) = \frac{1}{E_T}\sqrt{(\delta E_T^{em})^2 + (\frac{E_T^{em}}{E_T})^2 ~(\delta E_T)^2}}
\label{errRatio2}
\end{equation}

\begin{equation}
{\delta (E_T/N_{ch}) = \frac{1}{N_{ch}}\sqrt{(\delta E_T)^2 + (\frac{E_T}{N_{ch}})^2
~(\delta N_{ch})^2}}
\label{errRatio1}
\end{equation}

The systematic error in $(dE_T/d\eta)/(0.5 N_{part})$ is estimated using the formula
\begin{equation}
{\delta [(dE_T/d\eta)/(0.5 N_{part})] = \frac{1}{0.5 N_{part}}\sqrt{(\delta 
(dE_T/d\eta))^2 + (\frac{dE_T/d\eta}{N_{part}})^2 ~(\delta N_{part})^2}}
\label{errRatio3}
\end{equation}
The systematic error in $\sigma_{E_T}/E_T$, which is used to study 
$E_T$-fluctuation, is also calculated in a similar way using Eq.~\ref{errorP}.

\section{Estimation of Systematic Errors in $\epsilon_{Bj}\tau$}
The Bjoken formula used to estimate the initial energy density is
\begin{equation}
{\epsilon_{Bj} = <\frac{dE_{T}}{dy}> ~\frac{1}{\tau \pi R^2}}
\label{bjEqn11}
\end{equation}
where, $\tau$ is the formation time and $\pi R^2$ is the transverse overlap 
area of the colliding nuclei. The nuclear radius $R =R_0 A^{1/3}$, where $A$ 
is the mass number of the nuclei. As discussed in Chapter 3, to study the centrality
behavior of $\epsilon_{Bj}\tau$, we have used the relation $A=N_{part}/2$. 
A factor of 1.18 for $\eta \rightarrow y$-phase space conversion is used. Taking
the above aspects into account, the formula for $\epsilon_{Bj}\tau$ simplifies to,
\begin{eqnarray}
\epsilon_{Bj}\tau = <\frac{dE_T}{d\eta}>~\times \frac{1.18}{\pi R_0^2 
(N_{part}/2)^{2/3}}
~= \frac{<dE_T/d\eta>}{N_{part}^{2/3}} \times 0.414
\label{bjEqn12}
\end{eqnarray}

The systematic error in $\epsilon_{Bj}\tau$ is given by
\begin{equation}
{\delta [\epsilon_{Bj}\tau] = \frac{0.414}{N_{part}^{2/3}}\sqrt{(\delta 
 (<dE_T/d\eta>))^2 + (\frac{2}{3}\frac{<dE_T/d\eta>}{N_{part}})^2 ~(\delta N_{part})^2}}
\label{BjErr}
\end{equation}

\vfill 
\eject

\end{document}